\documentclass[11pt]{report}
\usepackage{setspace}
\usepackage{epsfig}
\begin{document}

\thispagestyle{empty}
\begin{center}
\[ \]
\[ \]
\[ \]
\[ \]
\LARGE{A RELATIVISTIC DESCRIPTION OF HADRONIC DECAYS OF THE EXOTIC MESON PI\,1}
\vspace{0.2in}
\[ \]
\[ \]
\[ \]
\[ \]
\Large{{\bf Nikodem~J.~Pop\l awski}}
\vspace{0.1in}
\[ \]
\[ \]
\[ \]
\[ \]
\[ \]
\large{Submitted to the faculty of the University Graduate School} \\
\large{in partial fulfillment of the requirements} \\ 
\large{for the degree} \\ 
\large{Doctor of Philosophy} \\
\large{in the Department of Physics,} \\
\large{Indiana University} \\
\large{October 2004} \\
\end{center}

\newpage
\pagenumbering{roman}
\setcounter{page}{2}
\[ \]
\[ \]
\[ \]

\begin{center}
\large{Accepted by the Graduate Faculty, Indiana University, in partial} \\
\large{fulfillment of the requirements for the degree of Doctor of Philosophy}. \\
\end{center}
\begin{flushright}
\vspace{0.95in}
\large{Adam P. Szczepaniak, Ph.D.} \\
\vspace{0.85in}
\large{Alex R. Dzierba, Ph.D.}
\end{flushright}
\begin{flushleft}
\vspace{0.12in}
\large{Doctoral} \\
\large{Committee}
\end{flushleft}
\begin{flushright}
\vspace{0.02in}
\large{Charles J. Horowitz, Ph.D.} \\
\vspace{0.85in}
\large{J. Timothy Londergan, Ph.D.} \\
\vspace{0.85in}
\large{Stuart L. Mufson, Ph.D.} \\
\end{flushright}
\begin{flushleft}
\large{September 29, 2004}
\end{flushleft}

\newpage
\doublespacing
\[ \]
\[ \]
\[ \]
\[ \]
\[ \]
\[ \]
\[ \]
\[ \]
\[ \]
\[ \]
\[ \]
\[ \]
\[ \]

\begin{flushright}
\large{\it I dedicate this thesis} \\
\large{\it to my Parents and my Grandparents}
\end{flushright}

\newpage
\[ \]
\begin{center}
  \Large{{\bf ACKNOWLEDGMENTS}} 
\end{center}
\[ \]

I would like to express my great appreciation to Prof.~Adam Szczepaniak, the chair of my Ph.D. committee, for being my teacher and advisor, for supervising my research and thesis, and for all his help during my graduate studies at Indiana University.
I would like to thank the members of my Ph.D. committee: Professors Alex Dzierba, Chuck Horowitz, Tim Londergan, and Stuart Mufson for reading the manuscript of my thesis, correcting mistakes, and providing useful and helpful suggestions. I would also like to thank my physics and astronomy professors for sharing their knowledge with me, and the Department of Physics for providing a great atmosphere for me as a graduate student. 
I am thankful to my teacher, Marek Golka, for sparking my interest in physics when I was in high school, and to Dr.~Maciej Swat for introducing me to the physics program at Indiana University. 

I want to thank my best friend, Chris Cox, and his family for their love and help during my time in the United States.
I would like to thank my grandparents, my brothers, and the rest of my family for their love and encouragement.

Most of all, I want to thank my Mom, Bo\.{z}ena, and my Dad, Janusz, for their love and support, for teaching me good things, for leading me to discover my interests and helping me to develop them, and for encouraging me to be always open to new challenges.

\vspace{0.4in}
\begin{flushright}
Nikodem Pop\l awski\\
\end{flushright}
Bloomington, Indiana, USA, 20 X MMIV

\newpage
\vspace{0.4in}
\begin{center}
  \Large{{\bf ABSTRACT}} 
\end{center}
\vspace{0.2in}
\begin{center}
  \large{Nikodem J. Pop\l awski}
\end{center}
\vspace{0.1in}
\begin{center}
  \large{A Relativistic Description of Hadronic Decays of the Exotic Meson $\pi_{1}$}
\end{center}
\vspace{0.2in}
Exotic mesons are striking predictions of quantum chromodynamics that go beyond the quark model. They can provide great insight into understanding phenomena such as asymptotic freedom, confinement, and dynamical symmetry breaking. This work analyzes hadronic decays of exotic mesons, with a focus on the lightest one, the $J^{PC}=1^{-+}\,\,\pi_{1}$, in a fully relativistic formalism. The relativistic spin wave functions of normal and exotic mesons are constructed based on unitary representations of the Poincar\'{e} group. The radial wave functions are obtained from phenomenological considerations of the mass operator. We find that fully relativistic results using Wigner rotations differ significantly from nonrelativistic ones. Moreover, the $S+P$ selection rule is also satisfied in relativistic formalism. Final state interactions do not change these results much.

\tableofcontents
\listoftables
\listoffigures

\chapter{Introduction}
\pagenumbering{arabic}
\setcounter{page}{1}

Atomic nuclei consist of nucleons (protons and neutrons) which are built of quarks and gluons. Quarks and gluons can also combine to form matter called mesons that are not commonly found on Earth, but are naturally created in processes that occur in outer space. Quarks and gluons interact with each other via the strong nuclear force to form hadrons such as nucleons and mesons. This strong force confines quarks and gluons inside hadrons. Exchange of light mesons, in particular $\pi$, $\rho$, $\sigma$, $\omega$, can be used to approximate the effective nucleon-nucleon interaction that binds protons and neutrons in atomic nuclei. In the so-called constituent quark model which describes matter to a good approximation, mesons are regarded as quark-antiquark pairs. In the same model, the proton, the neutron, and other baryons are triplets of quarks. 

Recently, there has been evidence for a new kind of particles dubbed ``exotic mesons". They are called exotic because they have unusual quantum numbers, which are not allowed for quark-antiquark pairs. A theoretical description of the mechanism of exotic meson decays is vital to our understanding of nature of the quark-gluon interaction. A complete model which describes the behavior of exotic mesons should be based on the theory of relativity, in which one deals with velocities close to speed of light. In this work we will attempt to construct a relativistic model of hadronic decays of exotic mesons, in particular the so-called $\pi_{1}$ exotic meson. The main goal is to determine how much relativistic description of the $\pi_{1}$ decays differ from the existing nonrelativistic predictions. 

This thesis is organized as follows.
A brief review of QCD and the constituent quark model will be given in the following sections. In Chapter~2 we will introduce the exotic mesons and review the experimental situation. In Chapter~3 we will discuss the foundations of our model. In Chapter~4 we will present relativistic dynamics for noninteracting and interacting particles, and prepare the spinor framework for relativistic exotic meson decays. In Chapter~5 and~6, a relativistic construction of normal and exotic meson wave function will be given, with a particular emphasis on relativistic effects, including spin-orbit coupling and phase space modification. Decays of normal mesons and of the $\pi_{1}$ exotic meson will be analyzed in Chapter~7 and~8, respectively. In Chapter~9 we will explore the effects caused by residual interactions between the decay products, the so-called final state interactions. The final conclusions will be summarized in Chapter~10.

\section{QCD and the constituent quark model}

The discovery of the pion in 1947 helped to understand the nature of the nucleon-nucleon force. However many other mesons and baryons were found shortly after, which implied that none of these particles were elementary, and that pions were not the quanta of the strong interaction. In order to extend the $N+\pi$ scheme to other hadrons and to account for certain decay patterns such as a long lifetime of the $\Sigma^{-}$, M.~Gell-Mann, T.~Nakano and K.~Nishijima independently proposed in 1953 the concept of strangeness. Strong decays with a short lifetime on the order of $10^{-24}-10^{-22}s$ had the property of conserving strangeness, whereas much longer weak decays violated this conservation.

In 1961, Gell-Mann and independently Y.~Ne'eman introduced the eightfold way, i.e., the SU(3) symmetry ordering of all subatomic particles analogous to the ordering of the chemical elements in the periodic table. It was a generalization of the SU(2) isospin symmetry of the nucleon into an SU(3) with strangeness as a second additive quantum number. A success of this theory was the discovery of the $\Omega^{-}$ baryon. The lightest mesons were also organized into a nonet.

As in the periodic table, a large number of hadrons suggested the existence of substructure. SU(3) symmetry and, in particular, its breaking led Gell-Mann and G.~Zweig to postulate the quark in 1963 \cite{cqm1,cqm2}. They suggested that mesons and baryons are composites of quarks or antiquarks having three flavors: $u$, $d$ and the heavier $s$. Since fractional charges have never been observed, the introduction of quarks was treated more as a mathematical explanation of flavor patterns than as a postulate of an actual physical model. In 1965 O.~W.~Greenberg, M.~Y.~Han and Y.~Nambu introduced the quark property of color charge in order to remedy a statistics problem in constructing the $\Delta^{++}$ wave function. All observed hadrons had to be neutral singlets of the color SU(3) symmetry.

In 1968-69, an experiment at SLAC in which electrons were scattered off protons (deep inelastic scattering) led J.~Bjorken and R.~P.~Feynman to realize that the data could be explained as evidence of small hard cores inside the proton called partons. This picture, however, had several problems, for example that $\sim50\%$ of the proton momentum was not in quarks.

In 1973, a quantum field theory of the strong interaction was formulated by H.~Fritzsch and Gell-Mann, based on the Yang-Mills color SU(3) nonabelian gauge symmetry which is different from the approximate flavor SU(3). In this theory, called quantum chromodynamics (QCD), quarks and massless gluons (quanta of the strong-interaction field) carry a color charge. The structure of QCD is similar to quantum electrodynamics, based on the U(1) symmetry, but much richer. Because gluons carry color charge they can interact with other gluons. The gauge-invariant Lagrangian, unlike that of QED, has cubic and quartic terms in the field potential leading to nonlinear classical equations of motion and interesting topological properties of the vacuum. 

In 1973, D.~Politzer, D.~Gross and F.~Wilczek discovered that QCD has a special property called asymptotic freedom, i.e., at short distances the coupling constant is small enough for perturbation theory to be valid. Unfortunately, at larger distances the coupling constant is on the order of 1 and the QCD Hamiltonian cannot be solved perturbatively. Quantum chromodynamics exhibits at this scale another distinct feature, quark color confinement, so that we may observe only colorless particles. At present we know six quark flavors: $u$, $d$, $s$, $c$, $b$ and $t$. Together with gluons they are included in the Standard Model of fundamental particles and interactions.

The QCD Lagrangian is in practice very difficult to solve. Thus, one is forced to deal with phenomenological models. One such model, the QCD sum-rule approach, was introduced in the late 1970's and applied to describe mesonic properties \cite{sumr1}. This technique was also extended to baryons \cite{sumr2}. The basic idea of QCD sum rules is to match a QCD description of an appropriate momentum-space correlation function with a phenomenological one, and establish a correspondence between hadronic and quark degrees of freedom \cite{sumr3}. This approach provides a connection between the QCD Lagrangian and hadron physics.

Meson and baryon spectra are well described in the constituent quark model (CQM). The CQM Hamiltonians written in the 1960's contained only the kinetic terms and short distance spin-spin interaction. They quite successfully predicted the magnetic moments of the ground state baryons, as long as the magnetic moments of constituent quarks had their classical values.  
The use of a nonrelativistic model was justified for $c\bar{c}$ and $b\bar{b}$ mesons, but did not work very well for light mesons.  
More sophisticated Hamiltonians treated spin-dependent interactions nonperturbatively, and based them on QCD \cite{cqm4}. It was possible to describe hadrons within a unified, relativized quark model with chromodynamics, in which the $q\bar{q}$ interaction is a sum of the Coulomb (one-gluon-exchange) potential and a linear confining term expected from QCD \cite{cqm5,cqm6}. 

High energy hadron-hadron and hadron-nucleus scattering at small and intermediate momentum transfers are well described by assuming that mesons and baryons are bound states of two and three constituent quarks, respectively \cite{cqm7}. Moreover, exclusive processes at small and intermediate momentum transfers agree well with the constituent quark model predictions of elastic and transition form factors. Therefore, the CQM approach provides a relevant description of nonperturbative QCD at low and intermediate momentum scales. Because it is hard to derive the constituent quark model from QCD, one may search for a relation between CQM and sum rules \cite{sumr4}.   

\section{Mesons}

The most convenient formulation of QCD is a constituent representation in which hadron states are dominated by a small number of constituents. It will be assumed that in this representation, interactions that change the number of particles as well as other relativistic effects are small. A natural choice is the Coulomb gauge because it operates with simple degrees of freedom in the nonrelativistic limit. This framework works especially in QED, where for example the hydrogen atom is very well described by the Coulomb potential. The Coulomb gauge will be discussed more in Chapter~3.   

Each hadron is composed of quarks and may contain valence gluons. The simplest configuration of quarks that gives a color singlet is a quark-antiquark pair (meson). The next possibility for a colorless strongly interacting particle is a bound state of three quarks (baryon). One could construct more complicated configurations, for example the so-called pentaquarks, but until now, there has been no strong evidence for objects built of more than three quarks.

Mesons can be classified based on their quantum numbers $J^{PC}$. Here $J$ is total angular momentum of a particle, $P$ is its parity, and $C$ denotes charge conjugation. The total angular momentum is given by
\begin{equation}
{\bf J}={\bf L}+{\bf S},
\label{angmomen1}
\end{equation} 
where $L$ is relative orbital angular momentum of a quark-antiquark pair and $S$ denotes total intrinsic spin of this pair,
\begin{equation}
{\bf S}={\bf S}_{1}+{\bf S}_{2}.
\label{angmomen2}
\end{equation}
Because quarks are fermions with spin 1/2, the values of $S$ can be either 0 or 1. Thus, the values of $J$ are integer and mesons are bosons. The orbital angular momentum $L$ can take any integer positive value or zero, and this determines all possible meson configurations.

Parity, which determines how the sign of the wave function of a particle behaves under a spatial reflection, can be obtained from
\begin{equation}
P=(-1)^{L+1},
\label{parity}
\end{equation}
whereas charge conjugation, describing the particle-antiparticle symmetry (well-defined only for neutral mesons composed of a quark and an antiquark of the same flavor), is given by
\begin{equation}
C=(-1)^{L+S}.
\label{chargeconj}
\end{equation}
In the case of two-boson systems, parity would be given by a slightly different formula, $P=(-1)^{L}$. 

The $u$ flavor is given the third component of isospin $I_{3}=1/2$, whereas the $d$ flavor has $I_{3}=-1/2$. The concept of isospin came from the idea of treating the proton and the neutron as two states of one particle, the nucleon, having two values of the $I_{3}$, like fermions have two values of spin quantized along a fixed axis. Light unflavored mesons, i.e., mesons containing only flavors $u$ and $d$, have $I_{3}$ equal to either 0 or 1. 
For example, the pion with $J^{PC}=0^{-+}$ and $I=1$ has three isospin components: $\pi^{+}$, $\pi^{0}$ and $\pi^{-}$ (isospin triplet), whereas for $I=0$ there is only one component: $\eta$ (singlet).
The $s$ quark has isospin 0; thus strange mesons have isospin $1/2$.
From three quark flavors we can build up nine mesons grouped into an octet and a singlet. 
In Table~\ref{mesons1} we present the classification of light unflavored mesons with respect to the above quantum numbers. In Table~\ref{mesons2} we show the flavor wave functions of the nine pseudoscalar ($J^{PC}=0^{-+}$) and nine vector ($J^{PC}=1^{--}$) mesons.

\begin{table}
\centering
\begin{tabular}{|c||c|c|c|c|}
\hline
L & 0 & 0 & 1 & 1 \\
\hline
S & 0 & 1 & 0 & 1 \\
\hline
J & 0 & 1 & 1 & 0,1,2 \\
\hline
PC & $-+$ & $--$ & $+-$ & $++$ \\
\hline\hline
I=1 & $\pi$ & $\rho$ & b & a \\
\hline
I=0 & $\eta$,\,$\eta'$ & $\omega$,\,$\phi$ & h,\,h' & f,f' \\
\hline
\end{tabular}
\caption{\label{mesons1} Classification of the simplest light unflavored mesons. }
\vspace{0.3in}
\centering
\begin{tabular}{|c||c|c|}
\hline
 & $J^{PC}=0^{-+}$ & $J^{PC}=1^{--}$ \\
\hline\hline
$-u\bar{d}$ & $\pi^{+}$ & $\rho^{+}$ \\
\hline
$d\bar{u}$ & $\pi^{-}$ & $\rho^{-}$ \\
\hline
$\frac{1}{\sqrt{2}}(u\bar{u}-d\bar{d})$ & $\pi^{0}$ & $\rho^{0}$ \\
\hline
$u\bar{s}$ & $K^{+}$ & $K^{\ast+}$ \\
\hline
$d\bar{s}$ & $K^{0}$ & $K^{\ast0}$ \\
\hline
$s\bar{u}$ & $K^{-}$ & $K^{\ast-}$ \\
\hline
$-s\bar{d}$ & $\bar{K}^{0}$ & $\bar{K}^{\ast0}$ \\
\hline
$\frac{1}{\sqrt{6}}(u\bar{u}+d\bar{d}-2s\bar{s})$ & $\eta_{8}$ & $\omega_{8}$ \\
\hline
$\frac{1}{\sqrt{3}}(u\bar{u}+d\bar{d}+s\bar{s})$ & $\eta_{0}$ & $\omega_{0}$ \\
\hline
\end{tabular}
\caption{\label{mesons2} Flavor SU(3) pseudoscalar and vector meson nonets. }
\end{table}

In reality however, we do not observe exact configurations corresponding to $\eta_{0}$ and $\eta_{8}$ but rather their linear combinations known as $\eta$ and $\eta'$. The transformation matrix between both pairs must be orthogonal and thus has one parameter, the mixing angle $\theta$. A similar situation occurs for $\omega_{0}$ and $\omega_{8}$, but in this case the mixing angle is such that the observed particles are given by
\begin{equation} 
\omega = \frac{1}{\sqrt{2}}(u\bar{u}+d\bar{d}),\,\,\phi=s\bar{s}.
\end{equation}

For a particular combination of $L$ and $S$ we have more than one flavor multiplet due to different radial quantum numbers. Normal pions and kaons have $n=1$, and are the lightest, whereas radially excited mesons with higher values of $n$ are heavier. For example, the $\pi(1300)$ is a candidate for a radially excited $\pi$, and the $\rho(1450)$ is a candidate for a radially excited $\rho$. In this work we will deal only with radial ground state mesons because their orbital wave functions should have the same size as the well-known pions and kaons.  
Knowledge of a meson's structure is the first step towards understanding its dynamics, which is responsible for the spectrum of all observed mesons and their decay widths. In fact, all mesons that can decay strongly are not bound states but resonances, and we can study QCD by analyzing how they decay. The quark model description of such decays assumes $q\bar{q}$ pair creation in the gluonic field of the decaying meson. A phenomenological model based on quark-antiquark pair production from the vacuum is referred to as the $^{3}P_{0}$ model \cite{3p0,3p0_1,3p0_2,3p0_3,dec1,dec2}. Although this decay mechanism gives quite successful values for meson widths, it is not rigorously related to QCD which allows $q\bar{q}$ creation only from a gluon.  
The recent benchmark predictions for decay widths of light mesons are given in Refs.~\cite{bench1,bench2}.

\chapter{Exotic mesons}

In the preceding chapter we presented classification of the low-lying unflavored mesons. There are, however, certain combinations of internal meson quantum numbers: spin $J$, parity $P$, and charge conjugation $C$, which are missing in this classification, such as $0^{--},\,0^{+-},\,1^{-+}$, or $2^{--}$. These quantum numbers cannot be obtained from adding the quantum numbers of the quark and the antiquark alone. The corresponding mesons are referred to as the exotic mesons. 

From lattice QCD and model calculations it follows that the lightest exotic mesons may be obtained by adding an extra constituent gluon with $J^{PC}=1^{--}$ to a quark-antiquark system. Such $q\bar{q}g$ states are referred to as hybrid mesons. In this work we will focus on mesons with the $J^{PC}=1^{-+}$ quantum numbers. The isovector multiplet with $J^{PC}=1^{-+}$: $\pi_{1}^{+},\,\pi_{1}^{-},\,\pi_{1}^{0}$, is predicted to be the lightest exotic~\cite{jkm1}. One must emphasize however, that $q\bar{q}g$ states can also have nonexotic quantum numbers. The hybrid components of normal mesons may be important in the mechanism of meson decays. Nonexotic hybrid mesons will be discussed in Chapter~7. 

Hadrons with excited gluonic degrees of freedom may supply new insight into quantum chromodynamics at low energies, where the gluon dynamics should be responsible for phenomena such as color confinement and dynamical symmetry breaking. Therefore, the discovery of exotic mesons is of a great importance. In this chapter we will briefly review the experimental situation in the search for the $\pi_{1}$, and present theoretical predictions for its mass and width. 

\section{Experimental situation}

\begin{figure}[ht]
\centering
\includegraphics[width=2.5in]{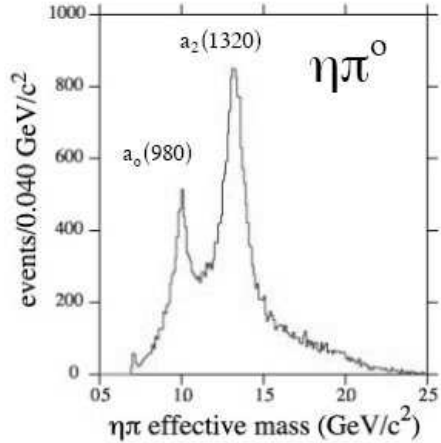}
\caption{\label{a0a2} Distribution of the $\eta\pi^{0}$ effective mass in the reaction $\pi^{-}p\rightarrow\eta\pi^{0}n$ at 18 GeV. The two resonances correspond to the $a_{0}(980)$ and $a_{2}(1320)$ mesons. }
\end{figure}

A resonance can be identified by analyzing the spectrum of its decay products. If a strong, narrow resonance is present, this dependence takes the form of a sharp peak, as shown in Fig.~\ref{a0a2}. In this picture presenting the BNL E852 data of the $\eta\pi^{0}$ channel in the charge exchange reaction $\pi^{-}p\rightarrow\eta\pi^{0}n$ \cite{ex9}, two resonances can be clearly seen. The corresponding particles are the $a_{0}(980)$ and the $a_{2}(1320)$. If a resonance is weakly produced, an amplitude analysis may be required which identifies the resonance by a phase motion of the amplitude as a function of the invariant mass of the decay products.
 
Using such amplitude analyses, several candidates for the $\pi_{1}$ have been recently reported. The $\pi_{1}$(1400) with mass $M=1370\pm16^{+50}_{-30}$ MeV and width $\Gamma=385\pm40^{+65}_{-105}$ MeV, was reported by the E852 Collaboration in the $\eta\pi^{-}$ channel of the process $\pi^{-}p\rightarrow\pi^{-}\eta p$ \cite{ex1,ex2}. This state was confirmed by the Crystal Barrel Collaboration in the $\pi\eta$ channel in the reactions $\bar{p}n\rightarrow\pi^{-}\pi^{0}\eta$ \cite{ex3} and $\bar{p}p\rightarrow\pi^{0}\pi^{0}\eta$ \cite{ex4}. In the $\eta\pi^{0}$ channel, two resonances shown in Fig.~\ref{a0a2} provide benchmarks for the amplitude analysis. A possible signal on the order of 1$\%$ of the dominant $a_{2}(1320)$ have been extracted in both $\eta\pi^{0}$ and $\eta\pi^{-}$ final states. 

\begin{figure}[t]
\centering
\includegraphics[width=3.5in]{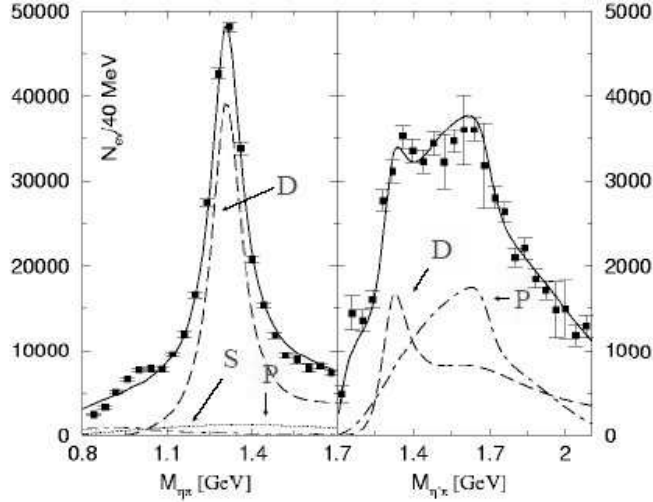}
\caption{\label{eta2} Comparison of the data on $\eta\pi^{-}$ (left) and $\eta'\pi^{-}$ (right) production in the $\pi^{-}p$ interaction at 18 GeV with the results of the amplitude analysis: the D-wave (dashed line), the S-wave (dash-dotted line), and the exotic P-wave (dotted line). The resonance peaks in the D-wave correspond to the $a_{2}(1320)$. }
\end{figure}

A Breit-Wigner (BW) parametrization of the S-wave and D-wave corresponding to the $a_{0}$ and $a_{2}$ mesons in the $\eta\pi^{0}$ and $\eta\pi^{-}$ channels is confirmed by the data, but the resonance interpretation of the P-wave is problematic. First, the left panel of Fig.~\ref{eta2} which represents the $\eta\pi^{-}$ spectrum shows that the signal for the $\pi_{1}(1400)$ is weak. Second, it is impossible to find a selfconsistent set of the BW parameters for the P-wave. As a result, its phase as a function of the invariant mass does not increase over $90^{\circ}$ which is required for a resonance~\cite{mac,ex9}.

The E852 Collaboration has also reported two $\pi_{1}(1600)$ states. One of these has $M=1597\pm10^{+45}_{-10}$ MeV and $\Gamma=340\pm40\pm50$ MeV, and decays into $\eta'\pi$ \cite{ex5}. In this channel, two strong amplitudes are extracted corresponding to the $a_{2}(1320)$ and $\pi_{1}(1600)$, as shown in the right panel of Fig.~\ref{eta2}. The exotic signal is here much stronger, as compared to those in the $\eta\pi^{0}$ and $\eta\pi^{-}$ channels.
 
\begin{figure}[ht]
\centering
\includegraphics[width=3.0in]{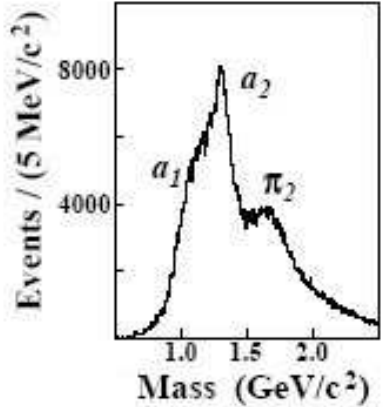}
\caption{\label{pirho} Distribution of the $\rho^{0}\pi^{-}$ effective mass in the reaction $\pi^{-}p\rightarrow\pi^{+}\pi^{-}\pi^{-}p$ at 18 GeV. }
\end{figure}

The other $\pi_{1}(1600)$ state with $M=1593\pm8^{+29}_{-47}$ MeV and $\Gamma=168\pm20^{+150}_{-12}$ MeV, was reported in the $\rho^{0}\pi^{-}$ channel~\cite{ex6,ex7}. In this case, all expected well-known states: $a_{1}(1260)$, $a_{2}(1320)$, and $\pi_{2}(1670)$ are observed, as shown in Fig.~\ref{pirho}~\cite{ex6}. In addition, the amplitude analysis shows that the amplitude with exotic numbers $J^{PC}=1^{-+}$ has structure which is consistent with a resonance at 1.6 GeV decaying into $\rho\pi$. Evidence for the $\pi_{1}(1600)$ has also been reported by the VES collaboration in three channels, $b_{1}\pi$, $\eta'\pi$ and $\rho\pi$ \cite{ex8}, with $M=1.61(2)$ GeV and $\Gamma=0.29(3)$ GeV. The $\pi_{1}(1600)$ signals in all these channels are somewhat different from one another, and therefore further experiments are needed to clarify the nature of these signals. 

Only the $\pi_{1}(1600)$ reported in the $\rho\pi$ channel has a width on the order of 100$-$200 MeV, i.e., comparable to other meson resonances. The broad structures in the $\eta\pi$ and $\eta'\pi$ channels can be accounted for by low-energy rescattering effects~\cite{ex10}. It is possible however, that the $1^{-+}$ exotic meson in the $\eta'\pi$ channel is the same as $\pi_{1}(1600)$ meson seen through its decay into $\rho\pi$. However, at this point this is only speculation~\cite{mac}.

\section{Theoretical predictions}

The mass of the $\pi_{1}$ can be obtained from calculations based on lattice QCD \cite{latt1,latt2,latt3,latt4}. They give values in the region $1.8-2.0$ GeV. Theoretical predictions for this mass are based on various models. The QCD sum-rule predictions vary widely between 1.5 and 2.5 GeV \cite{sum1,sum2,sum3}. The MIT bag model places this mass in the region $1.3-1.8$ GeV \cite{bag1,bag2,bag3}. According to the constituent gluon model, light exotics should have masses in the $1.8-2.2$ GeV range \cite{cgm}. The diquark cluster model predicts the $\pi_{1}$ state at 1.4 GeV \cite{diq}. Finally, the flux tube model predicts the $1^{-+}$ mass similar to the lattice results \cite{IP1,ftm1,CP}.

In Tables~\ref{exotic1} and \ref{exotic2} we show $\pi_{1}$ width predictions calculated using various nonrelativistic models: IKP~\cite{IKP}, CP~\cite{CP} and PSS~\cite{PSS}. At $\pi_{1}$ mass equal to 1.6 GeV, the dominant modes are $\pi b_{1}(1235)$ and $\pi f_{1}(1285)$. For larger values of this mass, the above modes are still dominant, together with the $K{\bar{K}}_{1}(1400)$.

\begin{table}[ht]
\centering
\begin{tabular}{|c||r|r|r|r|}
\hline
$\pi_{1}\rightarrow$ & $\pi b_{1}(1235)$ & $\pi f_{1}(1285)$ & $\pi\rho(770)$ & $\pi\eta(1295)$ \\
\hline\hline
IKP & 59 & 14 & 8 & 1 \\
\hline
PSS & 24 & 5 & 9 & 2 \\ 
\hline
\end{tabular}
\caption{\label{exotic1} Decay widths in MeV of various nonrelativistic models for the $\pi_{1}$ with mass 1.6 GeV. }
\vspace{0.3in}
\centering
\begin{tabular}{|c||r|r|r|}
\hline
$\pi_{1}\rightarrow$ & $\pi b_{1}(1235)$ & $\pi f_{1}(1285)$ & $\pi\rho(770)$ \\
\hline\hline
IKP & 58 & 38 & 16 \\
\hline
PSS & 43 & 10 & 16 \\ 
\hline
CP & 170 & 60 & 5$-$20 \\
\hline
\end{tabular}
\end{table}
\begin{table}[ht]
\centering
\begin{tabular}{|c||r|r|r|r|r|}
\hline
$\pi_{1}\rightarrow$ & $\pi\eta(1295)$ & $K{\bar{K}}_{1}(1400)$ & $K{\bar{K}}_{1}(1270)$ & $\pi\rho(1450)$ & $\eta a_{1}(1260)$ \\
\hline\hline
IKP & 21 & 75 & 19 & 12 & 13 \\
\hline
PSS & 27 & 33 & 7 & 12 & 7 \\ 
\hline
\end{tabular}
\caption{\label{exotic2} Decay widths in MeV of various nonrelativistic models for the $\pi_{1}$ with mass 2.0 GeV. }
\end{table}

It should be noted that in each channel, one outgoing meson has orbital angular momentum $L=0$ (these mesons such as $\pi$ or $K$ are called S-mesons). The other has $L=1$ (these mesons such as $b_{1}$, $f_{1}$ or $K_{1}$ are called P-mesons). The above models favor modes that satisfy the so-called $S+P$ selection rule. It states that a hybrid meson prefers to decay into one S-meson and one P-meson. There is a chance that relativistic corrections could significantly change this situation and favor the $\eta\pi$, $\eta'\pi$, and $\rho\pi$ modes. This work aims to explore that possibility.

Theoretical predictions indicate the importance of searching for the $\pi_{1}$ in the $\rho\pi$, $b_{1}\pi$, and $f_{1}\pi$ channels. They also suggest a search for the $K{\bar{K}}_{1}$ channel. In order to compare these predictions with experiment however, more knowledge of branching ratios is necessary.

\chapter{Dynamical foundations}

In the preceding chapter we described exotic mesons as quark-antiquark-gluon bound states. In order to proceed to their dynamics, we need to know how to obtain the exotic meson wave functions. In the nonrelativistic case, this can be done by using the Born-Oppenheimer approximation which works quite successfully for normal mesons treated as $q\bar{q}$ states. Furthermore, this procedure together with lattice simulations will provide some important information about the composition of the lightest hybrid meson.

The constituent gluon plays a central role in the structure of an exotic meson. As the photon in QED, the gluon needs to be described in a particular gauge. A natural framework for introducing the constituent quark model and providing insights into calculating meson decays is the Coulomb gauge, which is free of unphysical degrees of freedom and has a good nonrelativistic quantum-mechanical limit. Relativization will be accomplished by using a relativistic phase space and transforming quark-antiquark states under Lorentz boosts. 

The model presented in this work is microscopic, i.e., at the level of quarks and gluons. However, mesons interact with each other via meson exchange. This force may contribute significantly to the dynamics of exotic mesons, and the quantitative analysis of this problem will be the subject of Chapter~9.

We will begin the present chapter with the Born-Oppenheimer approximation. Then we will review the Coulomb gauge for QCD and describe the Coulomb gauge picture of normal and hybrid mesons. Finally we will proceed to relativistic effects in the Coulomb-gauge constituent quark model.    

\section{Heavy quarkonia and the Born-Oppenheimer approximation}

Lattice simulations are quite successful in predicting mass spectra for mesons and baryons. Hadronic decays, however, provide the real difficulty for such estimates. Thus we are left with phenomenological models of the coupling between mesons and hybrids. 
In general, there are two approaches for describing hadronic decays of hybrid mesons. The first regards a hybrid as a quark-antiquark state with an additional constituent gluon \cite{qqg1}. Such a meson would decay through gluon dissociation into a $q\bar{q}$ pair \cite{qqg2,qqg3}. The second approach assumes that a hybrid is a quark-antiquark pair moving on an adiabatic surface generated by an excited gluonic flux-tube \cite{IP1,IP2}. In this case a hybrid meson would decay because of phenomenological pair production described by the $^{3}P_{0}$ model \cite{IKP,KI,CP,PSS}. Recently, an extended version of the flux tube model has been introduced \cite{ases3}.

If constituent quarks composing mesons are heavy, such systems (heavy quarkonia) can be studied using the Born-Oppenheimer approximation \cite{jkm1,jkm2,jkm3}. In this approach it is assumed that formation of gluonic field distributions decouples from the dynamics of the slowly moving quarks, and therefore hadronic decays can be described within nonrelativistic quantum mechanics. This approximation can be justified for light quarks, because dynamical chiral symmetry breaking leads to massive consituent quarks.

In the Born-Oppenheimer model, a hybrid meson is treated analogously to a diatomic molecule in which the heavy quarks correspond to the nuclei and the gluon field corresponds to the electrons. 
Initially, a quark and an antiquark are treated as spatially fixed color sources and this determines the glue energy levels as a function of the $q\bar{q}$ separation. Each energy level defines an adiabatic potential $V_{q\bar{q}}(r)$. 
The quark motion is restored by solving the radial Schr\"{o}dinger equation for each of these potentials. 

The lowest static potential gives a normal meson spectrum, whereas the excited potentials lead to hybrid mesons. The static potentials are determined from lattice simulations.
The gluonic configurations can be classified according to symmetries of the $q\bar{q}$ ``molecule''. The strong interaction is invariant under rotations around the $q\bar{q}$ axis, a reflection in a plane containing the pair, and with respect to the product $PC$. Each configuration can be thus labeled by the corresponding eigenvalues, denoted by $\Lambda$ (the magnitude of the projection of the total gluon angular momentum onto the molecular axis), $Y=\pm1$ (the sign of this projection), and $PC=\pm1$, respectively.

\begin{figure}[t]
\centering
\includegraphics[width=2.5in]{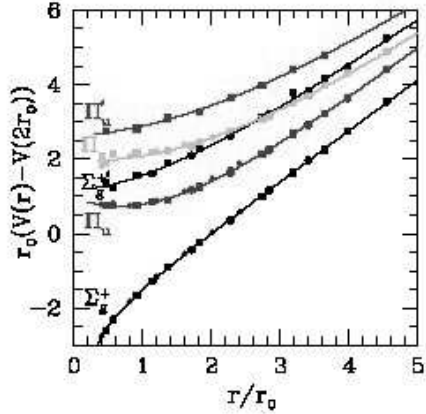}
\caption{\label{adiabat} The static quark potential $V_{\Sigma_{g}^{+}}(r)$ and selected gluonic excitations. }
\end{figure}

States with $\Lambda=0,1,2,...$ are denoted by $\Sigma,\Pi,\Delta,...$, respectively. States which are even (odd) under the combined $PC$ operation are denoted by $g$ ($u$). Lattice simulations for the ground state configuration and the lowest gluonic excitations are shown in Fig.~\ref{adiabat}~\cite{jkm1}. The parameter $r_{0}$ is on the order of 0.5 fm. In the ground state (normal meson) $\Lambda=0$ ($\Sigma_{g}^{+}$) and for the first excited state $\Lambda=1$ ($\Pi_{u}$).

If the gluon is in a relative S-wave with respect to a $q\bar{q}$ pair, it has $PC=+1$. Lattice results show, however, that the lowest excited configuration has the gluon with $PC=-1$ so the gluon orbital angular momentum with respect to a $q\bar{q}$ pair must be odd. The simplest choice is $L=1$. Therefore, in Chapter~5, in order to construct the $\pi_{1}$ spin wave function we will couple a transverse gluon to the $q\bar{q}$ state with the $\rho$ quantum numbers, in a relative P-wave.

\section{The Coulomb gauge}

In quantum electrodynamics, the electromagnetic field arises naturally from demanding an invariance of the action under the local gauge U(1) transformation. If $\phi$ is a complex scalar field then the corresponding Lagrangian
\begin{equation}
L=(\partial_{\mu}\phi)(\partial^{\mu}\phi)-m^{2}\phi^{\ast}\phi
\end{equation}
is invariant under the transformation
\begin{equation}
\phi\rightarrow e^{-i\Lambda}\phi,\,\,\,\,\phi^{\ast}\rightarrow e^{i\Lambda}\phi^{\ast},
\end{equation}
where $\Lambda$ is an arbitrary real constant. If $\Lambda$ depends on the spacetime coordinates, however, then the derivative of the field does not transform covariantly, i.e., in the same way as $\phi$. In order to remedy this problem one introduces the covariant derivative (like in the general theory of relativity)
\begin{equation}
D_{\mu}=\partial_{\mu}+ieA_{\mu},
\end{equation}
where $e$ is a real constant (the electric charge) and $A_{\mu}$ is the electromagnetic potential. This potential must transform according to
\begin{equation}
A_{\mu}\rightarrow A_{\mu}+\frac{1}{e}\partial_{\mu}\Lambda.
\label{gauge1}
\end{equation}   
The quantity
\begin{equation}
F_{\mu\nu}=\partial_{\mu}A_{\nu}-\partial_{\nu}A_{\mu}
\end{equation}
is the electromagnetic field tensor and its six nonzero components correspond to the fields ${\bf E}$ and ${\bf B}$. The simplest Lagrangian built up from the gauge-invariant quantities is thus
\begin{equation}
L=-\frac{1}{4}F_{\mu\nu}F^{\mu\nu},
\end{equation}
and its variation with respect to $A_{\mu}$ leads to the Maxwell equations in vacuum.
Because of the gauge invariance we may introduce one constraint on the components of the field potential. In the Coulomb gauge this constraint is given by
\begin{equation}
\nabla\cdot{\bf A}=0.
\end{equation}

In quantum chromodynamics, the local gauge transformations form the SU(3) group
\begin{equation}
\phi_{i}\rightarrow S_{ij}\phi_{j}=(e^{\frac{i}{2}\lambda^{k}\Lambda^{k}})_{ij}\phi_{j},
\end{equation}
where $i,j=1,2,3$ and $k=1..8$. The matrices $\lambda$ are the hermitian and traceless generators of SU(3) (Gell-Mann matrices). In this case the expression for the covariant derivative is given by
\begin{equation}
D_{\mu}=\partial_{\mu}+igA_{\mu}^{k}\frac{\lambda^{k}}{2},
\end{equation}
whereas the potential transforms according to
\begin{equation}
A_{\mu}^{k}\frac{\lambda^{k}}{2}\rightarrow S\,A_{\mu}^{k}\frac{\lambda^{k}}{2}\,S^{-1}-\frac{i}{g}(\partial_{\mu}S)S^{-1}. 
\label{gauge2}
\end{equation}  
The gauge-invariant field tensor is given by
\begin{equation}
G_{\mu\nu}^{a}=\partial_{\mu}A_{\nu}^{a}-\partial_{\nu}A_{\mu}^{a}+gf^{abc}A_{\mu}^{b}A_{\nu}^{c},
\label{field}
\end{equation} 
where $f_{abc}$ are the SU(3) structure constants, and the simplest field Lagrangian is thus
\begin{equation}
L=-\frac{1}{4}G_{\mu\nu}^{a}G^{\mu\nu}_{a}.
\end{equation}      

The chromoelectric field corresponds to the $G_{0\alpha}$ components of the field tensor
\begin{equation}
{\bf E}^{a}=-\dot{{\bf A}}^{a}-\nabla A^{0a}+gf^{abc}A^{0b}{\bf A}^{c},
\end{equation}
and satisfies the Gauss law
\begin{equation}
\nabla\cdot{\bf E}^{a}+gf^{abc}{\bf A}^{b}\cdot{\bf E}^{c}=g\rho^{a}_{q}.
\end{equation}
Here $\rho^{a}_{q}=\psi^{\dag}\frac{\lambda^{a}}{2}\psi$ is the quark color charge density. Introducing the covariant derivative in the adjoint representation
\begin{equation}
D^{ab}_{\mu}=\delta^{ab}\partial_{\mu}+igT_{ab}^{c}A^{c}_{\mu},
\end{equation}
where $T^{c}_{ab}=if^{cab}$, leads to
\begin{equation}
{\bf D}^{ab}\cdot{\bf E}^{b}=g\rho^{a}_{q}.
\end{equation}
If ${\bf E}_{\parallel}=-\nabla\phi$ is the longitudinal part of of the chromoelectric field then we obtain
\begin{equation}
-({\bf D}^{ab}\cdot\nabla)\phi^{b}=g\rho^{a},
\end{equation} 
where $\rho=\rho_{q}+\rho_{g}$ is a total color charge density. Here the transverse gluon color charge density is given by
\begin{equation}
\rho^{a}_{g}=f^{abc}{\bf E}_{\perp}^{b}\cdot{\bf A}^{c},
\end{equation}
where ${\bf E}_{\perp}$ is the transverse part of the chromoelectric field.
Combining the above equations leads to:
\begin{eqnarray}
& & \phi^{a}=\frac{1}{\nabla\cdot{\bf D}}g\rho^{a}, \nonumber \\
& & A^{0a}=\frac{1}{\nabla\cdot{\bf D}}(-\nabla^{2})\frac{1}{\nabla\cdot{\bf D}}g\rho^{a}. 
\end{eqnarray}
The last equation results in the instantaneous nonabelian Coulomb interaction Hamiltonian
\begin{equation}
H_{C}=\frac{1}{2}\int d^{3}x\,d^{3}y\,\rho^{a}(x)K_{ab}(x,y;A)\rho^{b}(y),
\end{equation}
where 
\begin{equation}
K_{ab}(x,y;A)=\langle x,a|\frac{g}{\nabla\cdot{\bf D}}(-\nabla^{2})\frac{g}{\nabla\cdot{\bf D}}|y,b\rangle.
\end{equation}
After quantization, the field ${\bf E}_{\perp}$ becomes the momentum conjugate to the vector potential.
The confining nonabelian Coulomb potential will be represented in the diagrams below by the dashed lines.
More details related to properties of the Coulomb-gauge QCD can be found in Ref.~\cite{ases1}. 

A simple phenomenological picture of hadrons and their decays in terms of quantum mechanical wave functions emerges naturally in a fixed gauge approach. In the Coulomb gauge, for example, the precursor of flux tube dynamics originates from the nonabelian Coulomb potential, which also determines the quark wave functions~\cite{ases1,ases2}. The string couples to a $q\bar{q}$ pair via transverse gluon emission and absorption and such a coupling carries the $^3S_1$ quantum numbers. 

\begin{figure}[h]
\centering
\includegraphics[width=2.5in]{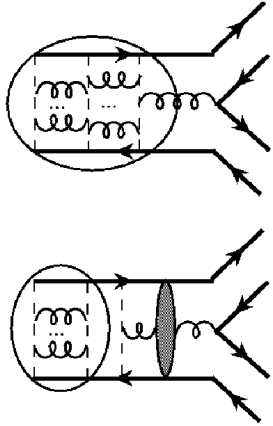}
\caption{\label{decays} Strong decay of a hybrid meson (top) and a normal meson (bottom). }
\end{figure}

In a description of decays based on Coulomb gauge quantization it is necessary to include the hybrid quark-antiquark-gluon configurations, since they appear as intermediate states in the decay of mesons. If such hybrid states also exist as asymptotic states, they would provide insight into the dynamics of confined gluons \cite{N}. Fig.~\ref{decays} shows diagrams corresponding to strong decays of $q\bar{q}g$ hybrid mesons (top) and $q\bar{q}$ normal mesons (bottom). The gluons connecting the Coulomb lines represent formation of the flux tube, e.g. the gluon string in the ground state. The overall initial state is enclosed by the solid oval. In the lower diagram the hybrid meson state appears as an intermediate state in a normal meson decay, which is assumed to proceed via mixing of a $q\bar{q}$ pair with a virtual excitation of a gluonic string and its subsequent decay.

\begin{figure}[t]
\centering
\includegraphics[width=2.5in]{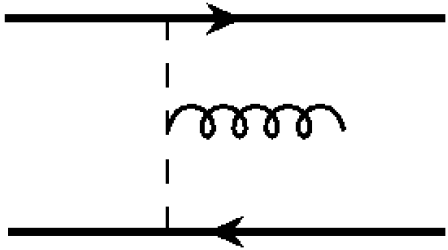}
\caption{\label{3body} A three-body potential between a transverse gluon and a static $q\bar{q}$ source. }
\end{figure}

In the Coulomb gauge the quantum numbers of the gluonic states can be associated with those of a transverse gluon in the presence of the static $q\bar{q}$ source. This is because transverse gluons are dressed~\cite{ases1,rein}, and on average behave like the constituent particles with the effective mass $m_g\sim500\mbox{ MeV}$~\cite{mg1,mg2}. Thus low-energy excited gluonic states are expected to have a small number of transverse gluons. The flux tube itself is expected to emerge from the strong coupling of transverse gluons to the Coulomb potential. The transverse gluon wave function can be obtained by diagonalizing the net quark-antiquark-gluon interactions shown in Fig.~\ref{3body} (in addition to the gluon kinetic energy).

The above three-body interaction plays an essential role in the dynamics of hybrid mesons. A transverse gluon has a gradient coupling to the Coulomb potential. Thus the P-wave transverse gluon receives no energy shift from this coupling and the energy of the S-wave gluon state is increased. In the Coulomb gauge picture, the shift of the S-wave state via this three-body interaction may be the cause of the inversion of the S$-$P levels seen on the lattice. Using only two-body potentials between quarks and gluons leads to the S-wave gluon in the lowest energy excited state, which disagrees with lattice data \cite{ases4}.

\section{Relativistic effects}

Another and very important issue is the question of relativistic effects. Even though a simple nonrelativistic description appears to be quite successful in predicting decay widths of mesons as heavy as $\sim1-2\,$GeV, the presence of light quarks raises the question of validity of this description. It has been shown that relativistic effects for hadronic form factors may be significant \cite{rel1,rel2,rel3,rel4}. It is possible that they are responsible for the discrepancy between experimental data and theoretical predictions. This work will try to estimate the size of relativistic effects applied to the $\pi_{1}$ decays.

In order to calculate relativistic decay amplitudes exactly in the Coulomb gauge, one needs to find the fundamental quantities (dynamical generators of the Poincar\'{e} group) in terms of the chromodynamical fields. This problem, however, is very difficult to state and in this work we will not solve the Coulomb gauge QCD Hamiltonian to obtain meson wave functions. Instead, we will use the general transformation properties under the remaining kinematical symmetries (rotations and translations) to construct the states. 

The relativistic meson and hybrid spin wave functions will be elements of irreducible, unitary representations of the Poincar\'{e} group for noninteracting particles. The interaction between particles should enter the dynamics by finding a new mass operator (the Bakamjian-Thomas model). Because an exact form of the strong potential between quarks is unknown, we will employ instead a simple parametrization of the meson orbital wave function. This is clearly an approximation which cannot be avoided without solving dynamical equations for the boost generators~\cite{rel2,rel3}.

\chapter{Relativistic dynamics}

In this chapter we will review the Lorentz group in vector and spinor representations, which is a framework for the ten fundamental quantities describing the dynamics of a system of noninteracting particles. Then we will introduce the Bakamjian-Thomas construction of these generators for interacting particles. Finally we will discuss a Wigner rotation, which plays an essential role in constructing relativistic and covariant spin wave functions for mesons and hybrids.   

\section{The Lorentz group}

The principle of relativity in the framework of general relativity requires that physical laws must be invariant under all transformations of the coordinates. Gravitational fields are automatically included if one deals with curvilinear coordinates, however, they are important only for large-scale phenomena. Yet in the physics of elementary particles, the curvature of the spacetime is small and can be neglected. Therefore one needs to deal only with the metric tensor of a flat spacetime.
In this case the principle of relativity requires that physical laws must be invariant under transformations from one inertial frame to another. Such transformations are called inhomogeneous Lorentz transformations, and the coordinates transform linearly according to
\begin{equation}
x'^{\mu}=\Lambda^{\mu}_{\,\,\nu}x^{\nu}+a^{\mu},
\label{Lor1}
\end{equation}
where $\Lambda^{\mu}_{\,\,\nu}$ is the Lorentz matrix, and $a^{\mu}$ is a constant four-vector. From the invariance of the finite interval $x^{'\mu}x^{'}_{\mu}=x^{\mu}x_{\mu}$ for $a^{\mu}=0$, it follows that the Lorentz matrix must be orthogonal,
\begin{equation}
\Lambda^{\mu}_{\,\,\rho}\Lambda^{\nu}_{\,\,\lambda}g_{\mu\nu}=g_{\rho\lambda}, \
\label{Lor2}
\end{equation}
or $\Lambda\Lambda^{T}=1$. Thus its determinant can be either 1 or -1.

All inhomogeneous Lorentz transformations can be divided into four categories, depending on the signs of the determinant of $\Lambda$ and the component $\Lambda^{0}_{\,\,0}$. We will be interested in the proper Lorentz transformations, having both signs positive. They can be built up from infinitesimal transformations involving boosts, rotations and translations, but cannot involve reflections. The proper inhomogeneous Lorentz transformations are continuous and form a Lie group called the Poincar\'{e} group. If $a^{\mu}=0$ then this group is called the Lorentz group. The principle of relativity will be satisfied if physical laws are invariant under infinitesimal transformations given by ($\ref{Lor1}$) in which
\begin{equation}
\Lambda^{\mu}_{\,\,\nu}=\delta^{\mu}_{\nu}+\omega^{\mu}_{\,\,\nu},
\label{Lor3}
\end{equation}
where $\omega^{\mu}_{\,\,\nu}$ are infinitesimal quantities that are antisymmetric $\omega_{\mu\nu}=-\omega_{\nu\mu}$.
This property results from the orthogonality of $\Lambda^{\mu}_{\,\,\nu}$.

Rotations are orthogonal transformations of the coordinates mixing their spatial components and form a subgroup O(3) of the Lorentz group.
They are described by a Lorentz matrix with $\Lambda^{0}_{\,\,0}=1$, $\Lambda^{0}_{\,\,i}=0$ and $\Lambda^{i}_{\,\,0}=0$. The remaining components are functions of three angles which may be chosen as the Eulerian angles of a rigid body. For example, a rotation by the angle $\phi$ about the z-axis corresponds to the Lorentz matrix,
\begin{equation} 
R_{z}(\phi)=\left( \begin{array}{cccc}
1 & 0 & 0 & 0 \\
0 & \cos\phi & \sin\phi & 0 \\
0 & -\sin\phi & \cos\phi & 0 \\
0 & 0 & 0 & 1 \end{array} \right),
\end{equation}
and similarly for two other axes.
For an infinitesimal angle of rotation we can write only linear terms in $\phi$,
\begin{equation}
R_{z}(\delta\phi)=1+iJ_{z}\delta\phi,
\end{equation}
and the passage to a finite rotation is given by
\begin{equation}
R_{z}(\phi)=\lim_{N\rightarrow\infty}(1+iJ_{z}\frac{\phi}{N})^{N}=e^{iJ_{z}\phi}.
\end{equation}
The matrix $J_{z}$ is called the generator of the rotation about the z-axis.
The rotation group is nonabelian, i.e. $[R_{x},R_{y}]\neq[R_{y},R_{x}]$. The corresponding generators satisfy the Lie algebra,
\begin{equation} 
[J_{i},J_{k}]=i\epsilon_{ikl}J_{l}.
\label{Lie1}
\end{equation}

Boosts are described by a Lorentz matrix with $\Lambda^{i}_{\,\,j}=0$. For example, a boost in the z-direction with velocity $v$ corresponds to
\begin{equation}
B_{z}(\psi)=\left( \begin{array}{cccc}
\cosh\psi & 0 & 0 & \sinh\psi \\
0 & 1 & 0 & 0 \\
0 & 0 & 1 & 0 \\
\sinh\psi & 0 & 0 & \cosh\psi \end{array} \right),
\end{equation}
where $\cosh\psi=\gamma=(1-v^{2})^{-1/2}$.
Writing 
\begin{equation}
B_{z}(\psi)=1+iK_{z}(\psi)
\end{equation}
for an infinitesimal value of $\psi$ leads to the Lie algebra of the homogenous Lorentz group (rotations and boosts), given by ($\ref{Lie1}$) and
\begin{eqnarray}
& & [J_{i},K_{j}]=i\epsilon_{ijk}K_{k}, \nonumber \\
& & [K_{i},K_{j}]=-i\epsilon_{ijk}J_{k}. 
\label{Lie2}
\end{eqnarray}

An arbitrary vector $(p^{0},{\bf p})$ transforms under the boost with velocity ${\bf v}$ according to
\begin{eqnarray}
& & p'^{0}=\gamma(p^{0}+{\bf v}\cdot{\bf p}), \nonumber \\
& & {\bf p}'_{\parallel}=\gamma({\bf p}_{\parallel}+{\bf v}p^{0}), \nonumber \\
& & {\bf p}'_{\perp}={\bf p}_{\perp},
\end{eqnarray}
where ${\bf p}_{\parallel}+{\bf p}_{\perp}={\bf p}$ and ${\bf p}_{\parallel}=({\bf v}\cdot{\bf p}){\bf v}/v^{2}$. Parametrization by $M$, ${\bf P}$ and $E$ such that ${\bf v}\gamma={\bf P}/M$ and $E=\sqrt{M^{2}+{\bf P}^{2}}$ leads to the following transformation laws:
\begin{eqnarray}
& & p'^{0}=p^{0}\frac{E}{M}+\frac{{\bf p}\cdot{\bf P}}{M}, \nonumber \\
& & {\bf p}'={\bf p}+p^{0}\frac{{\bf P}}{M}+\frac{({\bf p}\cdot{\bf P}){\bf P}}{M(E+M)}. 
\label{vecboost}
\end{eqnarray}

If we introduce the four-dimensional antisymmetric generators $J_{\mu\nu}$ defined by
\begin{equation} 
J_{ij}=-\epsilon_{ijk}J_{k},\,\,\,\,J_{0i}=K_{i},
\end{equation}
then the Lie algebra may be written as one equation,
\begin{equation}
[J_{\mu\nu},J_{\rho\sigma}]=i(g_{\nu\rho}J_{\mu\sigma}+g_{\mu\sigma}J_{\nu\rho}-g_{\mu\rho}J_{\nu\sigma}-g_{\nu\sigma}J_{\mu\rho}).
\label{Lie3}
\end{equation}
The components $J_{\mu\nu}$ are proportional to the quantities $\omega_{\mu\nu}$ given in ($\ref{Lor3}$), and infinitesimal constants of proportionality are either $\delta\phi$ or $\delta\psi$.

The above operators may be expressed as differential operators instead of matrices. This will enable us to introduce the generators of translations. For an infinitesimal rotation about the z-axis we have
\begin{eqnarray} 
& & J_{z}f(x,y,z)=\lim_{\phi\rightarrow 0}i\frac{f(x',y',z')-f(x,y,z)}{\phi}= \nonumber \\
& & =\lim_{\phi\rightarrow 0}i\frac{f(x+y\phi,y-x\phi,z)-f(x,y,z)}{\phi}=i(y\frac{\partial}{\partial x}-x\frac{\partial}{\partial y})f.
\end{eqnarray}
For a boost in the z-direction we obtain similarly,
\begin{equation} 
K_{z}=i(t\frac{\partial}{\partial z}+z\frac{\partial}{\partial t}).
\end{equation}
One can easily check that such defined operators satisfy the Lie algebra of the homogenous Lorentz group, ($\ref{Lie1}$) and $\ref{Lie2}$).
For translations we can write
\begin{equation}
T_{z}f(t,x,y,z)=\lim_{\zeta\rightarrow 0}i\frac{f(t,x,y,z+\zeta)-f(t,x,y,z)}{\zeta}=i\frac{\partial}{\partial z}f,
\end{equation}
and this leads to
\begin{eqnarray}
& & [T_{\mu},T_{\nu}]=0, \nonumber \\
& & [T_{\rho},J_{\mu\nu}]=i(T_{\nu}g_{\mu\rho}-T_{\mu}g_{\nu\rho}).
\label{Lie4}
\end{eqnarray}
Equations ($\ref{Lie3}$) and ($\ref{Lie4}$) constitute the complete Lie algebra of the Poincar\'{e} group.

\section{The ten fundamental quantities}

Another requirement for a dynamical theory is that the equations of motion should be expressible in Hamiltonian form. This is necessary in order to make a transition from classical to quantum theory. The dynamics of a system is described by quantities called dynamical variables, which for particles can be taken as their coordinates and momenta, and for fields as their four-coordinates in spacetime. Any two dynamical variables $\xi$ and $\eta$ must have a Poisson bracket $[\xi,\eta]$, and its form must not change under an infinitesimal Lorentz transformation. From this it follows that each dynamical variable $\xi$ will change according to
\begin{equation}
\xi'=\xi+[\xi,F],
\end{equation}
where $F$ is an infinitesimal dynamical variable independent of $\xi$ and depends on the change in the coordinate system. Thus it must depend linearly on the infinitesimal quantities $\omega_{\mu\nu}$ and $a_{\mu}$. Therefore we can write
\begin{equation}
F=-P^{\mu}a_{\mu}+\frac{1}{2}M^{\mu\nu}\omega_{\mu\nu},
\label{Gen1}
\end{equation}
where $P^{\mu}$ and $M^{\mu\nu}=-M^{\nu\mu}$ are finite dynamical variables called the fundamental quantities \cite{Dirac}.

Each of the ten fundamental quantities is associated with an infinitesimal transformation of the Poincar\'{e} group. $P_{0}$ is the total energy of the system and is related to a translation in time, $P_{i}$ form the three-dimensional total momentum and are related to translations in space, and $M_{ij}$ correspond to the total angular momentum and are related to three-dimensional rotations. The quantities $M_{0i}$ correspond to boosts but do not form any additive constants of motion. From the commutation relations between infinitesimal transformations ($\ref{Lie3}$) and ($\ref{Lie4}$), it follows that the Poincar\'{e} algebra is given by:
\begin{eqnarray}
& & [P_{\mu},P_{\nu}]=0, \nonumber \\
& & [P_{\rho},M_{\mu\nu}]=g_{\nu\rho}P_{\mu}-g_{\mu\rho}P_{\nu}, \nonumber \\
& & [M_{\mu\nu},M_{\rho\sigma}]=-g_{\mu\rho}M_{\nu\sigma}-g_{\nu\sigma}M_{\mu\rho}+g_{\nu\rho}M_{\mu\sigma}+g_{\mu\sigma}M_{\nu\rho}.
\label{Gen2}
\end{eqnarray}
In order to describe a dynamical system one must find a solution of these equations, i.e. $P_{\mu}$ and $M_{\mu\nu}$. This is the central issue in relativistic quantum mechanics.

A simple solution of ($\ref{Gen2}$) for a single point particle is given by
\begin{equation}
P_{\mu}=p_{\mu},\,\,\,\,M_{\mu\nu}=q_{\mu}p_{\nu}=q_{\nu}p_{\mu},
\label{Gen3}
\end{equation}
where $q_{\mu}$ are the coordinates of a point in spacetime and $p_{\mu}$ are their conjugate momenta,
\begin{equation} 
[q_{\mu},q_{\nu}]=0,\,\,\,\,[p_{\mu},p_{\nu}]=0,\,\,\,\,[p_{\mu},q_{\nu}]=g_{\mu\nu}.
\end{equation}
One usually works with dynamical variables referring to a particular instant of time. The fundamental quantities associated with transformations that leave this instant invariant (spatial translations and rotations) appear to be simple, whereas the remaining $P_{0}$ and $M_{i0}$ called Hamiltonians are not.
Without loss of generality we may take $q_{0}=0$. Therefore $p_{0}$ no longer has a meaning. But we can modify formulae ($\ref{Gen3}$) in order to eliminate $p_{0}$ from them. Let us take
\begin{eqnarray} 
& & P_{\mu}=p_{\mu}+\lambda_{\mu}(p^{\rho}p_{\rho}-m^{2}), \nonumber \\
& & M_{\mu\nu}=q_{\mu}p_{\nu}-q_{\nu}p_{\mu}+\lambda_{\mu\nu}(p^{\rho}p_{\rho}-m^{2}),
\end{eqnarray}
where $m$ is a constant, with an appropriate choice of $\lambda_{\mu}$ and $\lambda_{\mu\nu}$. This leads to
\begin{eqnarray}
& & P_{i}=p_{i},\,\,\,\,M_{ij}=q_{i}p_{j}-q_{j}p_{i}, \nonumber \\
& & P_{0}=\sqrt{p_{j}p_{j}+m^{2}},\,\,\,\,M_{i0}=q_{i}\sqrt{p_{j}p_{j}+m^{2}}.
\label{Gen4}
\end{eqnarray}
These are the fundamental quantities for a particle with mass $m$ in the so-called instant form of dynamics. There are two other forms: the point form and the front form, but the quantities appearing there are not as intuitive as in the instant form \cite{Dirac}.

If for a single particle we replace $q_{i}$ by the operators $i\frac{\partial}{\partial p_{i}}$ and $M_{i0}$ by the so-called velocity operators $V_{i}=\frac{1}{2}(q_{i}H+Hq_{i})$, where $H=P_{0}$, we will transit to quantum dynamics. In vector notation we can write
\begin{eqnarray} 
& & {\bf P}={\bf p},\,\,\,\,{\bf M}={\bf q}\times{\bf p}, \nonumber \\
& & H=\sqrt{m^{2}+{\bf p}^{2}},\,\,\,\,{\bf V}=\frac{1}{2}({\bf q}H+H{\bf q}).
\label{BT1}
\end{eqnarray}
For two noninteracting particles the ten operators are given by sums,
\begin{eqnarray} 
& & {\bf P}={\bf p}_{1}+{\bf p}_{2},\,\,\,\,{\bf M}={\bf q}_{1}\times{\bf p}_{1}+{\bf q}_{2}\times{\bf p}_{2}, \nonumber \\
& & H=\sqrt{m^{2}_{1}+{\bf p}^{2}_{1}}+\sqrt{m^{2}_{2}+{\bf p}^{2}_{2}},\,\,\,\,{\bf V}={\bf q}_{1}\sqrt{m^{2}_{1}+{\bf p}^{2}_{1}}+{\bf q}_{2}\sqrt{m^{2}_{2}+{\bf p}^{2}_{2}}.
\label{BT2}
\end{eqnarray}
The expression
\begin{equation}
M=\sqrt{H^{2}-{\bf P}^{2}}
\label{mop}
\end{equation}
is the mass operator of the system viewed as a single entity, and commutes with all ten operators ($\ref{BT2}$).

The last topic we will discuss in this section is related to the Casimir operators of the Lorentz group, i.e., the quantities that commute with all ten fundamental quantities $P_{\mu}$ and $M_{\mu\nu}$. Using the equations of the Poincar\'{e} algebra ($\ref{Gen2}$) one can show that the only operators that have this property are
\begin{equation} 
C_{1}=P^{\mu}P_{\mu},\,\,\,\,C_{2}=W^{\mu}W_{\mu},
\end{equation}
where $W^{\mu}$ is the Pauli-Lubanski pseudovector,
\begin{equation} 
W^{\mu}=\frac{1}{2}\epsilon^{\mu\nu\lambda\rho}M_{\nu\lambda}P_{\rho}.
\end{equation}
The first Casimir operator is just equal to $M^{2}$, and is related to the mass of a system viewed as a single entity, or to the mass of a particle.
In the rest frame of a massive particle with mass $m$, the operator $C_{2}$ behaves like the square of the angular momentum operator ${\bf M}^{2}$, and for spin $s$ has $2s+1$ eigenvalues. For a massless particle, however, there exist only two eigenvalues $\pm s$ \cite{Wigner}.

\section{Spinor representation of the Lorentz group}

We defined the Lorentz group via the transformation properties of the coordinates $x^{\mu}$. Quantities that transform under the Lorentz tranformations in the same way as the coordinates are called vectors, and the matrix $\Lambda^{\mu}_{\,\,\nu}$ is referred to as the vector representation of the Lorentz group. This representation is suitable when dealing with vector particles having integer values of spin. 
However, for particles with spin $1/2$ (fermions) it is much more useful to introduce the spinor representation of the Lorentz group \cite{Ryder}. This will be the subject of the present section.

Consider the group SU(2), consisting of $2\times2$ unitary matrices with unit determinant. These conditions imply
\begin{equation} 
U=\left( \begin{array}{cc}
a & b \\
-b^{\ast} & a^{\ast} \end{array} \right),\,\,\,\,|a|^{2}+|b|^{2}=1.
\end{equation}
It can be shown that if a matrix $H$ is hermitian and traceless, so is $H'$ obtained by the transformation $H'=UHU^{\dag}$.
Let us introduce the matrix $X$ given by
\begin{equation}
X={r^{i}\sigma_{i}}=\left( \begin{array}{cc}
z & x-iy \\
x+iy & -z \end{array} \right),
\end{equation}
where $\sigma_{i}$ are the Pauli matrices. Since $X$ is hermitian and traceless, so is $X'=UXU^{\dag}$. We also have ${\mbox{det}}X'={\mbox{det}}X$ which gives
\begin{equation} 
x'^{2}+y'^{2}+z'^{2}=x^{2}+y^{2}+z^{2}.
\end{equation}
This is the condition for a rotation of the position vector ${\bf r}$.
Therefore, we arrive at the conclusion that the SU(2) transformation is related to the O(3) rotation.

We would like to find the explicit form of the matrix $U$ that corresponds to an arbitrary rotation. For a rotation about the z-axis we have
\begin{equation} 
x'=x\,\cos\phi+y\,\sin\phi,\,\,\,\,y'=-x\,\sin\phi+y\cos\phi,\,\,\,\,z'=z,
\end{equation}
and substituting this into $X'U=UX$ gives $b=0$ and $a=e^{i\phi/2}$. Thus
\begin{equation} 
U_{z}(\phi)=\left( \begin{array}{cc}
e^{i\phi/2} & 0 \\
0 & e^{-i\phi/2} \end{array} \right)=e^{i\sigma_{z}\phi/2}.
\end{equation}
This result can be generalized to a rotation about the axis with the unit vector ${\bf n}$,
\begin{equation}
U_{{\bf n}}(\phi)=e^{i\sigma\cdot{\bf n}\,\phi/2}=\cos{\phi/2}+i\sin{\phi/2}\,\sigma\cdot{\bf n}.
\label{su2rot}
\end{equation}
The above relation is very similar to the corresponding expression for the Lorentz matrix for a rotation, $R_{{\bf n}}(\phi)=e^{i{\bf J}\cdot{\bf n}}$, and this is related to the fact that the Pauli matrices satisfy the same commutation relations as the matrices $J_{i}$:
\begin{equation} 
[\sigma_{i},\sigma_{j}]=i\epsilon_{ijk}\sigma_{k}.
\end{equation}
When a vector rotates by the full angle $2\pi$, a spinor rotates only by the angle $\pi$ and changes sign with respect to the original value. Thus both matrices $U$ and $-U$ correspond to the same rotation matrix $R$.

The matrix $U$ is regarded as the transformation matrix of a two-dimensional complex object $\xi=\left(\matrix{\xi_{1}\cr\xi_{2}\cr}\right)$,
\begin{equation} 
\xi'=U\xi,\,\,\,\,\xi^{\dag}=U^{\dag}\xi^{\dag}.
\end{equation}
The quantities having the above transformation property are called spinors.
We see that $\xi$ and $\xi^{\ast}$ transform in different ways, but we may show that $\left(\matrix{\xi_{1}\cr\xi_{2}\cr}\right)$ and $i\sigma_{2}\xi=\left(\matrix{\xi_{2}^{\ast}\cr-\xi_{1}^{\ast}\cr}\right)$ transform in the same way under SU(2).
We also notice that $\xi^{\dag}i\sigma_{2}\xi$ is a scalar under rotations, whereas $\xi(i\sigma_{2}\xi)^{\dag}$ transforms like a vector.

Now we proceed to transformations of spinors under boosts. From the Lie algebra of the Lorentz group ($\ref{Lie1}$) and ($\ref{Lie2}$) it follows that the matrices $K_{i}=\pm iJ_{i}$ are its solutions. Therefore, spinors should transform under boosts according to formula ($\ref{su2rot}$) with the replacement $\sigma^{i}\rightarrow\pm i\sigma^{i}$.
We may define two types of spinors $\xi$ and $\eta$, transforming with a plus or a minus sign, respectively. For the first one we have
\begin{equation} 
{\bf J}^{(1/2)}=\sigma/2,\,\,\,\,{\bf K}^{(1/2)}=-i\sigma/2,
\end{equation}
and if $(\phi,\psi)$ are the parameters of a pure rotation and a pure boost this spinor transforms according to
\begin{equation} 
\xi'=e^{i\sigma/2\cdot(\phi-i\psi)}\xi=C\xi.
\end{equation}
For the second one we have
\begin{equation} 
{\bf J}^{(1/2)}=\sigma/2,\,\,\,\,{\bf K}^{(1/2)}=i\sigma/2,
\end{equation}
and the Lorentz transformation is given by
\begin{equation}
\eta'=e^{i\sigma/2\cdot(\phi+i\psi)}\eta=D\eta.
\end{equation}
These are inequivalent representations of the Lorentz group and there is no matrix $S$ such that $D=SCS^{-1}$. Instead, we have $D=\sigma_{2}C^{\ast}\sigma_{2}$.
The matrices $C$ and $D$ are no longer unitary, but still unimodular. Such matrices build the group SL(2,C) which is related to the Lorentz group like SU(2) was related to the rotation group. The matrix $X$ is now given by
\begin{equation}
X=x^{\mu}\sigma_{\mu},
\end{equation}
where $\sigma_{0}$ is the 2$\times$2 unit matrix, and the transformation law has the form
\begin{equation} 
X'=GXG^{-1},
\end{equation}
where $G$ belongs to the SL(2,C).

If we define the parity transformation ${\bf v}\rightarrow-{\bf v}$, which changes the sign of ${\bf K}$ but leaves the sign of ${\bf J}$, then the spinors $\xi$ and $\eta$ will interchange. Therefore we may define the four-spinor $\Psi=\left(\matrix{\xi\cr\eta\cr}\right)$,
transforming under rotations and boosts according to
\begin{equation}
\Psi'=\left( \begin{array}{cc}
e^{i/2\sigma\cdot(\phi-i\psi)} & 0 \\
0 & e^{i/2\sigma\cdot(\phi+i\psi)} \end{array} \right)\Psi,
\label{unit}
\end{equation}
and under parity like
\begin{equation}
\Psi'=\left( \begin{array}{cc}
0 & 1 \\
1 & 0 \end{array} \right)\Psi.
\end{equation}
The bispinor $\Psi$ is an irreducible representation of the Lorentz group extended by parity.

For pure boosts we can write
\begin{eqnarray}
& & \xi'=[\cosh{\psi/2}+\sigma\cdot{\bf n}\sinh{\psi/2}]\xi, \nonumber \\
& & \eta'=[\cosh{\psi/2}-\sigma\cdot{\bf n}\sinh{\psi/2}]\eta,
\end{eqnarray}
where ${\bf n}$ is a unit vector in the direction of the boost.
If $\xi$ and $\eta$ refer to a particle at rest, then $\xi'=\xi({\bf p})$ and $\eta'=\eta({\bf p})$, where $\cosh{\psi}=\gamma$ and ${\bf p}=m{\bf v}\gamma$.
One can show that in a moving frame of reference these spinors satisfy the equation
\begin{equation}
\left( \begin{array}{cc}
-m & E+\sigma\cdot{\bf p} \\
E-\sigma\cdot{\bf p} & -m \end{array} \right)\Psi({\bf p})=0,
\end{equation}
where $E=\sqrt{m^{2}+{\bf p}^{2}}$.
Introducing the 4$\times$4 Dirac matrices in the chiral representation,
\begin{equation} 
\gamma^{0}=\left( \begin{array}{cc}
0 & 1 \\
1 & 0 \end{array} \right),\,\,\,\,\gamma^{i}=\left( \begin{array}{cc}
0 & -\sigma^{i} \\
\sigma^{i} & 0 \end{array} \right),
\end{equation}
leads to the Dirac equation
\begin{equation}
(\gamma^{\mu}p_{\mu}-m)\Psi(p)=0.
\label{Dir1}
\end{equation}
The matrices $\gamma^{\mu}$ satisfy the anticommutation relation
\begin{equation}
\gamma^{\mu}\gamma^{\nu}+\gamma^{\nu}\gamma^{\mu}=2g^{\mu\nu},
\end{equation}
which is actually their definition.

We will work in the standard representation of the Dirac matrices, in which $\gamma^{0}$ is diagonal,
\begin{equation} 
\gamma^{0}=\left( \begin{array}{cc}
1 & 0 \\
0 & -1 \end{array} \right).
\end{equation}
It can be obtained from the chiral representation by 
\begin{equation}
\gamma^{0}_{s}=T\gamma^{0}_{c}T^{-1},
\end{equation}
where
\begin{equation}
T=\frac{1}{\sqrt{2}}\left( \begin{array}{cc}
1 & 1 \\
1 & -1 \end{array} \right).
\end{equation}
Therefore the bispinor becomes
\begin{equation}
\Psi=\frac{1}{\sqrt{2}}\left(\matrix{\xi+\eta\cr\xi-\eta\cr}\right),
\end{equation}
and the spinor representation of the boost is given by the matrix
\begin{equation}
S=\left( \begin{array}{cc}
\cosh{\psi/2} & \sinh{\psi/2}\sigma\cdot{\bf n} \\
\sinh{\psi/2}\sigma\cdot{\bf n} & \cosh{\psi/2} \end{array} \right),
\end{equation}
or finally
\begin{equation}
S(0\rightarrow{\bf p})=S(m,{\bf p})=\frac{1}{\sqrt{2m(E(m,{\bf p})+m)}}\left(\matrix{E(m,{\bf p})+m & {\bf \sigma}\cdot{\bf p} \cr {\bf \sigma}\cdot{\bf p} & E(m,{\bf p})+m\cr}\right). 
\label{spinboost}
\end{equation}

The spinor representation for rotations and boosts can be also derived from the assumption that the Dirac equation in the position space
\begin{equation}
(i\gamma^{\mu}\partial_{\mu}-m)\Psi=0
\label{Dir}
\end{equation}
is invariant under Lorentz transformations $x^{\mu}\rightarrow x'^{\mu}$. For the bispinor we will assume that it transforms according to
\begin{equation}
\Psi'=S\Psi,
\end{equation}
where $S$ is a unimodular matrix corresponding to the Lorentz transformation. Its form can be derived from the requirement that the Dirac equation will remain unchanged, 
\begin{equation}
(i\gamma^{\mu'}\partial_{\mu'}-m)\Psi'=0,
\end{equation}
leading to
\begin{equation}
S^{-1}\gamma^{\mu}S=\Lambda^{\mu}_{\,\,\nu}\gamma^{\nu}.
\label{Dir2}
\end{equation}

In order to derive the form of $S$, we will first consider the infinitesimal Lorentz transformation ($\ref{Lor3}$). Consequently, the matrix $S$ will be a linear function of the generators $\omega_{\mu\nu}$, and the simplest guess is
\begin{equation}
S=1+a\omega_{\mu\nu}\gamma^{\mu}\gamma^{\nu},
\end{equation}
where $a$ is some constant.
The inverse matrix $S^{-1}$ in the linear approximation which suffices for our considerations is
\begin{equation} 
S^{-1}=1-a\omega_{\mu\nu}\gamma^{\mu}\gamma^{\nu}.
\end{equation}
Substitution of $S$ and $S^{-1}$ into ($\ref{Dir}$) gives $a=1/4$. The passage to finite transformations is again done by exponentiation,
\begin{equation}
S=e^{\frac{1}{4}\omega_{\mu\nu}\gamma^{\mu}\gamma^{\nu}}. 
\label{Dir3}
\end{equation}
One can show that this expression is equivalent to ($\ref{spinboost}$).

The explicit form of bispinors in the frame of reference in which a particle has momentum ${\bf p}$ can be obtained by acting with ($\ref{spinboost}$) on the solutions of the Dirac equations in the rest frame, given by
\begin{equation}
u(\lambda)=\left(\matrix{\chi(\lambda)\cr 0\cr}\right),\,\,\,\,v(\lambda)=\left(\matrix{0\cr i\sigma_{2}\chi(\lambda)\cr}\right),
\end{equation}
where $\lambda=\pm1/2$ are the values of spin.
The term $i\sigma_{2}$ in $v$ guarantees that spinors $v$ and $u^{\dag}$ transform in the same way under SU(2), and 
\begin{equation}
\chi^{T}(+1/2)=(1,0),\,\,\,\chi^{T}(-1/2)=(0,1).
\end{equation}
Thus we get
\begin{equation}
u({\bf p},\lambda)=\frac{1}{\sqrt{E(m,{\bf p})+m}}\left(\matrix{(E(m,{\bf p})+m)\chi(\lambda)\cr ({\bf \sigma}\cdot{\bf p})\chi(\lambda)\cr}\right)
\end{equation}
and
\begin{equation}
v({\bf p},\lambda)=\frac{1}{\sqrt{E(m,{\bf p})+m}}\left(\matrix{({\bf \sigma}\cdot{\bf p})i\sigma_{2}\chi(\lambda)\cr (E(m,{\bf p})+m)i\sigma_{2}\chi(\lambda)\cr}\right).
\end{equation}

The nonrelativistic limit of the Dirac equation is free of paradoxical properties only in the first approximation. It is possible, however, to find a representation in which it is clear how to associate operators with classical dynamical variables so that these operators tend to their expected nonrelativistic form \cite{fw1}. In the presence of an external field the nonrelativistic reduction is most conveniently obtained by an infinite set of canonical transformations related to a free field transformation. This is referred to as the Foldy-Wouthuysen representation, and can also be extended to Klein-Gordon and Proca particles \cite{fw2}. 
In this work the nonrelativistic limit will be reduced to the first approximation. Therefore, the Dirac representation will be suitable for our purposes.

\section{The Bakamjian-Thomas model for interacting particles}

For a system with a fixed number of particles, $P_{i}$ and $M_{ij}$ will be sums of their values for separate particles,
\begin{equation}
P_{i}=\sum p_{i},\,\,\,\,M_{ij}=\sum (q_{i}p_{j}-q_{j}p_{i}).
\label{Gen5}
\end{equation}
For the Hamiltonians one must add the interaction terms,
\begin{eqnarray}
& & P_{0}=\sum \sqrt{p_{j}p_{j}+m^{2}}+U, \nonumber \\
& & M_{i0}=\sum q_{i}\sqrt{p_{j}p_{j}+m^{2}}+U_{i}.
\label{Gen6}
\end{eqnarray}
From the commutation relations ($\ref{Gen2}$) it follows that $U$ is a three-dimensional scalar, $U_{i}$ is a three-dimensional vector, and
\begin{equation} 
U_{i}=q_{i}U+b_{i},
\end{equation}
where $b_{i}$ is a constant three-dimensional vector. The remaining conditions for $U$ and $U_{i}$ are quadratic and therefore a construction of a complete dynamical theory of a relativistic theory is very difficult.

An interaction enters only in $H$ and ${\bf V}$. A practical method of constructing the generators in this case was developed in \cite{BT}, where the set of new operators satisfying simpler commutation relations was introduced. In this set, the interaction appears only in the mass operator ($\ref{mop}$).
Suppose we can make a transformation from ${\bf q}_{1},{\bf q}_{2}$ and ${\bf p}_{1},{\bf p}_{2}$ to the total momentum ${\bf P}$, the coordinates of the center-of-mass ${\bf R}$, the relative momentum ${\bf p}$ and the relative coordinate vector ${\bf r}$. The commutation relations are not disturbed if:
\begin{enumerate}
\item ${\bf M}={\bf R}\times{\bf P}+{\bf r}\times{\bf p}$, 
\item $M$ depends on ${\bf p}$ only,
\item ${\bf V}$ can be expressed in terms of $M$, ${\bf P}$, ${\bf M}$ and ${\bf R}$ only.
\end{enumerate}

The interaction will be included if we replace $M$ by any other function of ${\bf p}$ and ${\bf r}$ which is a scalar for space rotations. The only nonzero commutators of the set $M,{\bf P},{\bf M},{\bf R}$ are
\begin{equation} 
[P_{i},R_{j}]=-i\delta_{ij},\,\,\,\,[M_{i},M_{j}]=i\epsilon_{ijk}M_{k},
\end{equation}
and the mass operator $M$ is Poincare invariant if it commutes with ${\bf P},{\bf M},{\bf R}$. Therefore it is only necessary to make sure that the above condition is satisfied. 
The macroscopic Hamiltonian of a system is given by 
\begin{equation}
H=\sqrt{{\bf P}^{2}+M^{2}({\bf p},{\bf r})},
\end{equation}
and is obtained from the microscopic one (\ref{Gen6}) via introducing the above relative variables. The above results can be generalized to systems with more than two particles, and to particles with intrinsic spin \cite{BT,osb}. An explicit construction for a unitary operator that insures the free motion of the center of mass of any system is given in \cite{gs}. However, for a given potential there is no unique way in which the relative variables may be defined \cite{kf}.
Unfortunately, an exact form of the potential is not known and one must use phenomenological forms of the mass operator. In this work we will assume a gaussian form of the orbital wave function and fit it to a few measured form factors. This approach will not allow for a deeper understanding of the relativistic quark dynamics, although it makes possible to estimate relativistic corrections to the $\pi_{1}$ decay widths.

\section{Wigner rotation}

In this section we will derive how spin transforms under the boost transformations. It will be necessary for a construction of covariant spin wave functions for quark-antiquark pairs (mesons).
Our goal is to solve 
\begin{equation}
\Lambda(0\rightarrow{\bf P})|{\bf p},\lambda\rangle=\Lambda(0\rightarrow{\bf P})\Lambda(0\rightarrow{\bf p})|0,\lambda\rangle,
\end{equation}
where $\Lambda({\bf p}\rightarrow{\bf q})$ boosts a particle with a momentum ${\bf p}$ to a frame in which the momentum is equal to ${\bf q}$, and $|{\bf p},\lambda\rangle$ is the spinor state.
Multiplying the above expression by the identity $\Lambda(0\rightarrow{\bf p}')\Lambda({\bf p}'\rightarrow0)$, where ${\bf p}'$ is obtained from ${\bf p}$ according to ($\ref{vecboost}$), leads to
\begin{equation} 
\Lambda(0\rightarrow{\bf P})|{\bf p},\lambda\rangle=\Lambda(0\rightarrow{\bf p}')R({\bf p},{\bf P})|0,\lambda\rangle.
\end{equation}
The quantity $R$ is given by
\begin{equation}
R({\bf p},{\bf P})=\Lambda({\bf p}'\rightarrow0)\Lambda({\bf p}\rightarrow{\bf p}')\Lambda(0\rightarrow{\bf p}),
\end{equation}
and we will find its explicit form.

In spinor representation we can write
\begin{equation} 
R({\bf p},{\bf P})=S(m,-{\bf p}')S(M,{\bf P})S(m,{\bf p}),
\end{equation}
where $M$ parametrizes the boost $\Lambda(0\rightarrow{\bf P})$ like before.
Substituting the expression ($\ref{spinboost}$) in the above equation leads after somewhat lengthy calculations to
\begin{equation}
R({\bf p},{\bf P})=\left( \begin{array}{cc}
D^{(1/2)}({\bf p},{\bf P}) & 0 \\
0 & D^{(1/2)}({\bf p},{\bf P}) \end{array} \right),
\label{Wig1}
\end{equation}
where 
\begin{equation}
D^{(1/2)}_{\lambda\lambda'}({\bf q},{\bf P})=\Bigl[\frac{(E(m,{\bf q})+m)(E(M,{\bf P})+M)+{\bf P}\cdot{\bf q}+i{\bf \sigma}\cdot({\bf P}\times{\bf q})}{\sqrt{2(E(m,{\bf q})+m)(E(M,{\bf P})+M)(E(m,{\bf q})E(M,{\bf P})+{\bf P}\cdot{\bf q}+mM)}}\Bigr]_{\lambda\lambda'}.
\label{Wig2}
\end{equation}
It can be shown that $D^{(1/2)}$ has the form $\cos{\phi/2}+i\sin{\phi/2}\sigma\cdot{\bf n}$ and thus the matrix ($\ref{Wig1}$) represents a pure rotation.
The matrix $D^{(1/2)}$ is called the Wigner rotation matrix.
In nonrelativistic quantum mechanics, spin should not change under boosts and this is reflected in the large-mass limit of formula ($\ref{Wig2}$), 
\begin{equation} 
D^{(1/2)}_{\lambda\lambda'}\rightarrow\delta_{\lambda\lambda'}.
\end{equation}

Finally, we obtain the transformation law for spinor states under boosts,
\begin{equation}
\Lambda(0\rightarrow{\bf P})|{\bf p},\lambda\rangle=\Lambda(0\rightarrow{\bf p}')D_{\lambda\lambda'}^{(1/2)}({\bf p},{\bf P})|0,\lambda'\rangle=D^{(1/2)}_{\lambda\lambda'}({\bf p},{\bf P})|\Lambda(0\rightarrow{\bf P}){\bf p},\lambda'\rangle.
\label{Wig3}
\end{equation}
The state of a system having more than one spin index transforms like a spin tensor, i.e., each index transforms independently with the $D^{(1/2)}$ matrix according to formula ($\ref{Wig3}$).

\long\def\symbolfootnote[#1]#2{\begingroup%
\def\thefootnote{\fnsymbol{footnote}}\footnote[#1]{#2}\endgroup}

\chapter{Relativistic spin wave function for mesons and hybrids}

Having described transformation laws for a single particle with spin $1/2$, we may proceed to systems of noninteracting particles. We will focus on quark-antiquark pairs, i.e. mesons. This is necessary in order to construct the spin wave functions for the outgoing mesons resulting from the decay of the $\pi_{1}$. Since these mesons have nonzero momenta, a relativistic model of hadronic decays will have to include a Wigner rotation of spin. A similar construction is also required for the $\pi_{1}$ spin wave function because the $q\bar{q}$ pair must be boosted to a moving frame before it couples with a gluon to a rest-frame hybrid. 
In the following section we will show how to build a relativistic spin wave function for each light unflavored meson (or a meson with equal masses of quarks). Following that we will add the gluon and build the $\pi_{1}$. Mesons with different masses of quarks will be considered later. In each case we will begin with a rest-frame function, and then use the results of the preceding chapter to obtain a general expression for any frame of reference\symbolfootnote[2]{This chapter is based on work by A.P.Szczepaniak}.

\section{Meson spin wave functions}

The spin wave function for a meson is constructed as an element of an irreducible representation of the Poincare group \cite{rel2,bkt}. In the rest frame of a meson, the quark momenta are given by
\begin{equation} 
l^{\mu}_{q}=(E(m_{q},{\bf q}),{\bf q}),\,\,\,\,l^{\mu}_{\bar{q}}=(E(m_{\bar{q}},-{\bf q}),-{\bf q}),
\end{equation}
and the normalized spin-0 and spin-1 wave function corresponding to $J^{PC}=0^{-+}$ and $J^{PC}=1^{--}$ are simply given by Clebsch-Gordan coefficients,

\begin{equation} 
\Psi_{q\bar{q}}({\bf q},{\bf l}_{q\bar{q}}=0,\sigma_{q},\sigma_{\bar{q}})=\langle \frac{1}{2},\sigma_{q};\frac{1}{2},\sigma_{\bar{q}}|0,0\rangle =\chi^{\dag}(\sigma_{q})\frac{i\sigma_{2}}{\sqrt{2}}\chi(\sigma_{\bar{q}}),
\label{CG0}
\end{equation}
and
\begin{equation}
\Psi^{\lambda_{q\bar{q}}}_{q\bar{q}}({\bf q},{\bf l}_{q\bar{q}}=0,\sigma_{q},\sigma_{\bar{q}})=\langle \frac{1}{2},\sigma_{q};\frac{1}{2},\sigma_{\bar{q}}|1,\lambda_{q\bar{q}}\rangle =\chi^{\dag}(\sigma_{q})\frac{\sigma^{i}i\sigma_{2}}{\sqrt{2}}\chi(\sigma_{\bar{q}})\epsilon^{i}(\lambda_{q\bar{q}}).
\label{CG1}
\end{equation} 
A factor $i\sigma_{2}$ accounts that the antiparticle spin doublet transforms under SU(2) in the same way as the particle doublet.
The canonical polarization vectors
\begin{equation}
{\bf \epsilon}(\pm1)=\frac{\mp1}{\sqrt{2}}\left(\matrix{1\cr\pm i\cr0\cr}\right),\,\,\,\,{\bf \epsilon}(0)=\left(\matrix{0\cr0\cr1\cr}\right),
\label{eq:polar}
\end{equation}
correspond to spin 1 quantized along the z-axis and satisfy the orthogonality relation
\begin{equation}
\sum_{\lambda}\epsilon^{i}(\lambda)\epsilon^{j\ast}(\lambda)=\delta^{ij}.
\end{equation}
The invariant mass of the $q\bar{q}$ pair is
\begin{equation}
m_{q\bar{q}}=E(m_{q},{\bf q})+E(m_{\bar{q}},-{\bf q}),
\end{equation}
where $E(m,{\bf p})=\sqrt{m^{2}+{\bf p}^{2}}$, and the total momentum of this system ${\bf l}_{q\bar{q}}$ is of course equal to zero.
In the following we will assume 
\begin{equation}
m_{q}=m_{\bar{q}}=m.
\end{equation}
This condition is satisfied to a good approximation by the quarks $u,d$ and may be used for a construction of light unflavored meson states.

The rest frame wave functions (\ref{CG0}) and (\ref{CG1}) may also be expressed in terms of Dirac spinors quantized along the z-axis,
\begin{equation}
\Psi_{q\bar{q}}({\bf q},{\bf l}_{q\bar{q}}=0,\sigma_{q},\sigma_{\bar{q}})=\frac{1}{\sqrt{2}m_{q\bar{q}}}\bar{u}({\bf q},\sigma_{q})\gamma^{5}v(-{\bf q},\sigma_{\bar{q}}),
\label{eq:00r}
\end{equation}
and
\begin{equation}
\Psi^{\lambda_{q\bar{q}}}_{q\bar{q}}({\bf q},{\bf l}_{q\bar{q}}=0,\sigma_{q},\sigma_{\bar{q}})=\frac{1}{\sqrt{2}m_{q\bar{q}}}\bar{u}({\bf q},\sigma_{q})\Bigl[\gamma^{i}-\frac{2q^{i}}{m_{q\bar{q}}+2m}\Bigr]v(-{\bf q},\sigma_{\bar{q}})\epsilon^{i}({\bf l}_{q\bar{q}}=0,\lambda_{q\bar{q}}).
\label{eq:01r} 
\end{equation}
A spin wave function written in this form is manifestly covariant and thus it is straightforward to find how it transforms under Lorentz transformations. In the above $\epsilon^{i}({\bf l}_{q\bar{q}}=0,\lambda_{q\bar{q}})$ are the spatial components of the polarization four-vector
\begin{equation}
\epsilon^{\mu}({\bf l}_{q\bar{q}}=0,\lambda_{q\bar{q}})=(0,{\bf \epsilon}(\lambda_{q\bar{q}})),
\end{equation}
whose time component is zero in order to satisfy the transversity condition $\epsilon^{\mu}({\bf k},\lambda)k_{\mu}=0$.

\begin{figure}[h]
\centering
\includegraphics[width=2.0in,angle=270]{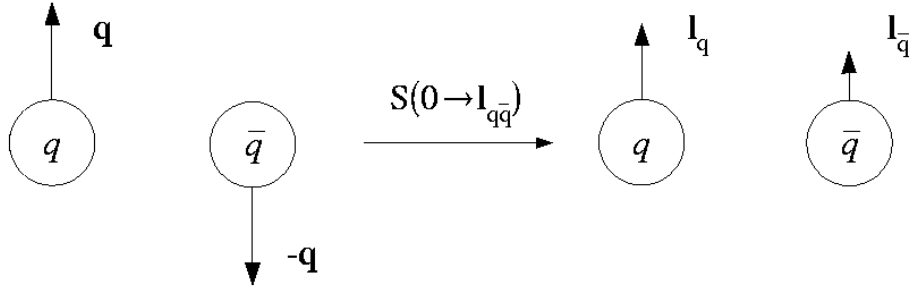}
\caption{\label{boo} Boost of a quark-antiquark pair from the rest frame to a moving frame. }
\end{figure}

Now we apply a boost from the rest frame of a $q\bar{q}$ pair to a frame of reference in which the momenta of the quark and the antiquark are ${\bf l}_{q}$ and ${\bf l}_{\bar{q}}$, respectively, and the total momentum is ${\bf l}_{q\bar{q}}={\bf l}_{q}+{\bf l}_{\bar{q}}$, as shown in Fig.~\ref{boo}. The new momenta are given by
\begin{equation}
{\bf l}_{q}={\bf q}+\frac{({\bf q}\cdot{\bf l}_{q\bar{q}}){\bf l}_{q\bar{q}}}{E(m_{q\bar{q}},{\bf l}_{q\bar{q}})[m_{q\bar{q}}+E(m_{q\bar{q}},{\bf l}_{q\bar{q}})]}+\frac{E(m,{\bf q})}{m_{q\bar{q}}}{\bf l}_{q\bar{q}},
\label{eq:b1}
\end{equation}
and
\begin{equation}
{\bf l}_{\bar{q}}=-{\bf q}-\frac{({\bf q}\cdot{\bf l}_{q\bar{q}}){\bf l}_{q\bar{q}}}
{E(m_{q\bar{q}},{\bf l}_{q\bar{q}})[m_{q\bar{q}}+E(m_{q\bar{q}},{\bf l}_{q\bar{q
}})]}+\frac{E(m,{\bf q})}{m_{q\bar{q}}}{\bf l}_{q\bar{q}}.
\label{eq:b2}
\end{equation}
The spin wave function of a meson in a moving frame is obtained from the rest frame wave function, as we stated at the end of Chapter~4, by acting with the Wigner rotation matrix ($\ref{Wig2}$) on each of both spin indices. This gives
\begin{equation}
\Psi^{\lambda_{q\bar{q}}}_{q\bar{q}}({\bf q},{\bf l}_{q\bar{q}},\lambda_{q},\lambda_{\bar{q}})=\sum_{\sigma_{q},\sigma_{\bar{q}}}\Psi^{\lambda_{q\bar{q}}}_{q\bar{q}}({\bf q},{\bf l}_{q\bar{q}}=0,\sigma_{q},\sigma_{\bar{q}})D^{\ast(1/2)}_{\lambda_{q}\sigma_{q}}({\bf q},{\bf l}_{q\bar{q}})D^{(1/2)}_{\lambda_{\bar{q}}\sigma_{\bar{q}}}(-{\bf q},{\bf l}_{q\bar{q}}).
\label{eq:d} 
\end{equation}
The above Wigner rotation matrix corresponds to the boost with ${\bf v}\gamma={\bf P}/M$. In spinor representation Eq.~(\ref{Wig3}) leads to the following transformation laws for spinors $u^{\dag}$ and $v$:
\begin{eqnarray}
& & \sum_{\sigma_{\bar{q}}}D^{(1/2)}_{\lambda_{\bar{q}}\sigma_{\bar{q}}}(-{\bf q},{\bf l}_{q\bar{q}})v(-{\bf q},\sigma_{\bar{q}})=S({\bf l}_{q\bar{q}}\rightarrow0)v({\bf l}_{\bar{q}},\lambda_{\bar{q}}), \nonumber \\
& & \sum_{\sigma_{q}}D^{\ast(1/2)}_{\lambda_{q}\sigma_{q}}({\bf q},{\bf l}_{q\bar{q}})u^{\dag}({\bf q},\sigma_{q})=u^{\dag}({\bf l}_{q},\lambda_{q})S^{\dag}({\bf l}_{q\bar{q}}\rightarrow0),
\end{eqnarray}
where $S({\bf l}_{q\bar{q}}\rightarrow0)$ is the Dirac representation of the boost taking ${\bf l}_{q}$ to ${\bf q}$ and ${\bf l}_{\bar{q}}$ to $-{\bf q}$, given by ($\ref{spinboost}$) with ${\bf p}=-{\bf l}_{q\bar{q}}$ and $M=m_{q\bar{q}}$. From these laws we obtain the general form of the spin-0 wave function:
\begin{eqnarray}
& & \Psi_{q\bar{q}}({\bf q},{\bf l}_{q\bar{q}},\lambda_{q},\lambda_{\bar{q}})=\frac{1}{\sqrt{2}m_{q\bar{q}}}u^{\dag}({\bf l}_{q},\lambda_{q})S^{\dag}({\bf l}_{q\bar{q}}\rightarrow 0)\gamma^{0}\gamma^{5}S({\bf l}_{q\bar{q}}\rightarrow 0)v({\bf l}_{\bar{q}},\lambda_{\bar{q}})= \nonumber \\
& & =\frac{1}{\sqrt{2}m_{q\bar{q}}}\bar{u}({\bf l}_{q},\lambda_{q})S^{-1}({\bf l}_{q\bar{q}}\rightarrow 0)\gamma^{5}S({\bf l}_{q\bar{q}}\rightarrow 0)v({\bf l}_{\bar{q}},\lambda_{\bar{q}})= \nonumber \\
& & =\frac{1}{\sqrt{2}m_{q\bar{q}}}\bar{u}({\bf l}_{q},\lambda_{q})\gamma^{5}v({\bf l}_{\bar{q}},\lambda_{\bar{q}}).
\label{eq:00}
\end{eqnarray}

Similarly we derive the spin-1 wave function:
\begin{eqnarray}
& & \Psi^{\lambda_{q\bar{q}}}_{q\bar{q}}({\bf q},{\bf l}_{q\bar{q}},\lambda_{q},\lambda_{\bar{q}})= \nonumber \\
& & =\frac{1}{\sqrt{2}m_{q\bar{q}}}u^{\dag}({\bf l}_{q},\lambda_{q})S^{\dag}({\bf l}_{q\bar{q}}\rightarrow 0)\gamma^{0}\Bigl(\gamma^{i}-\frac{2q^{i}}{m_{q\bar{q}}+2m}\Bigr)S({\bf l}_{q\bar{q}}\rightarrow 0)v({\bf l}_{\bar{q}},\lambda_{\bar{q}})\epsilon^{i}(\lambda_{q\bar{q}})= \nonumber \\
& & =\frac{1}{\sqrt{2}m_{q\bar{q}}}\bar{u}({\bf l}_{q},\lambda_{q})S^{-1}({\bf l}_{q\bar{q}}\rightarrow 0)\Bigl(\gamma^{i}-\frac{2q^{i}}{m_{q\bar{q}}+2m}\Bigr)S({\bf l}_{q\bar{q}}\rightarrow 0)v({\bf l}_{\bar{q}},\lambda_{\bar{q}})\epsilon^{i}(\lambda_{q\bar{q}})= \nonumber \\
& & =\frac{1}{\sqrt{2}m_{q\bar{q}}}\bar{u}({\bf l}_{q},\lambda_{q})\Bigl[\Lambda^{i}_{\,\,\nu}({\bf l}_{q\bar{q}}\rightarrow 0)\gamma^{\nu}-\frac{2q^{i}}{m_{q\bar{q}}+2m}\Bigr]v({\bf l}_{\bar{q}},\lambda_{\bar{q}})\epsilon^{i}(\lambda_{q\bar{q}})= \nonumber \\
& & =\frac{1}{\sqrt{2}m_{q\bar{q}}}\bar{u}({\bf l}_{q},\lambda_{q})\Bigl[\gamma^{\nu}-\frac{p_{q}^{\nu}-p_{\bar{q}}^{\nu}}{m_{q\bar{q}}+2m}\Bigr]v({\bf l}_{\bar{q}},\lambda_{\bar{q}})\Lambda^{i}_{\,\,\nu}({\bf l}_{q\bar{q}}\rightarrow 0)\epsilon^{i}(\lambda_{q\bar{q}})= \nonumber \\
& & =-\frac{1}{\sqrt{2}m_{q\bar{q}}}\bar{u}({\bf l}_{q},\lambda_{q})\Bigl[\gamma^{\mu}-\frac{l^{\mu}_{q}-l^{\mu}_{\bar{q}}}{m_{q\bar{q}}+2m}\Bigr]v({\bf l}_{\bar{q}},\lambda_{\bar{q}})\epsilon_{\mu}({\bf l}_{q\bar{q}},\lambda_{q\bar{q}}),\label{eq:01}
\end{eqnarray}
where $\epsilon^{\mu}({\bf l}_{q\bar{q}},\lambda_{q\bar{q}})$ are obtained from ($\ref{eq:polar}$) through the boost with ${\bf \beta}\gamma={\bf l}_{q\bar{q}}/m_{q\bar{q}}$:
\begin{eqnarray}
& & \epsilon^{0}({\bf l}_{q\bar{q}},\lambda_{q\bar{q}})=\frac{{\bf l}_{q\bar{q}}\cdot{\bf \epsilon}(\lambda_{q\bar{q}})}{m_{q\bar{q}}}, \nonumber \\
& & {\bf \epsilon}({\bf l}_{q\bar{q}},\lambda_{q\bar{q}})={\bf \epsilon}(\lambda_{q\bar{q}})+\frac{({\bf l}_{q\bar{q}}\cdot{\bf \epsilon}(\lambda_{q\bar{q}})){\bf l}_{q\bar{q}}}{m_{q\bar{q}}(E(m_{q\bar{q}},{\bf l}_{q\bar{q}})+m_{q\bar{q}})}.
\label{polarb}
\end{eqnarray}
The invariant mass of the $q\bar{q}$ system is now
\begin{equation}
m_{q\bar{q}}=m_{q\bar{q}}({\bf l}_{q},{\bf l}_{\bar{q}})=\sqrt{(E(m,{\bf l}_{q})+E(m,{\bf l}_{\bar{q}}))^{2}-({\bf l}_{q}+{\bf l}_{\bar{q}})^{2}}.
\label{eq:mi1}
\end{equation}
The wave functions ($\ref{eq:00}$) and ($\ref{eq:01}$) are still normalized:
\begin{eqnarray} 
& & \sum_{\lambda_{q},\lambda_{\bar{q}}}\Psi^{\ast}_{q\bar{q}}({\bf q},{\bf l}_{q\bar{q}},\lambda_{q},\lambda_{\bar{q}})\Psi_{q\bar{q}}({\bf q},{\bf l}_{q\bar{q}},\lambda_{q},\lambda_{\bar{q}})=1, \nonumber \\
& & \sum_{\lambda_{q},\lambda_{\bar{q}}}\Psi^{\ast\lambda_{q\bar{q}}}_{q\bar{q}}({\bf q},{\bf l}_{q\bar{q}},\lambda_{q},\lambda_{\bar{q}})\Psi^{\lambda_{q\bar{q}}'}_{q\bar{q}}({\bf q},{\bf l}_{q\bar{q}},\lambda_{q},\lambda_{\bar{q}})=\delta_{\lambda_{q\bar{q}}\lambda_{q\bar{q}}'}.
\end{eqnarray}

By coupling the spin wave function ($\ref{eq:00r}$) or ($\ref{eq:01r}$), respectively, with one unit of the orbital angular momentum $L=1$, one obtains the rest frame spin wave functions for the quark-antiquark pair with quantum numbers $J^{PC}=1^{+-}$ or $0^{++}$, $1^{++}$ and $2^{++}$. Explicitly we have
\begin{equation}
\Psi^{\lambda_{q\bar{q}}}_{q\bar{q}}({\bf q},{\bf l}_{q\bar{q}}=0,\sigma_{q},\sigma_{\bar{q}})=\frac{1}{\sqrt{2}m_{q\bar{q}}({\bf q},-{\bf q})}\bar{u}({\bf q},\sigma_{q})\gamma^{5}v(-{\bf q},\sigma_{\bar{q}})Y_{1\lambda_{q\bar{q}}}(\bar{{\bf q}})
\label{eq:10r}
\end{equation}
for the $1^{+-}$ meson, and
\begin{eqnarray}
& & \Psi^{\lambda_{q\bar{q}}}_{q\bar{q}}({\bf q},{\bf l}_{q\bar{q}}=0,\sigma_{q},\sigma_{\bar{q}})=\sum_{\lambda,l}\frac{1}{\sqrt{2}m_{q\bar{q}}}\bar{u}({\bf q},\sigma_{q})\Bigl[\gamma^{i}-\frac{2q^{i}}{m_{q\bar{q}}+2m}\Bigr]v(-{\bf q},\sigma_{\bar{q}})\epsilon^{i}(0,\lambda) \nonumber \\
& & \times\,Y_{1l}(\bar{{\bf q}})\langle 1,\lambda;1,l|J,\lambda_{q\bar{q}}\rangle
\label{eq:11r}
\end{eqnarray}
for the $J^{++}$ ($J=0,1,2$). Here $Y_{L\lambda}(\bar{{\bf q}})$ is a spherical harmonic and $\bar{{\bf q}}={\bf q}/|{\bf q}|$.
Using ($\ref{eq:d}$) one can show that the wave functions for the $q\bar{q}$ pair moving with the total momentum ${\bf l}_{q\bar{q}}$ are given by
\begin{equation}
\Psi^{\lambda_{q\bar{q}}}_{q\bar{q}}({\bf l}_{q},{\bf l}_{\bar{q}},\lambda_{q},\lambda_{\bar{q}})=\frac{1}{\sqrt{2}m_{q\bar{q}}({\bf l}_{q},{\bf l}_{\bar{q}})}\bar{u}({\bf l}_{q},\lambda_{q})\gamma^{5}v({\bf l}_{\bar{q}},\lambda_{\bar{q}})Y_{1\lambda_{q\bar{q}}}(\bar{{\bf q}}),
\label{eq:10}
\end{equation}
and
\begin{eqnarray}
& & \Psi^{\lambda_{q\bar{q}}}_{q\bar{q}}({\bf l}_{q},{\bf l}_{\bar{q}},\lambda_{q},\lambda_{\bar{q}})=-\sum_{\lambda,l}\frac{1}{\sqrt{2}m_{q\bar{q}}}\bar{u}({\bf l}_{q},\lambda_{q})\Bigl[\gamma^{\mu}-\frac{l^{\mu}_{q}-l^{\mu}_{\bar{q}}}{m_{q\bar{q}}+2m}\Bigr]v({\bf l}_{\bar{q}},\lambda_{\bar{q}})\epsilon_{\mu}({\bf l}_{q\bar{q}},\lambda) \nonumber \\
& & \times\,Y_{1l}(\bar{{\bf q}})\langle 1,\lambda;1,l|J,\lambda_{q\bar{q}}\rangle,
\label{eq:11}
\end{eqnarray}
respectively, where $m_{q\bar{q}}=m_{q\bar{q}}({\bf l}_{q},{\bf l}_{\bar{q}})$ and ${\bf q}$ remains the same but must be written in terms of new variables:
\begin{equation}
{\bf q}=\Lambda({\bf l}_{q\bar{q}}\rightarrow0){\bf l}_{q}={\bf l}_{q}-\frac{E(m,{\bf l}_{q}){\bf l}_{q\bar{q}}}{m_{q\bar{q}}}+\frac{({\bf l}_{q\bar{q}}\cdot{\bf l}_{q}){\bf l}_{q\bar{q}}}{m_{q\bar{q}}(E(m_{q\bar{q}},-{\bf l}_{q\bar{q}})+m_{q\bar{q}})}.
\label{eq:q}
\end{equation}

In order to construct meson spin wave functions for higher orbital angular momenta $L$ one need only to replace $Y_{1l}$ with $Y_{Ll}$ in ($\ref{eq:10}$) and ($\ref{eq:11}$). In $L=0$ [formulae ($\ref{eq:00r}$), ($\ref{eq:01r}$), ($\ref{eq:00}$) and ($\ref{eq:01}$)] we skipped a constant factor $Y_{00}$ to make the spin wave function normalized to $1$. But from now on, for consistency, we will assume this constant being implicitly included. 
In the nonrelativistic limit, where Wigner rotations may be ignored, all the spin wave functions for mesons simply reduce to the Clebsch-Gordan coefficients that we started from, coupled to appropriate spherical harmonics.

\section{The $\pi_{1}$ spin wave function}

As we showed in Chapter~3, in the lightest hybrid meson $\pi_{1}$ wave function, a constituent gluon is expected to have one unit of orbital angular momentum with respect to a $q{\bar q}$ pair. Thus, the quantum numbers $P,C$ require a quark and an antiquark to have parallel spins ($S=1$)

In the rest frame of a 3-body system with a $q\bar{q}$ pair moving with momentum $-{\bf Q}$ and a transverse gluon with momentum ${\bf Q}$, the total spin wave function of the hybrid is obtained by coupling the $q\bar{q}$ spin-1 wave function ($\ref{eq:01}$) and the gluon wave function ($J^{PC}=1^{--}$) to a total spin $S=0,1,2$ and $J^{PC}=0^{++},1^{++},2^{++}$ states, respectively. Here, we will derive the expressions for each value of $S$ separately, although the physical $\pi_{1}$ state should be a superposition of all three components. The way of calculating the corresponding coefficients in this linear combination will be given in Chapter~7.   
The total $J^{PC}=1^{-+}$ exotic meson wave function is then obtained by adding one unit of orbital angular momentum between the gluon and the $q{\bar q}$:
\begin{eqnarray} 
& & \Psi^{\lambda_{ex}}_{q\bar{q}g(S)}(\lambda_{q},\lambda_{\bar{q}},\lambda_{g})=\sum_{\lambda_{q\bar{q}},\sigma=\pm1,M,l}\Psi^{\lambda_{q\bar{q}}}_{q\bar{q}}({\bf q},{\bf l}_{q\bar{q}}=-{\bf Q},\lambda_{q},\lambda_{\bar{q}})\langle 1,\lambda_{q\bar{q}};1,\sigma|S,M\rangle D^{(1)\ast}_{\lambda_{g}\sigma}(\bar{{\bf Q}}) \nonumber \\
& & \times\,Y_{1l}(\bar{{\bf Q}})\langle S,M;1,l|1,\lambda_{ex}\rangle.
\label{eq:exot}
\end{eqnarray}

\begin{figure}[h]
\centering
\includegraphics[width=1.5in,angle=270]{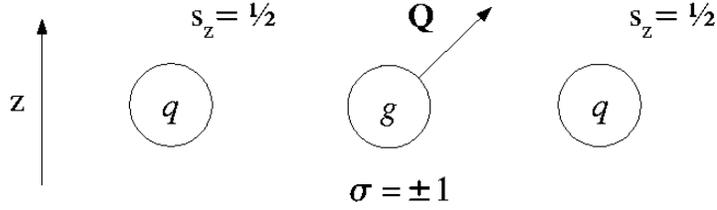}
\caption{\label{ro} A constituent picture of a hybrid: the quark-antiquark pair quantized along the z-axis and the transverse gluon quantized in the helicity basis. }
\end{figure}

\begin{figure}[ht]
\centering
\includegraphics[width=1.5in,angle=270]{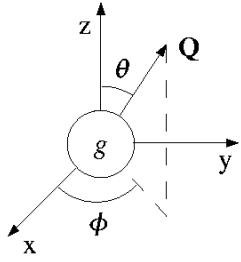}
\caption{\label{angl} A polar angle and an azimuth of the gluon momentum vector. }
\end{figure}

The spin-1 rotation matrix $D^{(1)}$ relates the transverse gluon states in the helicity basis $\sigma$ (i.e., along its momentum) to the basis described by spin $\lambda_{g}$ quantized along a fixed z-axis, as shown in Fig.~\ref{ro}. Thus, the helicity basis is rotated so that a coupling between a gluon and a $q\bar{q}$ pair may be done in the same basis with Clebsch-Gordan coefficients, since we are in the rest frame of a hybrid. Explicitly we have 
\begin{equation}
|{\bf Q},\lambda_{g}\rangle =\sum_{\sigma}D^{(1)\ast}_{\lambda_{g}\sigma}(\phi,\theta,-\phi)|{\bf Q},\sigma\rangle,
\end{equation}
where $\theta$ and $\phi$ are the polar angle and the azimuth of the direction of the gluon momentum ${\bf Q}$, as shown in Fig.~\ref{angl}. 
For the gluon polarization vector with spin quantized along the z-axis we can write
\begin{equation}
\epsilon^{i}_{c}({\bf Q},\lambda_{g})=\sum_{\sigma=\pm1}D^{(1)\ast}_{\lambda_{g}\sigma}(\phi,\theta,-\phi)\epsilon^{i}_{h}({\bf Q},\sigma), 
\end{equation}
where the helicity polarization vectors are given by
\begin{equation}
\epsilon^{i}_{h}({\bf Q},\sigma)=\sum_{\lambda_{g}}D^{(1)}_{\lambda_{g}\sigma}(\phi,\theta,-\phi)\epsilon^{i}(\lambda_{g}).
\end{equation}
Using the unitarity of the matrix $D^{(1)}$ one can show
\begin{equation}
\epsilon^{i}_{c}({\bf Q},\lambda_{g})\epsilon^{\ast i}_{h}({\bf Q},\sigma)=D^{(1)\ast}_{\lambda_{g}\sigma},
\end{equation}
and with the help of the identity $\epsilon^{\ast i}_{h}({\bf Q},\sigma)\epsilon^{j}_{h}({\bf Q},\sigma)=\delta^{ij}-\bar{Q}^{i}\bar{Q}^{j}$ we finally obtain
\begin{equation}
\epsilon^{i}_{c}({\bf Q},\lambda_{g})=\epsilon^{j}(\lambda_{g})(\delta^{ij}-\bar{Q}^{i}\bar{Q}^{j}),
\label{polart}
\end{equation}
where $\bar{{\bf Q}}^{i}={\bf Q}^{i}/|{\bf Q}|$.

The Clebsch-Gordan coefficients and the spherical harmonic in ($\ref{eq:exot}$) can be expressed in terms of the polarization vectors ($\ref{eq:polar}$). For example:
\begin{eqnarray}
& & \langle 1,\lambda';0,0|1,\lambda\rangle =\epsilon^{\ast}(\lambda')\cdot\epsilon(\lambda), \nonumber \\
& & \langle 1,\lambda';1,\lambda|0,0\rangle = -\frac{1}{\sqrt{3}}\epsilon^{\ast}(\lambda')\cdot\epsilon^{\ast}(\lambda), \nonumber \\
& & \langle 1,\lambda';1,\lambda''|1,\lambda\rangle = \frac{i}{\sqrt{2}}[\epsilon^{\ast}(\lambda')\times\epsilon^{\ast}(\lambda'')]\cdot\epsilon(\lambda), \nonumber \\
& & Y_{1l}(\bar{{\bf Q}})=\sqrt{\frac{3}{4\pi}}\epsilon(l)\cdot\bar{{\bf Q}}.
\label{app}
\end{eqnarray}
Therefore we obtain:
\begin{eqnarray}
& & \sum_{l} \langle 1,\lambda_{q\bar{q}};1,\lambda_{g}|0,0 \rangle Y_{1l}(\bar{{\bf Q}}) \langle 0,0;1,l|1,\lambda_{ex} \rangle\propto[\epsilon^{\ast}(\lambda_{q\bar{q}})\cdot\epsilon^{\ast}(\lambda_{g})][\bar{{\bf Q}}\cdot\epsilon(\lambda_{ex})], \nonumber \\
& & \sum_{l,s} \langle 1,\lambda_{q\bar{q}};1,\lambda_{g}|1,s \rangle Y_{1l}(\bar{{\bf Q}}) \langle 1,s;1,l|1,\lambda_{ex} \rangle\propto[\epsilon^{\ast}(\lambda_{q\bar{q}})\times\epsilon^{\ast}(\lambda_{g})]\cdot[\bar{{\bf Q}}\times\epsilon(\lambda_{ex})], \nonumber \\
& & \sum_{l,s} \langle 1,\lambda_{q\bar{q}};1,\lambda_{g}|2,s \rangle Y_{1l}(\bar{{\bf Q}}) \langle 2,s;1,l|1,\lambda_{ex} \rangle\propto\bar{{\bf Q}}\cdot[\epsilon^{\ast}(\lambda_{q\bar{q}})\otimes\epsilon^{\ast}(\lambda_{g})]\cdot\epsilon(\lambda_{ex}),
\end{eqnarray}
and the action of the rotation matrix $D^{(1)}$ on the gluon states results in replacing $\epsilon^{i}(\lambda_{g})$ with $\epsilon^{i}_{c}({\bf Q},\lambda_{g})$. The normalized hybrid wave functions are then given by:
\begin{eqnarray}
& & \Psi^{\lambda_{ex}}_{q\bar{q}g(S=0)}=\sqrt{\frac{3}{8\pi}}\sum_{\lambda_{q\bar{q}}}\Psi^{\lambda_{q\bar{q}}}_{q\bar{q}}({\bf q},{\bf l}_{q\bar{q}}=-{\bf Q},\lambda_{q},\lambda_{\bar{q}})[{\bf \epsilon}^{\ast}(\lambda_{q\bar{q}})\cdot{\bf \epsilon}_{c}^{\ast}({\bf Q},\lambda_{g})][\bar{{\bf Q}}\cdot{\bf \epsilon}(\lambda_{ex})], \nonumber \\
& & \Psi^{\lambda_{ex}}_{q\bar{q}g(S=1)}=\sqrt{\frac{3}{8\pi}}\sum_{\lambda_{q\bar{q}}}\Psi^{\lambda_{q\bar{q}}}_{q\bar{q}}({\bf q},{\bf l}_{q\bar{q}}=-{\bf Q},\lambda_{q},\lambda_{\bar{q}})[{\bf \epsilon}^{\ast}(\lambda_{q\bar{q}})\times{\bf \epsilon}_{c}^{\ast}({\bf Q},\lambda_{g})]\cdot[\bar{{\bf Q}}\times{\bf \epsilon}(\lambda_{ex})], \nonumber \\
& & \Psi^{\lambda_{ex}}_{q\bar{q}g(S=2)}=\sqrt{\frac{27}{104\pi}}\sum_{\lambda_{q\bar{q}}}\Psi^{\lambda_{q\bar{q}}}_{q\bar{q}}({\bf q},{\bf l}_{q\bar{q}}=-{\bf Q},\lambda_{q},\lambda_{\bar{q}}) \nonumber \\
& & \times\,\bar{{\bf Q}}\cdot[{\bf \epsilon}^{\ast}(\lambda_{q\bar{q}})\otimes{\bf \epsilon}_{c}^{\ast}({\bf Q},\lambda_{g})]\cdot{\bf \epsilon}(\lambda_{ex}),
\label{eq:sex} 
\end{eqnarray}
where
\begin{equation}
(A\otimes B)_{ij}=A_{i}B_{j}+A_{j}B_{i}-\frac{2}{3}\delta_{ij}({\bf A}\cdot{\bf B}). 
\end{equation}

Writing the $q\bar{q}$ spin wave function more explicitly in terms of the quark momenta ${\bf p}_{q}$ and ${\bf p}_{\bar{q}}$ gives
\begin{eqnarray}
& & \Psi_{q\bar{q}g(S)}^{\lambda_{ex}}({\bf p}_{q},{\bf p}_{\bar{q}},\lambda_{q},\lambda_{\bar{q}},\lambda_{g})=-\frac{1}{\sqrt{2}m_{q\bar{q}}}\bar{u}({\bf p}_{q},\lambda_{q})\Bigl[\gamma^{\mu}-\frac{p^{\mu}_{q}-p^{\mu}_{\bar{q}}}{m_{q\bar{q}}+2m}\Bigr]v({\bf p}_{\bar{q}},\lambda_{\bar{q}}) \nonumber \\
& & \times\,\psi_{\mu(S)}(-{\bf p}_{q}-{\bf p}_{\bar{q}},\lambda_{g},\lambda_{ex}),
\label{eq:s}
\end{eqnarray}
where the gluon terms are respectively:
\begin{eqnarray}
& & \psi_{\mu(S=0)}({\bf Q},\lambda_{g},\lambda_{ex})=\sqrt{\frac{3}{8\pi}}\epsilon^{\ast}_{c\mu}({\bf Q},\lambda_{g})\bar{Q}^{l}\epsilon^{l}(\lambda_{ex}), \nonumber \\
& & \psi_{\mu(S=1)}({\bf Q},\lambda_{g},\lambda_{ex})=-\sqrt{\frac{3}{8\pi}}\Bigl[g^{k}_{\mu}-\frac{K_{\mu}K^{k}}{m_{q\bar{q}}(E_{q\bar{q}}+m_{q\bar{q}})}\Bigr]\epsilon^{\ast l}_{c}({\bf Q},\lambda_{g})\bar{Q}^{k}\epsilon^{l}(\lambda_{ex}), \nonumber \\
& & \psi_{\mu(S=2)}({\bf Q},\lambda_{g},\lambda_{ex})=\frac{3}{\sqrt{13}}\Bigl(\psi_{\mu(S=1)}({\bf Q},\lambda_{g},\lambda_{ex})-\frac{2}{3}\psi_{\mu(S=0)}({\bf Q},\lambda_{g},\lambda_{ex})\Bigr), 
\label{eq:ss}
\end{eqnarray}
and
\begin{eqnarray}
& & m_{q\bar{q}}=m_{q\bar{q}}({\bf p}_{q},{\bf p}_{\bar{q}}),\,\,\,\,E_{q\bar{q}}=E(m_{q\bar{q}},{\bf Q}), \nonumber \\
& & {\bf K}=-{\bf Q},\,\,\,\,K^{0}=E_{q\bar{q}}+m_{q\bar{q}},\,\,\,\,\epsilon_{c}^{0}({\bf Q},\lambda_{g})=0.
\end{eqnarray}
The loss of linear independence between all three functions $\psi_{\mu(S)}$ came from the replacement of $\epsilon^{i}(\lambda_{g})$ with $\epsilon^{i}_{c}({\bf Q},\lambda_{g})$, which is perpendicular to the momentum vector ${\bf Q}$.

The spin wave functions for other hybrid mesons can be constructed in similar fashion. In particular, in Chapter~7 we will describe the $\rho$ and $b_{1}$ mesons as gluonic bound states, and construct corresponding wave functions.

\chapter{Meson and hybrid states}

In the preceding chapter we constructed the spin wave functions for normal and hybrid mesons assuming that quarks do not interact, so each spin index can transform separately under a Wigner rotation.
The interaction between a quark and an antiquark enters through the Hamiltonian $H=P^{0}$ and the boost generators of the Poincar\'{e} group $M^{0i}$. 
It is possible to produce models of interaction for a fixed number of constituents that preserve the Poincar\'{e} algebra for noninteracting particles following the prescription of Bakamjian and Thomas, as we discussed in Chapter~4.
Unfortunately, such a construction does not guarantee that physical observables such as current matrix elements or decay amplitudes will be relativistically covariant.
Thus we must deal with phenomenological models of the quark dynamics, and in this chapter we will follow the common practice of employing a simple parametrization of the orbital wave functions.

\section{Mesons as $q\bar{q}$ bound states}

Unitary representations of noncompact groups are infinite-dimensional \cite{Wigner}. The rotation group is compact, because rotating by the angle $2\pi$ (or $4\pi$ for spinors) returns the transformed quantity back to the original state. However, this is not the case for boosts and therefore they do not form a compact group. This is reflected in the fact that the spinor representation of the Lorentz group ($\ref{unit}$) is not unitary.

In quantum mechanics we are only interested in a unitary representation of a symmetry group, because the transition probabilities between states do not depend on the choice of a frame of reference. The problem of non-unitarity of the Lorentz group is solved by introducing the Fock space in which states are described by kets $|{\bf p},\lambda\rangle$ with momentum ${\bf p}$ and spin $\lambda$. This representation is infinite-dimensional because the spectrum of values of ${\bf p}$ is continuous, and thus it is unitary. It is also irreducible, because the states have well-defined values of mass $m$ and spin $s$. 
In this section we will construct states for all mesons whose spin wave functions we have built in Chapter~5.

The $\pi\,(I=1)$ and $\eta\,(I=0)$ states ($J^{PC}=0^{-+}$), characterized by momentum ${\bf P}$ and spin $\lambda_{q\bar{q}}$, are constructed in terms of the annihilation and creation operators:
\begin{eqnarray}
& & |0^{-+}({\bf P},I,I_{3})\rangle =\sum_{\lambda,c,f}\int\frac{d^{3}{\bf p}_{q}}{(2\pi)^{3}2E(m,{\bf p}_{q})}\frac{d^{3}{\bf p}_{\bar{q}}}{(2\pi)^{3}2E(m,{\bf p}_{\bar{q}})}2(E(m,{\bf p}_{q})+E(m,{\bf p}_{\bar{q}})) \nonumber \\
& & \times\,(2\pi)^{3}\delta^{3}({\bf p}_{q}+{\bf p}_{\bar{q}}-{\bf P})\frac{1}{\sqrt{3}}\delta_{c_{q}c_{\bar{q}}}\langle\frac{1}{2},f_{q};\frac{1}{2},f_{\bar{q}}|I,I_{3}\rangle\Psi_{q\bar{q}}({\bf p}_{q},{\bf p}_{\bar{q}},\lambda_{q},\lambda_{\bar{q}}) \nonumber \\
& & \times\,\frac{1}{N(P)}\psi_{L}(m_{q\bar{q}}({\bf p}_{q},{\bf p}_{\bar{q}})/\mu)\,b^{\dag}_{{\bf p}_{q}\lambda_{q}f_{q}c_{q}}d^{\dag}_{{\bf p}_{\bar{q}}\lambda_{\bar{q}}f_{\bar{q}}c_{\bar{q}}}|0\rangle,
\label{eq:0-+}
\end{eqnarray}
where the operators satisfy the anticommutation relations:
\begin{eqnarray}
& & \{b_{{\bf p}\lambda fc},b^{\dag}_{{\bf p}'\lambda' f'c'}\}=\{d_{{\bf p}\lambda fc},d^{\dag}_{{\bf p}'\lambda' f'c'}\}=(2\pi)^{3}2E(m,{\bf p})\delta^{3}({\bf p}-{\bf p}')\delta_{\lambda\lambda'}\delta_{ff'}\delta_{cc'}, \nonumber \\
& & \{b_{{\bf p}\lambda fc},b_{{\bf p}'\lambda' f'c'}\}=\{d_{{\bf p}\lambda fc},d_{{\bf p}'\lambda' f'c'}\}=\{b^{\dag}_{{\bf p}\lambda fc},b^{\dag}_{{\bf p}'\lambda' f'c'}\}=\{d^{\dag}_{{\bf p}\lambda fc},d^{\dag}_{{\bf p}'\lambda' f'c'}\} =0, \nonumber \\
& & \{b_{{\bf p}\lambda fc},d_{{\bf p}'\lambda' f'c'}\}=\{b_{{\bf p}\lambda fc},d^{\dag}_{{\bf p}'\lambda' f'c'}\}=\{b^{\dag}_{{\bf p}\lambda fc},d_{{\bf p}'\lambda' f'c'}\}=\{b^{\dag}_{{\bf p}\lambda fc},d^{\dag}_{{\bf p}'\lambda' f'c'}\}=0. 
\label{anticom}
\end{eqnarray}
In the above, $\Psi_{q\bar{q}}$ represents the spin-0 wave function ($\ref{eq:00}$), written explicitly in terms of the momenta ${\bf p}_{q}$ and ${\bf p}_{\bar{q}}$ instead of the relativistic relative momentum ${\bf q}$ and the center-of-mass momentum ${\bf P}={\bf l}_{q\bar{q}}$ (Eq.~(\ref{eq:b1}) with ${\bf l}_{q}={\bf p}_{q}$ and Eq.~(\ref{eq:b2}) with ${\bf l}_{\bar{q}}={\bf p}_{\bar{q}}$).
The third component of isospin, flavor and color are respectively denoted by $I_{3}$, $f$ and $c$. The factor $\delta_{c_{q}c_{\bar{q}}}$ guarantees that the meson state is colorless.

The orbital wave function $\psi_{L}$ results from the strong and electroweak interaction between quarks that leads to a bound state (meson). Such a function depends on momenta only through the invariant mass of a quark-antiquark pair ($\ref{eq:mi1}$). Normalization constants are denoted by $N$ (with $P=|{\bf P}|$) and the $\mu$'s are free parameters, being scalar functions of meson quantum numbers. Finally, the isospin Clebsch-Gordan coefficient can be written as
\begin{eqnarray}
& & \langle\frac{1}{2},f_{q};\frac{1}{2},f_{\bar{q}}|1,I_{3}\rangle=\frac{1}{\sqrt{2}}\sigma^{i}_{f_{q}f_{\bar{q}}}\epsilon^{i}(I_{3}), \nonumber \\
& & \langle\frac{1}{2},f_{q};\frac{1}{2},f_{\bar{q}}|0,0\rangle=\frac{1}{\sqrt{2}}\delta_{f_{q}f_{\bar{q}}},
\label{isospmatr}
\end{eqnarray}
where $f=1$ for the $u$ quark (antiquark) and $f=2$ for the $d$.
The flavor structure of the $\eta$ state (as well as other isospin zero mesons) was chosen as a linear combination $\frac{1}{\sqrt{2}}(|u\bar{u}\rangle +|d\bar{d}\rangle)$, although in general these states are linear combinations $\cos(\phi)[|u\bar{u}\rangle+|d\bar{d}\rangle]/\sqrt{2}+\sin(\phi)|s\bar{s}\rangle$. The $|s\bar{s}\rangle$ does not contribute to the $\pi_{1}$ decay amplitude and therefore may be neglected in calculations, provided this amplitude is multiplied by a factor $\cos(\phi)$.

Similarly the $\rho\,(I=1)$ and $\omega,\,\phi\,(I=0)$ states ($J^{PC}=1^{--}$) are given by
\begin{eqnarray}
& & |1^{--}({\bf P},I,I_{3},\lambda)\rangle =\sum_{\lambda,c,f}\int\frac{d^{3}{\bf p}_{q}}{(2\pi)^{3}2E(m,{\bf p}_{q})}\frac{d^{3}{\bf p}_{\bar{q}}}{(2\pi)^{3}2E(m,{\bf p}_{\bar{q}})}2(E(m,{\bf p}_{q})+E(m,{\bf p}_{\bar{q}})) \nonumber \\
& & \times\,(2\pi)^{3}\delta^{3}({\bf p}_{q}+{\bf p}_{\bar{q}}-{\bf P})\frac{1}{\sqrt{3}}\delta_{c_{q}c_{\bar{q}}}\langle\frac{1}{2},f_{q};\frac{1}{2},f_{\bar{q}}|I,I_{3}\rangle\Psi_{q\bar{q}}^{\lambda}({\bf p}_{q},{\bf p}_{\bar{q}},\lambda_{q},\lambda_{\bar{q}}) \nonumber \\
& & \times\,\frac{1}{N(P)}\psi_{L}(m_{q\bar{q}}({\bf p}_{q},{\bf p}_{\bar{q}})/\mu)\,b^{\dag}_{{\bf p}_{q}\lambda_{q}f_{q}c_{q}}d^{\dag}_{{\bf p}_{\bar{q}}\lambda_{\bar{q}}f_{\bar{q}}c_{\bar{q}}}|0\rangle,
\label{eq:1--}
\end{eqnarray}
where $\Psi_{q\bar{q}}^{\lambda}$ denotes the spin-1 wave function ($\ref{eq:01}$).

The $b_{1}\,(I=1)$ and $h_{1}\,(I=0)$ mesons ($J^{PC}=1^{+-}$) have an additional orbital angular momentum $L=1$ represented by $Y_{1l}(\bar{{\bf q}})$, where ${\bf q}$ is the momentum of the constituent quark in the meson rest frame
\begin{equation}
{\bf q}({\bf p}_{q},{\bf P})=\Lambda({\bf P}\rightarrow0){\bf p}_{q}={\bf p}_{q}-\frac{E(m,{\bf p}_{q}){\bf P}}{m_{q\bar{q}}}+\frac{({\bf P}\cdot{\bf p}_{q}){\bf P}}{m_{q\bar{q}}(E(m_{q\bar{q}},-{\bf P})+m_{q\bar{q}})},
\label{eq:qq}
\end{equation}
with
\begin{equation}
m_{q\bar{q}}=m_{q\bar{q}}({\bf p}_{q},{\bf P}-{\bf p}_{q}),\,\,\,\,\bar{{\bf q}}={\bf q}/|{\bf q}|.
\end{equation}
The corresponding states are given by ($\ref{eq:1--}$), although the spin wave function $\Psi_{q\bar{q}}^{\lambda}$ is given now by ($\ref{eq:10}$). Finally, the $a\,(I=1)$ and $f\,(I=0)$ states ($J^{PC}=0,1,2^{++}$) are described by ($\ref{eq:1--}$) with the spin wave function ($\ref{eq:11}$).

The states are normalized
\begin{equation}
\langle {\bf P},\lambda,I_{3}|{\bf P}',\lambda',I'_{3}\rangle =(2\pi)^{3}2E(m_{M},{\bf P})\delta^{3}({\bf P}-{\bf P}')\delta_{\lambda\lambda'}\delta_{I_{3}I_{3}'},
\label{eq:norm1}
\end{equation}
where $m_{M}$ is the meson mass. That fixes the normalization constants,
\begin{eqnarray}
& & N^{2}_{M}(P)=(2E(m_{M},{\bf P}))^{-1}\int\frac{d^{3}{\bf k}}{(2\pi)^{3}}\frac{(E(m,{\bf k})+E(m,{\bf P}-{\bf k}))^{2}}{E(m,{\bf k})E(m,{\bf P}-{\bf k})}[Y_{L0}(\bar{{\bf q}}({\bf k},{\bf P}))]^{2} \nonumber \\
& & \times\,[\psi_{L}(m_{q\bar{q}}({\bf k},{\bf P}-{\bf k})/\mu_{M})]^{2},
\label{eq:norm2}
\end{eqnarray}
with ${\bf q}$ given in ($\ref{eq:qq}$) and $L$ being the orbital angular momentum of the meson. Without loss of generality we have taken ${\bf P}=P{\bf e}_{z}$.

The orbital angular momentum wave function for a meson depends on the potential between a quark and an antiquark. An explicit form of such a potential is not known exactly and such a function must be modeled. Because of Lorentz invariance it may depend on momenta only through the invariant mass of a $q\bar{q}$ pair. Moreover, it must tend to zero for large momenta fast enough to make the amplitude convergent. The simplest choice is the gaussian function
\begin{equation}
\psi_{L}(m_{q\bar{q}}({\bf p}_{q},{\bf p}_{\bar{q}})/\mu)=e^{-m^{2}_{q\bar{q}}({\bf p}_{q},{\bf p}_{\bar{q}})/8\mu^{2}}. 
\end{equation}
 The integrals in the decay amplitudes are not elementary and must be computed numerically. In the nonrelativistic limit (for a large $m$), however, they can be expressed in terms of the error function.

The free parameters of the model presented are: the quark masses $m$, the size parameters of the orbital wave functions $\mu$ and the strong coupling $g$. The pion decay constant $f_{\pi}$ and the elastic form factor $F_{\pi}$, defined by
\begin{equation}
\langle0|A^{\mu,i}({\bf 0})|\pi^{k}({\bf p})\rangle=f_{\pi}p^{\mu}\delta_{ik},
\label{eq:formf1a}
\end{equation}
and
\begin{equation}
\langle\pi^{i}({\bf p}')|V^{\mu,j}({\bf 0})|\pi^{k}({\bf p})\rangle=F_{\pi}(p^{\mu}+p'^{\mu})i\epsilon_{ijk},
\label{eq:formf1b}
\end{equation}
are used to constrain the $\pi$ wave function parameters (with the $\pi$ state given by ($\ref{eq:0-+}$)). The axial and the vector currents are defined by
\begin{equation}
A^{\mu,i}({\bf 0})=\bar{\psi}_{cf}({\bf 0})\gamma^{\mu}\gamma_{5}\frac{\sigma^{i}}{2}\psi_{cf},
\label{eq:curr1}
\end{equation}
and
\begin{equation}
V^{\mu,j}({\bf 0})=\bar{\psi}_{cf}({\bf 0})\gamma^{\mu}\frac{\sigma^{j}}{2}\psi_{cf},
\label{eq:curr2}
\end{equation}
with $\psi_{cf}({\bf x})$ given in ($\ref{eq:psi}$). By virtue of Lorentz invariance $f_{\pi}$ is a constant, whereas $F_{\pi}$ is a function of $Q^{2}=-({\bf p}-{\bf p}')^{2}$.

As mentioned previously, it is not possible to construct the wave functions with a fixed number of constituents in a Lorentz covariant way. Thus the current matrix elements are expected not to be exactly Lorentz covariant. This will be reflected, for example, in different values of $f_\pi$ obtained from spatial and time components of the axial current (rotational symmetry is not broken). Even if we replaced the factor $E(m_{M},{\bf P})$ in ($\ref{eq:norm1}$) by $1$, it would be very difficult to find the generators of the Poincare group that satisfy the commutation relations.
Thus, our model with the exponential orbital wave functions will not be exactly covariant. The resulting form factors will depend on the frame of reference. In order to obtain $F_{\pi}(Q^{2}=0)=1$, one typically employs the time component $\mu=0$ and works in the Breit frame of reference. In this case we obtain
\begin{equation}
f_{\pi}(P)=\frac{\sqrt{3}m}{N_{\pi}(P)E(m_{\pi},{\bf P})}\int\frac{d^{3}{\bf p}}{(2\pi)^{3}}\frac{(p^{0}+q^{0})^{2}}{p^{0}q^{0}m_{q\bar{q}}}e^{-\frac{m_{q\bar{q}}^{2}}{8\mu_{\pi}^{2}}},
\label{eq:formf2}
\end{equation}and
\begin{eqnarray} 
& & F_{\pi}({\bf P},{\bf P}')=\frac{1}{N_{\pi}(P)N_{\pi}(P')(E(m_{\pi},{\bf P})+E(m_{\pi},{\bf P}'))}\int\frac{d^{3}{\bf p}}{(2\pi)^{3}}\frac{(p^{0}+q^{0})(p^{0}+r^{0})}{m_{q\bar{q}}m_{q\bar{q}}'p^{0}q^{0}r^{0}} \nonumber \\
& & \times\,[(p\cdot q)r^{0}+(p\cdot r)q^{0}-(q\cdot r)p^{0}+m^{2}(p^{0}+q^{0}+r^{0})]e^{-\frac{m^{2}_{q\bar{q}}+m'^{2}_{q\bar{q}}}{8\mu_{\pi}^{2}}},
\label{eq:formf3}
\end{eqnarray}
where:
\begin{eqnarray} 
& & {\bf q}={\bf P}-{\bf p},\,\,\,\,{\bf r}={\bf P}'-{\bf p},\,\,\,\,p^{0}=E(m,{\bf p}),\,\,\,\,q^{0}=E(m,{\bf q}),\,\,\,\,r^{0}=E(m,{\bf r}), \nonumber \\
& & m_{q\bar{q}}({\bf p},{\bf q})=[(E(m,{\bf p})+E(m,{\bf q}))^{2}-({\bf p}+{\bf q})^{2}]^{1/2}, \nonumber \\
& & m_{q\bar{q}}=m_{q\bar{q}}({\bf p},{\bf q}),\,\,\,\,m'_{q\bar{q}}=m_{q\bar{q}}({\bf p},{\bf r}),
\end{eqnarray}
and the pion normalization constant is given in ($\ref{eq:norm2}$).
For other light unflavored mesons there are not enough experimental data to constrain their parameters $\mu$. However, they are expected to be on the same order as $\mu_{\pi}$.

By taking $m$ large as compared to the $\mu$'s and $P_{0}$, one obtains the nonrelativistic limit in which quarks are heavy. Their motion may be described by nonrelativistic quantum mechanics and, as we demonstrated at the end of Chapter~5, spin does not change via Wigner rotations. Therefore, all spin wave functions are just described by Clebsch-Gordan coefficients and spherical harmonics, and the spin factors in the decay amplitudes reduce to traces of products of Pauli matrices. All energy terms $E(m,{\bf p})$ tend to $m$, whereas the invariant masses ($\ref{eq:mi1}$) and ($\ref{eq:mi2}$) tend to $2m$ and $2m+E(m_{g},{\bf Q})$, respectively. In the orbital wave functions, however, we must keep the next leading terms depending on momenta, otherwise the amplitude would become divergent:
\begin{eqnarray}
& & m_{q\bar{q}}({\bf p}_{q},{\bf p}_{\bar{q}})\rightarrow 2m+\frac{({\bf p}_{q}-{\bf p}_{\bar{q}})^{2}}{4m}, \nonumber \\
& & m_{q\bar{q}g}({\bf p}_{q},{\bf p}_{\bar{q}},{\bf Q})\rightarrow 2m+m_{g}+\frac{{\bf p}^{2}_{q}+{\bf p}^{\bar{q}}_{2}}{2m}+E(m_{g},{\bf Q}).
\label{eq:imn}
\end{eqnarray}
In the above we have ${\bf Q}=-{\bf p}_{q}-{\bf p}_{\bar{q}}$ because the $q\bar{q}g$ state is at rest.
The normalization constants are given in this limit by
\begin{equation}
N^{2}_{M}(P)=2E^{-1}(m_{M},{\bf P})\int\frac{d^{3}{\bf k}}{(2\pi)^{3}}[Y_{L0}(\bar{{\bf q}}({\bf k},{\bf P}))]^{2}[\psi_{L}(m_{q\bar{q}}({\bf k},{\bf P}-{\bf k})/\mu_{M})]^{2},
\label{eq:normNR1}
\end{equation}
where $L$ is the orbital angular momentum of a meson.
For the decay amplitudes we will not derive the nonrelativistic formulae from the beginning, but instead, we will go with $m$ to very large values and keep only the leading terms.

\section{Exotic mesons as $q\bar{q}g$ bound states}

In our model a hybrid is regarded as a bound state of a quark, an antiquark and a gluon. Therefore, we can construct it in terms of the annihilation and creation operators.
The $\pi_{1}$ state $I^{G}(J^{PC})=1^{-}(1^{-+})$ in its rest frame is given by
\begin{eqnarray}
& & |ex(I_{3},\lambda_{ex})\rangle =\sum_{\lambda,c,f}\frac{1}{N_{ex}}\int\frac{d^{3}{\bf p}_{q}}{(2\pi)^{3}2E(m,{\bf p}_{q})}\frac{d^{3}{\bf p}_{\bar{q}}}{(2\pi)^{3}2E(m,{\bf p}_{\bar{q}})}\frac{d^{3}{\bf Q}}{(2\pi)^{3}2E(m_{g},{\bf Q})} \nonumber \\
& & \times\,2(E(m,{\bf p}_{q})+E(m,{\bf p}_{\bar{q}})+E(m_{g},{\bf Q}))(2\pi)^{3}\delta^{3}({\bf p}_{q}+{\bf p}_{\bar{q}}+{\bf Q}) \nonumber \\
& & \times\,\Psi^{\lambda_{ex}}_{q\bar{q}g}({\bf p}_{q},{\bf p}_{\bar{q}},\lambda_{q},\lambda_{\bar{q}},\lambda_{g})\psi_{L}'(m_{q\bar{q}}({\bf p}_{q},{\bf p}_{\bar{q}})/\mu_{ex},m_{q\bar{q}g}({\bf p}_{q},{\bf p}_{\bar{q}},{\bf Q})/\mu_{ex'}) \nonumber \\
& & \times\,\frac{1}{2}\lambda^{c_{g}}_{c_{q}c_{\bar{q}}}\langle\frac{1}{2},f_{q};\frac{1}{2},f_{\bar{q}}|I,I_{3}\rangle b^{\dag}_{{\bf p}_{q}\lambda_{q}f_{q}c_{q}}d^{\dag}_{{\bf p}_{\bar{q}}\lambda_{\bar{q}}f_{\bar{q}}c_{\bar{q}}}a^{\dag}_{{\bf Q}\lambda_{g}c_g}|0\rangle,
\label{eq:1-+}
\end{eqnarray}
where the spin wave function $\Psi_{q\bar{q}g}$ was given in ($\ref{eq:s}$) for $S=0,1,2$. The orbital wave function $\psi_{L}'$ depends only on the invariant mass of a quark-antiquark pair $m_{q\bar{q}}$ and the invariant mass of a 3-body system,
\begin{equation}
m_{q\bar{q}g}({\bf p}_{q},{\bf p}_{\bar{q}},{\bf Q})=E(m,{\bf p}_{q})+E(m,{\bf p}_{\bar{q}})+E(m_{g},{\bf Q}).
\label{eq:mi2}
\end{equation}
Here $m_{g}$ is the dynamical mass of a gluon in the Coulomb gauge (arising from the strong interaction with virtual particles), and $\lambda^{c_{g}}_{c_{q}c_{\bar{q}}}$ are the SU(3) Gell-Mann matrices. They guarantee that a hybrid meson is colorless. The gluon operators satisfy the commutation relations:
\begin{eqnarray}
& & [a_{{\bf p}\lambda c},a^{\dag}_{{\bf p}'\lambda' c'}]=(2\pi)^{3}2E(m_{g},{\bf p})\delta^{3}({\bf p}-{\bf p}')\delta_{\lambda\lambda'}\delta_{cc'}, \nonumber \\
& & [a_{{\bf p}\lambda c},a_{{\bf p}'\lambda' c'}]=[a^{\dag}_{{\bf p}\lambda c},a^{\dag}_{{\bf p}'\lambda' c'}]= 0.
\end{eqnarray}
The normalization ($\ref{eq:norm1}$) leads to
\begin{eqnarray}
& & N^{2}_{ex}=\frac{3}{4\pi}(2m_{ex})^{-1}\int\frac{d^{3}{\bf k}}{(2\pi)^{3}}\frac{d^{3}{\bf l}}{(2\pi)^{3}}\frac{(E(m,{\bf k})+E(m,{\bf l})+E(m_{g},-{\bf k}-{\bf l}))^{2}}{2E(m,{\bf k})E(m,{\bf l})E(m_{g},-{\bf k}-{\bf l})}\frac{(k_{z}+l_{z})^{2}}{({\bf k}+{\bf l})^{2}} \nonumber \\
& & \times\,[\psi_{L}'(m_{q\bar{q}}({\bf k},{\bf l})/\mu_{ex},m_{q\bar{q}g}({\bf k},{\bf l},-{\bf k}-{\bf l})/\mu_{ex'})]^{2}. 
\label{eq:norm3}
\end{eqnarray}
Hybrid mesons with other quantum numbers can be constructed in similar fashion.

The covariant orbital wave function of the $\pi_{1}$ may depend only on the invariant masses $m_{q\bar{q}}$ and $m_{q\bar{q}g}$. A natural choice is a product of two gaussian functions, 
\begin{equation}
\psi'_{L}(m_{q\bar{q}}({\bf p}_{q},{\bf p}_{\bar{q}})/\mu_{ex},m_{q\bar{q}g}({\bf p}_{q},{\bf p}_{\bar{q}},{\bf Q})/\mu'_{ex})=e^{-m^{2}_{q\bar{q}}({\bf p}_{q},{\bf p}_{\bar{q}})/8\mu^{2}_{ex}}e^{-m^{2}_{q\bar{q}g}({\bf p}_{q},{\bf p}_{\bar{q}},{\bf Q})/8\mu'^{2}_{ex}}. 
\end{equation}
In the nonrelativistic limit the $\pi_{1}$ normalization constant is given by
\begin{equation}
N^{2}_{ex}=\frac{3}{4\pi}\frac{(2m+m_{g})^{2}}{4m^{2}m_{g}m_{ex}}\int\frac{d^{3}{\bf k}}{(2\pi)^{3}}\frac{d^{3}{\bf l}}{(2\pi)^{3}}\frac{(k_{z}+l_{z})^{2}}{({\bf k}+{\bf l})^{2}}[\psi_{L}'(m_{q\bar{q}}({\bf k},{\bf l})/\mu_{ex},m_{q\bar{q}g}({\bf k},{\bf l},-{\bf k}-{\bf l})/\mu_{ex'})]^{2}.
\label{eq:normNR2}
\end{equation}

\chapter{Decay of normal mesons}

\begin{figure}[h]
\centering
\includegraphics[width=4.0in]{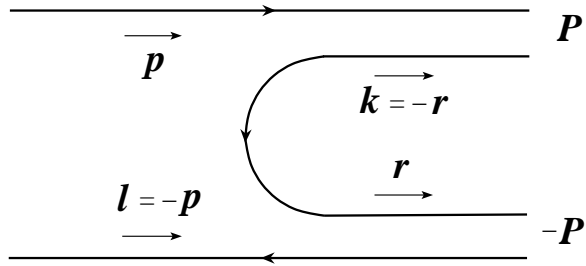}
\caption{\label{3p0} The $^{3}P_0$ decay of a normal meson. }
\end{figure}

\begin{figure}
\centering
\includegraphics[width=4.0in]{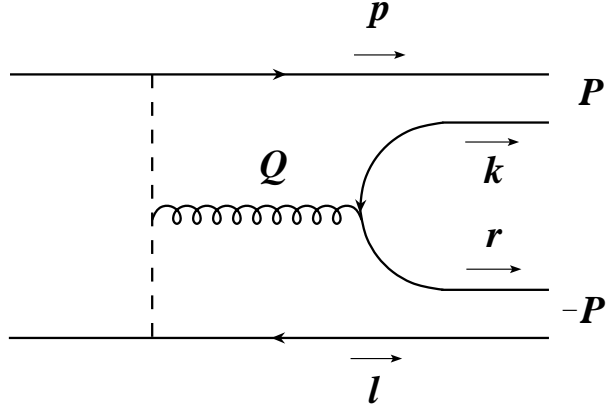}
\caption{\label{mesondec} The $^{3}S_1$ decay of a normal meson via its $q\bar{q}g$ component. }
\end{figure}

The common approach to normal meson decays is based on the $^3P_0$ model where $q{\bar q}$ pair creation is described by an effective operator that creates this pair from the vacuum in the presence of the normal $q{\bar q}$ component of the decaying meson, as shown in Fig.~\ref{3p0}
In the QCD-motivated Coulomb gauge, however, the decay of a normal meson is expected to proceed via mixing of the $q{\bar q}$ state with the $q{\bar q}g$ hybrid component followed by gluon dissociation to a $q{\bar q}$ pair, as shown in Fig.~\ref{mesondec}. The dashed line represents the confining non-abelian Coulomb potential. 
The hybrid component of the wave function is obtained by integrating the $q\bar{q}$ wave function over the amplitude of the transverse gluon emission from the Coulomb line~\cite{ases1}. The quantum numbers $P,C$ of such a $q\bar{q}$ state are determined from the corresponding conservation laws, whereas its spin may have more than one value (denoted in this chapter by $J$). 

We will be interested in estimating the size of relativistic effects in meson decays, and not in giving the absolute width predictions. Therefore, we can make calculations for each value of $J$ separately. The relative contributions from various $J$ and the total width may be obtained from the above gluon emission amplitude.
Since the quark pair is emitted in the $S=1$, $L=0$ state, this decay mechanism is also referred to as the $^{3}S_{1}$ model.

The Hamiltonian $H$ of gluon dissociation is equal to
\begin{equation}
H=\sum_{c,f}\int d^{3}{\bf x}\,\bar{\psi}_{c_{1}f_{1}}({\bf x})(g{\bf \gamma}\cdot{\bf A}^{c_{g}}({\bf x}))\psi_{c_{2}f_{2}}({\bf x})\delta_{f_{1}f_{2}}\frac{1}{2}\lambda^{c_{g}}_{c_{1}c_{2}}.
\label{H}
\end{equation}
In the constituent basis used here, the single-particle quark and antiquark wave functions correspond to the states of massive particles with a relativistic dispersion relation, in which the running quark mass is approximated by a constant constituent mass $m$,
\begin{equation}
\psi_{cf}({\bf x})=\sum_{\lambda}\int\frac{d^{3}{\bf k}}{(2\pi)^{3}2E(m,{\bf k})}[u({\bf k},\lambda)b_{{\bf k}\lambda cf}+v(-{\bf k},\lambda)d^{\dag}_{-{\bf k}\lambda cf}]e^{i{\bf k}\cdot{\bf x}}.
\label{eq:psi}
\end{equation}
Similarly, the gluon field ${\bf A}^{c_{g}}$ is expanded in a basis of transverse polarization vectors, with a single-particle wave function characterizing a state with mass $m_g$,
\begin{equation}
{\bf A}^{c_{g}}({\bf x})=\sum_{\lambda}\int\frac{d^{3}{\bf k}}{(2\pi)^{3}2E(m_{g},{\bf k})}[{\bf \epsilon}_{c}({\bf k},\lambda)a^{c_{g}}_{{\bf k}\lambda}+{\bf \epsilon}_{c}^{\ast}(-{\bf k},\lambda)a^{\dag c_{g}}_{-{\bf k}\lambda}]e^{i{\bf k}\cdot{\bf x}}.
\label{eq:g}
\end{equation}
Here $g$ is the strong coupling constant. The Hamiltonian part contributing to the decay amplitude is
\begin{eqnarray}
& & H=\sum_{\lambda,c,f}\int\frac{d^{3}{\bf k}_{1}}{(2\pi)^{3}2E(m,{\bf k}_{1})}\frac{d^{3}{\bf k}_{2}}{(2\pi)^{3}2E(m,{\bf k}_{2})}\frac{d^{3}{\bf k}}{(2\pi)^{3}2E(m_{g},{\bf k})}(2\pi)^{3}\delta^{3}({\bf k}-{\bf k}_{1}+{\bf k}_{2})\delta_{f_{1}f_{2}} \nonumber \\
& & \times\,\frac{1}{2}\lambda^{c_{g}}_{c_{1}c_{2}}\bar{u}({\bf k}_{1},\lambda_{1})(g\gamma^{j}\epsilon_{c}^{j}({\bf k},\lambda_{g}))v(-{\bf k}_{2},\lambda_{2})b^{\dag}_{{\bf k}_{1}\lambda_{1}c_{1}f_{1}}d^{\dag}_{-{\bf k}_{2}\lambda_{2}c_{2}f_{2}}a^{c_{g}}_{{\bf k}\lambda_{g}}.
\end{eqnarray}

In this chapter we will study the decays of the $\rho$ and $b_{1}$ since they are dominated (almost $100\%$) by a single mode, and their widths are well-known from experiment. Therefore they can be used to test the model presented in this work. Because we are interested in widths, all calculations will be done in the rest frame of a decaying meson.

\section{Decay $\rho\rightarrow2\pi$}

We will start from the $^{3}P_{0}$ Hamiltonian:
\begin{equation}
H=\Lambda\sum_{c,f}\int d^{3}{\bf x}\,\bar{\psi}_{c_{1}f_{1}}({\bf x})\psi_{c_{2}f_{2}}({\bf x})\delta_{f_{1}f_{2}}\delta_{c_{1}c_{2}},
\label{eq:H2}
\end{equation}
with $\psi$ defined in ($\ref{eq:psi}$) and $\Lambda$ being a mass scale which can be fixed by the absolute decay width and is expected to be of the order of the average quark momentum.
The amplitude of this mode is determined by the matrix element
\begin{equation}
\langle\pi({\bf P}),\pi({-\bf P})|H|\rho\rangle.
\end{equation}
The corresponding states are given by Eqs.~(\ref{eq:0-+}) and (\ref{eq:1--}).

In this matrix element we have two operators $b$ and two $d$ coming from the outgoing meson states, whereas the decaying meson state provides one $b^{\dag}$ and one $d^{\dag}$. Moreover, we get one $b^{\dag}$ and one $d^{\dag}$ from the Hamiltonian. Thus each pair $b,\,b^{\dag}$ and $d,\,d^{\dag}$ appears twice and Wick's rearrangement leads to two nonzero terms. The anticommutation relations (\ref{anticom}) give Dirac delta functions that guarantee the conservation of momentum and Kronecker deltas acting in flavor and color space. This simplifies the integration over momenta and reduces summations over flavor and color indices to calculating traces of corresponding matrix products.
   
The two terms in the above matrix element will be equal after integrating (up to a sign) because of symmetry, so it is enough to deal with only one and multiply the final expression for the amplitude by 2. A short proof of this statement will be given for the $\pi_{1}$ decay in the next chapter. Summing over color gives a factor $1/\sqrt{3}$,
whereas summing over flavor leads to a trace of a product of Pauli matrices appearing in the isospin factors (\ref{isospmatr}),
\begin{equation}
2^{-3/2}Tr(\sigma^{\rho}_{i}\sigma^{\pi}_{j}\sigma^{\pi}_{k})\epsilon_{i}(I_{3}^{\rho})\epsilon_{j}^{\ast}(I_{3}^{\pi})\epsilon_{k}^{\ast}(I_{3}^{\pi}).
\end{equation}
For all possible isospin channels the flavor factor is equal to $\pm 1/\sqrt{2}$.
For spin we obtain
\begin{equation}
W^{\lambda_{\rho}}({\bf p},{\bf r},{\bf k},{\bf l})=\frac{Tr\Bigl[(\not{k}+m)(\not{p}+m)\Bigl(\gamma^{i}-\frac{p^{i}-l^{i}}{m_{q\bar{q}}({\bf p},{\bf l})+2m}\Bigr)(\not{l}-m)(\not{r}-m)\Bigr]\epsilon^{i}(\lambda_{\rho})}{2^{3/2}m_{q\bar{q}}({\bf p},{\bf k})m_{q\bar{q}}({\bf r},{\bf l})m_{q\bar{q}}({\bf p},{\bf l})},
\label{eq:trrho}
\end{equation}
where ${\bf p}$ and ${\bf l}$ denote respectively the momenta of a quark and an antiquark in $\rho$, and ${\bf r}$ and ${\bf k}$ denote respectively the momenta of a quark and an antiquark created from the vacuum, as shown in Fig.~\ref{3p0}. We assume that all quarks are on-shell particles, i.e.:
\begin{equation}
p^{0}=E(m,{\bf p}),\,\,\,\,r^{0}=E(m,{\bf r}),\,\,\,\,k^{0}=E(m,{\bf k}),\,\,\,\,l^{0}=E(m,{\bf l}). 
\end{equation}
   
Integration over momenta gives $(2\pi)^{3}\delta^{3}({\bf 0})A({\bf P})$, where $A$ denotes the amplitude of this decay,
\begin{eqnarray}
& & A({\bf P},\lambda_{\rho})=\frac{2\Lambda}{\sqrt{6}N^{2}_{\pi}(P)N_{\rho}(0)}\int\frac{d^{3}{\bf p}}{(2\pi)^{3}}\frac{(E(m,{\bf p})+E(m,{\bf P}-{\bf p}))^{2}}{E(m,{\bf p})E^{2}(m,{\bf P}-{\bf p})} \nonumber \\
& & \times\,W^{\lambda_{\rho}}({\bf p},{\bf p}-{\bf P},{\bf P}-{\bf p},-{\bf p})[\psi_{L}(m_{q\bar{q}}({\bf p},{\bf P}-{\bf p})/\mu_{\pi})]^{2} \nonumber \\
& & \times\,\psi_{L}((E(m,{\bf p})+E(m,{\bf P}-{\bf p}))/\mu_{\rho}).
\label{eq:intrho}
\end{eqnarray}
The corresponding width is obtained from
\begin{equation}
d\Gamma=\frac{1}{32\pi^{2}}\frac{P_{0}|A({\bf P}_{0})|^{2}}{m^{2}_{\rho}}d\Omega,
\end{equation}
with ${\bf P}_{0}$ satisfying
\begin{equation}
P_{0}=\sqrt{\frac{m^{2}_{\rho}}{4}-m^{2}_{\pi}}
\end{equation}
and $P_{0}=|{\bf P}_{0}|$. The amplitude must be, according to the Wigner-Eckart theorem, of the form:
\begin{equation}
A({\bf P},\lambda_{\rho})=a_{1}(P)Y_{1\lambda_{\rho}}({\bf P}/P),
\end{equation}
where $a_{1}$ is the P-wave partial amplitude. Thus
\begin{equation}
\Gamma=\frac{P_{0}}{32\pi^{2}m^{2}_{\rho}}a^{2}_{1}(P_{0}).
\end{equation}
Taking ${\bf P}=Pe_{z}$ and $\lambda_{\rho}=0$ gives
\begin{equation}
a_{1}(P)=\sqrt{\frac{4\pi}{3}}A(Pe_{z},0),
\end{equation}
therefore the width is given by
\begin{equation}
\Gamma_{\rho\rightarrow2\pi}^{(p-wave)}=\frac{P_{0}}{24\pi m^{2}_{\rho}}A^{2}(P_{0}e_{z},0).
\label{eq:widrho}
\end{equation}
For large $m$ (in the nonrelativistic limit) the trace term tends to the value
\begin{equation}
\sqrt{2}(p^{i}-P^{i})\epsilon^{i}(\lambda_{\rho}),
\end{equation}
and the amplitude ($\ref{eq:intrho}$) simplifies to
\begin{equation}
A({\bf P},\lambda_{\rho})=\frac{8\Lambda}{\sqrt{3}mN^{2}_{\pi}(P)N_{\rho}(0)}\int\frac{d^{3}{\bf p}}{(2\pi)^{3}}(p^{i}-P^{i})\epsilon^{i}(\lambda_{\rho})[\psi_{L}^{\pi}]^{2}\psi_{L}^{\rho}.
\end{equation}
Here the normalization constants are given by ($\ref{eq:normNR1}$), and in the orbital wave functions $\psi_{L}$ we need to expand the invariant masses and energies only up to terms quadratic in momenta.

Now we proceed to the $^{3}S_{1}$ model.
For the $\rho$ meson the $q{\bar q}g$ component can be expanded in a basis of the $a_0$, $a_1$, $a_2$ wave functions, all having spin 1 and one unit of the orbital angular momentum between the quark and the antiquark ($\ref{eq:11}$), all coupled with a transverse gluon wave function to give the $J^{PC}=1^{--}$ state.
The wave functions for the $\rho$ are thus:
\begin{equation}
\Psi^{\lambda_{\rho}}_{q\bar{q}g(J)}(\lambda_{q},\lambda_{\bar{q}},\lambda_{g})=\sum_{\lambda_{q\bar{q}},\sigma=\pm1}\Psi^{J,\lambda_{q\bar{q}}}_{q\bar{q}}({\bf q},{\bf l}_{q\bar{q}}=-{\bf Q},\lambda_{q},\lambda_{\bar{q}})D^{(1)\ast}_{\lambda_{g}\sigma}(\bar{{\bf Q}})\langle J,\lambda_{q\bar{q}};1,\sigma|1,\lambda_{\rho}\rangle.
\label{eq:exotrho}
\end{equation}
where $\Psi^{J,\lambda_{q\bar{q}}}_{q\bar{q}}$ are the $a_0, a_1$ and $a_2$ $q{\bar q}$ wave functions for $J=0,1,2$ respectively.
The normalized wave functions for the $\rho$ are then given, similarly to those for the $\pi_{1}$ ($\ref{eq:sex}$), by:
\begin{eqnarray}
& & \Psi^{\lambda_{\rho}}_{q\bar{q}g(J=0)}=\sqrt{\frac{3}{8\pi}}\sum_{\lambda_{q\bar{q}}}\Psi^{\lambda_{q\bar{q}}}_{q\bar{q}}({\bf q},{\bf l}_{q\bar{q}}=-{\bf Q},\lambda_{q},\lambda_{\bar{q}})[{\bf \epsilon}^{\ast}(\lambda_{q\bar{q}})\cdot\bar{{\bf q}}][{\bf \epsilon}_{c}^{\ast}({\bf Q},\lambda_{g})\cdot{\bf \epsilon}(\lambda_{\rho})], \nonumber \\
& & \Psi^{\lambda_{\rho}}_{q\bar{q}g(J=1)}=\sqrt{\frac{9}{32\pi}}\sum_{\lambda_{q\bar{q}}}\Psi^{\lambda_{q\bar{q}}}_{q\bar{q}}({\bf q},{\bf l}_{q\bar{q}}=-{\bf Q},\lambda_{q},\lambda_{\bar{q}})[{\bf \epsilon}^{\ast}(\lambda_{q\bar{q}})\times\bar{{\bf q}}]\cdot[{\bf \epsilon}_{c}^{\ast}({\bf Q},\lambda_{g})\times{\bf \epsilon}(\lambda_{\rho})], \nonumber \\
& & \Psi^{\lambda_{\rho}}_{q\bar{q}g(J=2)}=\sqrt{\frac{27}{160\pi}}\sum_{\lambda_{q\bar{q}}}\Psi^{\lambda_{q\bar{q}}}_{q\bar{q}}({\bf q},{\bf l}_{q\bar{q}}=-{\bf Q},\lambda_{q},\lambda_{\bar{q}}) \nonumber \\
& & \times\,{\bf \epsilon}_{c}^{\ast}({\bf Q},\lambda_{g})\cdot[{\bf \epsilon}^{\ast}(\lambda_{q\bar{q}})\otimes\bar{{\bf q}}]\cdot{\bf \epsilon}(\lambda_{\rho}),
\end{eqnarray}
where $\Psi^{\lambda_{q\bar{q}}}_{q\bar{q}}$ is the spin-1 wave function ($\ref{eq:01}$). Writing this function more explicitly in terms of the quark momenta ${\bf p}_{q}$ and ${\bf p}_{\bar{q}}$ gives
\begin{eqnarray}
& & \Psi_{q\bar{q}g(J)}^{\lambda_{\rho}}({\bf p}_{q},{\bf p}_{\bar{q}},\lambda_{q},\lambda_{\bar{q}},\lambda_{g})=-\frac{1}{\sqrt{2}m_{q\bar{q}}}\bar{u}({\bf p}_{q},\lambda_{q})\Bigl[\gamma^{\mu}-\frac{p^{\mu}_{q}-p^{\mu}_{\bar{q}}}{m_{q\bar{q}}+2m}\Bigr]v({\bf p}_{\bar{q}},\lambda_{\bar{q}}) \nonumber \\
& & \times\,\psi_{\mu(J)}(-{\bf p}_{q}-{\bf p}_{\bar{q}},\lambda_{g},\lambda_{\rho}),
\label{eq:rex1}
\end{eqnarray}
where the gluon terms are respectively:
\begin{eqnarray}
& & \psi_{\mu(J=0)}({\bf Q},\lambda_{g},\lambda_{\rho})=-\sqrt{\frac{3}{8\pi}}\Bigl[g^{k}_{\mu}-\frac{K_{\mu}K^{k}}{m_{q\bar{q}}(E_{q\bar{q}}+m_{q\bar{q}})}\Bigr](\delta^{kl}\delta^{mn}){\bar{q}}^{l}\epsilon^{\ast m}_{c}({\bf Q},\lambda_{g})\epsilon^{n}(\lambda_{\rho}), \nonumber \\
& & \psi_{\mu(J=1)}({\bf Q},\lambda_{g},\lambda_{\rho})=-\sqrt{\frac{9}{32\pi}}\Bigl[g^{k}_{\mu}-\frac{K_{\mu}K^{k}}{m_{q\bar{q}}(E_{q\bar{q}}+m_{q\bar{q}})}\Bigr](\delta^{km}\delta^{ln}-\delta^{kn}\delta^{lm}){\bar{q}}^{l} \nonumber \\
& & \times\,\epsilon^{\ast m}_{c}({\bf Q},\lambda_{g})\epsilon^{n}(\lambda_{\rho}), \nonumber \\
& & \psi_{\mu(J=2)}({\bf Q},\lambda_{g},\lambda_{\rho})=-\sqrt{\frac{27}{160\pi}}\Bigl[g^{k}_{\mu}-\frac{K_{\mu}K^{k}}{m_{q\bar{q}}(E_{q\bar{q}}+m_{q\bar{q}})}\Bigr](\delta^{km}\delta^{ln}+\delta^{kn}\delta^{lm}-\frac{2}{3}\delta^{kl}\delta^{mn}) \nonumber \\
& & \times\,{\bar{q}}^{l}\epsilon^{\ast m}_{c}({\bf Q},\lambda_{g})\epsilon^{n}(\lambda_{\rho}),
\label{eq:rex2}
\end{eqnarray}
and
\begin{eqnarray}
& & m_{q\bar{q}}=m_{q\bar{q}}({\bf p}_{q},{\bf p}_{\bar{q}}),\,\,\,\,E_{q\bar{q}}=E(m_{q\bar{q}},{\bf Q}),\,\,\,\,{\bf K}=-{\bf Q}, \nonumber \\
& & K^{0}=E_{q\bar{q}}+m_{q\bar{q}},\,\,\,\,\epsilon_{c}^{0}({\bf Q},\lambda_{g})=0.
\end{eqnarray}
As before, ${\bf q}={\bf q}({\bf p}_{q},-{\bf Q})$ denotes the quark momentum in the rest frame of the $q\bar{q}$ pair ($\ref{eq:qq}$). The most general wave function will be given by a linear combination of the three components listed above, and the coefficients in this are provided by the $q\bar{q}g$ component mentioned at the beginning of this chapter.

The Hamiltonian matrix element leads again to two equal terms. Summation over flavor indices gives $\pm1/\sqrt{2}$ as before, whereas for color one obtains
\begin{equation}
\frac{1}{12}\lambda^{a}_{bc}\lambda^{a}_{cb}=\frac{4}{3}.
\end{equation}
Summation over spin gives
\begin{equation}
B^{\mu}_{\,\,\,j}\psi_{\mu j}^{(J)},
\label{spinf1}
\end{equation}
where
\begin{equation}
B^{\mu j}=\frac{Tr\Bigl[(\not{k}-m)(\not{p}-m)\Bigl(\gamma^{\mu}+\frac{p^{\mu}-l^{\mu}}{m_{q\bar{q}}({\bf p},{\bf l})+2m}\Bigr)(\not{l}+m)(\not{r}+m)\gamma^{j}\Bigr]}{2^{3/2}m_{q\bar{q}}({\bf p},{\bf k})m_{q\bar{q}}({\bf r},{\bf l})m_{q\bar{q}}({\bf p},{\bf l})},
\label{eq:tr1}
\end{equation}
and
\begin{equation}
\psi_{\mu}^{\,\,j(J)}(\lambda_{\rho})=\sum_{\lambda_{g}}\psi_{\mu(J)}({\bf Q},\lambda_{g},\lambda_{\rho})\epsilon^{j}_{c}({\bf Q},\lambda_{g}).
\label{eq:trs}
\end{equation}
The tensor $B^{\mu}_{\,\,\,j}$ corresponds to the first term contributing to the amplitude, and the notation is the same as in Fig.~\ref{mesondec}. The functions (\ref{eq:trs}) are given by:
\begin{eqnarray}
& & \psi_{\mu(J=0)}^{j}({\bf Q},\lambda_{g},\lambda_{\rho})=-\sqrt{\frac{3}{8\pi}}\Bigl[g^{k}_{\mu}-\frac{K_{\mu}K^{k}}{m_{q\bar{q}}(E_{q\bar{q}}+m_{q\bar{q}})}\Bigr]{\bar{q}}^{l}(\delta^{jm}-\bar{Q}^{j}\bar{Q}^{m})\epsilon^{n}(\lambda_{\rho})(\delta^{kl}\delta^{mn}), \nonumber \\
& & \psi_{\mu(J=1)}^{j}({\bf Q},\lambda_{g},\lambda_{\rho})=-\sqrt{\frac{9}{32\pi}}\Bigl[g^{k}_{\mu}-\frac{K_{\mu}K^{k}}{m_{q\bar{q}}(E_{q\bar{q}}+m_{q\bar{q}})}\Bigr]{\bar{q}}^{l}(\delta^{jm}-\bar{Q}^{j}\bar{Q}^{m})\epsilon^{n}(\lambda_{\rho}) \nonumber \\
& & \times\,(\delta^{km}\delta^{ln}-\delta^{kn}\delta^{lm}), \nonumber \\
& & \psi_{\mu(J=2)}^{j}({\bf Q},\lambda_{g},\lambda_{\rho})=-\sqrt{\frac{27}{160\pi}}\Bigl[g^{k}_{\mu}-\frac{K_{\mu}K^{k}}{m_{q\bar{q}}(E_{q\bar{q}}+m_{q\bar{q}})}\Bigr]{\bar{q}}^{l}(\delta^{jm}-\bar{Q}^{j}\bar{Q}^{m})\epsilon^{n}(\lambda_{\rho}) \nonumber \\
& & \times\,(\delta^{km}\delta^{ln}+\delta^{kn}\delta^{lm}-\frac{2}{3}\delta^{kl}\delta^{mn}).
\label{eq:rex3}
\end{eqnarray}
Consequently, the width can be determined from ($\ref{eq:widrho}$). In the nonrelativistic limit
\begin{equation}
B^{ij}\rightarrow-\sqrt{2}m\delta^{ij},
\label{tr1n}
\end{equation}
and the other components are of higher order in small quantities. Therefore:
\begin{eqnarray}
& & B^{\mu}_{\,\,\,j}\psi_{\mu j}^{(0)}\rightarrow-\sqrt{\frac{3}{4\pi}}m\bar{q}^{i}\epsilon^{j}(\lambda_{\rho})(\delta^{ij}-\bar{Q}^{i}\bar{Q}^{j}), \nonumber \\
& & B^{\mu}_{\,\,\,j}\psi_{\mu j}^{(1)}\rightarrow-\sqrt{\frac{9}{16\pi}}m\bar{q}^{i}\epsilon^{j}(\lambda_{\rho})(\delta^{ij}+\bar{Q}^{i}\bar{Q}^{j}), \nonumber \\
& & B^{\mu}_{\,\,\,j}\psi_{\mu j}^{(2)}\rightarrow-\sqrt{\frac{27}{80\pi}}m\bar{q}^{i}\epsilon^{j}(\lambda_{\rho})\frac{1}{3}(7\delta^{ij}-\bar{Q}^{i}\bar{Q}^{j}).
\end{eqnarray}
None of these functions vanishes. However, only two of them remain linearly independent.

\section{Decay $b_{1}\rightarrow\pi\omega$}

This process is a better test for this model because the ratio of the D-wave to the S-wave width rates is independent of the values of $\Lambda$ and $g$. 
We begin with the $b_{1}$ as a $q\bar{q}$ bound state and the decay Hamiltonian ($\ref{eq:H2}$).
The amplitude of this mode is determined by the matrix element
\begin{equation}
\langle\pi({\bf P}),\omega({-\bf P})|H|b_{1}\rangle.
\end{equation}
This will lead to two equal terms, as before. Summation over color and flavor gives respectively $1/\sqrt{3}$ and $\pm1/\sqrt{2}$, whereas the spin factor is given by
\begin{equation}
W^{\lambda_{\omega}}=\frac{Tr\Bigl[(\not{r}+m)(\not{k}-m)(\not{p}-m)(\not{l}-m)\Bigl(\gamma^{\mu}-\frac{r^{\mu}-l^{\mu}}{m_{q\bar{q}}({\bf r},{\bf l})+2m}\Bigr)\Bigr]\epsilon^{\ast}_{\mu}(-{\bf P},\lambda_{\omega})}{2^{3/2}m_{q\bar{q}}({\bf p},{\bf k})m_{q\bar{q}}({\bf r},{\bf l})m_{q\bar{q}}({\bf p},{\bf l})},
\end{equation}
with the same notation as in the preceding section. In the nonrelativistic limit this expression tends to $\sqrt{2}(p^{i}-P^{i})\epsilon^{i\ast}(\lambda_{\omega})$.
The amplitude for this decay is
\begin{eqnarray}
& & A({\bf P},\lambda_{\omega},\lambda_{b_{1}})=\frac{2\Lambda}{\sqrt{6}N_{\pi}(P)N_{\omega}(P)N_{b_{1}}(0)}\int\frac{d^{3}{\bf p}}{(2\pi)^{3}}\frac{(E(m,{\bf p})+E(m,{\bf P}-{\bf p}))^{2}}{E(m,{\bf p})E^{2}(m,{\bf P}-{\bf p})} \nonumber \\
& & \times\,W^{\lambda_{\omega}}({\bf p},{\bf p}-{\bf P},{\bf P}-{\bf p},-{\bf p})Y_{1\lambda_{b_{1}}}({\bf p}/|{\bf p}|)\psi_{L}((E(m,{\bf p})+E(m,{\bf P}-{\bf p}))/\mu_{b_{1}}) \nonumber \\
& & \times\,\psi_{L}(m_{q\bar{q}}({\bf p},{\bf P}-{\bf p})/\mu_{\pi})\psi_{L}(m_{q\bar{q}}({\bf p},{\bf P}-{\bf p})/\mu_{\omega}),
\label{eq:intb1}
\end{eqnarray}
which goes in the nonrelativistic limit to
\begin{equation}
A({\bf P},\lambda_{\omega},\lambda_{b_{1}})=\frac{4\Lambda}{\sqrt{3}mN_{\pi}(P)N_{\omega}(P)N_{b_{1}}(0)}\int\frac{d^{3}{\bf p}}{(2\pi)^{3}}(p^{i}-P^{i})\epsilon^{i}(\lambda_{\rho})\psi_{L}^{\pi}\psi_{L}^{\omega}\psi_{L}^{b_{1}}.
\end{equation}
This amplitude can be expanded into the partial waves
\begin{equation}
A({\bf P},\lambda_{\omega},\lambda_{b_{1}})=\sum_{L,l}a_{L}(P)Y_{Ll}({\bf P}/|{\bf P}|)\langle L,l;1,\lambda_{\omega}|1,\lambda_{b_{1}}\rangle,
\end{equation}
where $L=0$ or $2$. The decay width is given by
\begin{equation}
d\Gamma=\frac{1}{32\pi^{2}}\frac{P_{0}|A({\bf P}_{0})|^{2}}{m^{2}_{b_{1}}}d\Omega,
\end{equation}
with ${\bf P}_{0}$ satisfying
\begin{equation}
E(m_{\rho},{\bf P}_{0})+E(m_{\omega},-{\bf P}_{0})=m_{b_{1}}.
\end{equation}
This leads to
\begin{equation}
\Gamma_{L}=\frac{P_{0}}{32\pi^{2}m^{2}_{b_{1}}}a^{2}_{L}(P_{0}),\,\,\,\Gamma=\sum_{L}\Gamma_{L},
\label{eq:wid}
\end{equation}
Taking ${\bf P}=Pe_{z}$ and $\lambda_{b_{1}}=\lambda_{ex}=0$ gives the first equation for the two partial amplitudes,
\begin{equation}
A(Pe_{z},0,0)=A_{||}(P)=\sqrt{\frac{1}{4\pi}}a_{0}(P)+\sqrt{\frac{1}{2\pi}}a_{2}(P).
\end{equation}
If ${\bf P}=Pe_{\perp}$, where $e_{\perp}$ is an arbitrary unit vector perpendicular to $e_{z}$, then the second equation for the partial amplitudes is
\begin{equation}
A(Pe_{\perp},0,0)=A_{\perp}(P)=\sqrt{\frac{1}{4\pi}}a_{0}(P)-\sqrt{\frac{1}{8\pi}}a_{2}(P).
\end{equation}
Finally, one obtains:
\begin{eqnarray}
& & \Gamma_{b_{1}\rightarrow\rho\omega}^{(s-wave)}=\frac{P_{0}}{72\pi m^{2}_{b_{1}}}[A_{||}(P_{0})+2A_{\perp}(P_{0})]^{2}, \nonumber \\
& & \Gamma_{b_{1}\rightarrow\rho\omega}^{(d-wave)}=\frac{P_{0}}{36\pi m^{2}_{b_{1}}}[A_{||}(P_{0})-A_{\perp}(P_{0})]^{2}.
\label{eq:widb1}
\end{eqnarray} 

Now we move to the $b_{1}$ treated as a gluonic bound state. The $q{\bar q}g$ wave function with $J^{PC}=1^{+-}$, $I=1$ quantum numbers requires the $q{\bar q}$ to have the $\pi$ or $\pi_2$ quantum numbers. 
The corresponding, total wave functions are given by
\begin{equation}
\Psi^{\lambda_{b_{1}}}_{q\bar{q}g(J)}(\lambda_{q},\lambda_{\bar{q}},\lambda_{g})=\sum_{\lambda_{q\bar{q}},\sigma=\pm1}\Psi^{J,\lambda_{q\bar{q}}}_{q\bar{q}}({\bf q},{\bf l}_{q\bar{q}}=-{\bf Q},\lambda_{q},\lambda_{\bar{q}})D^{(1)\ast}_{\lambda_{g}\sigma}(\bar{{\bf Q}}) \langle J,\lambda_{q\bar{q}};1,\sigma|1,\lambda_{b_1} \rangle,
\label{eq:exotb1}
\end{equation}
with $\Psi^{J,\lambda_{q\bar{q}}}_{q\bar{q}}$ being the $\pi$ ($\pi_2$) $q{\bar q}$ wave function for $J=0$ ($J=2$).
The normalized spin wave function is thus given by
\begin{equation}
\Psi^{\lambda_{b_1}}_{q\bar{q}g(J)}=\sum_{\lambda}\Psi^{\lambda}_{q\bar{q}}({\bf q},-{\bf Q},\lambda_{q},\lambda_{\bar{q}})\zeta_{(J)}(\bar{{\bf Q}},\lambda,\lambda_{g},\lambda_{\rho}),
\end{equation}
where
\begin{eqnarray}
& & \zeta_{(J=0)}=\sqrt{\frac{3}{8\pi}}[\epsilon_{c}^{\ast}({\bf Q},\lambda_{g})\cdot\epsilon(\lambda_{b_{1}})] \nonumber \\
& & \zeta_{(J=2)}=\sqrt{\frac{27}{64\pi}}\bar{{\bf q}}\cdot[\epsilon_{c}^{\ast}({\bf Q},\lambda_{g})\otimes\epsilon(\lambda_{b_{1}})]\cdot\bar{{\bf q}},
\end{eqnarray}
with ${\bf Q}$ being the gluon momentum and ${\bf q}={\bf q}({\bf p}_{q},-{\bf Q})$ being the relative momentum in the $q\bar{q}$ pair ($\ref{eq:qq}$).
The spin factor is given by ($\ref{spinf1}$), but with a different tensor $B^{\mu j}$:
\begin{equation}
B^{\mu j}=\frac{Tr\Bigl[(\not{k}-m)(\not{p}-m)(\not{l}-m)\Bigl(\gamma^{\mu}-\frac{r^{\mu}-l^{\mu}}{m_{q\bar{q}}({\bf r},{\bf l})+2m}\Bigr)(\not{r}+m)\gamma^{j}\Bigr]}{2^{3/2}m_{q\bar{q}}({\bf p},{\bf k})m_{q\bar{q}}({\bf r},{\bf l})m_{q\bar{q}}({\bf p},{\bf l})}.
\end{equation}
The notation used above is the same as in Fig.~\ref{mesondec}.
In the nonrelativistic limit, however, this tensor tends to the same values as the $B^{\mu j}$ defined in ($\ref{eq:tr1}$).
The widths can be calculated, as in the $^{3}P_{0}$ model, from (\ref{eq:widb1}).

\section{Numerical results}

In a simple constituent quark model one assumes that the mass difference between the mesons $\pi$ and $\rho$ arises only from spin. Therefore, we can write
\begin{equation}
m_{M}={\bar{m}}_{M}+k(s_{1}\cdot s_{2}),
\end{equation}
where $M$ denotes either meson and ${\bar{m}}_{M}$ is its ``averaged'' mass. Using the identity $s(s+1)=s_{1}(s_{1}+1)+s_{2}(s_{2}+1)+2s_{1}\cdot s_{2}$, we can solve for ${\bar{m}}_{M}$. We have $s=0$ for the $\pi$, and $s=1$ for the $\rho$. We have also $s_{1}=s_{2}=1/2$. 
Substituting $m_{\pi}=$140~MeV and $m_{\rho}=$770~MeV gives ${\bar{m}}_{M}=$612~MeV. Thus $m_{u}=m_{d}={\bar{m}}_{M}/2=$306~MeV. A similar relation can be used for the $K$ and $K^{\ast}$ (decays to strange mesons will be described in the next chapter), leading to ${\bar{m}}_{K}=$792~MeV and $m_{s}={\bar{m}}_{K}-m_{u}=$486~MeV. 

The averaged mass of the $\pi$ and $\rho$ mesons should rather be used instead of their physical masses in the normalization constant $(\ref{eq:norm2})$. In similar fashion, the averaged mass of the $K$ and $K^{\ast}$ should be used in $(\ref{eq:normK})$. This procedure can also be followed for the $b_{1}$ and $a_{J}$ mesons, or for the $h_{1}$ and $f_{J}$ mesons. In this case, however, there is an additional term proportional to the spin-orbit interaction, $S\cdot L$. The averaged masses for the $b_{1}$ and $f_{1}$ are found to be close to their physical values. Therefore the physical values will be used in normalization for these mesons.  

\begin{figure}[t]
\centering
\includegraphics[width=2.5in,height=3.5in,angle=270]{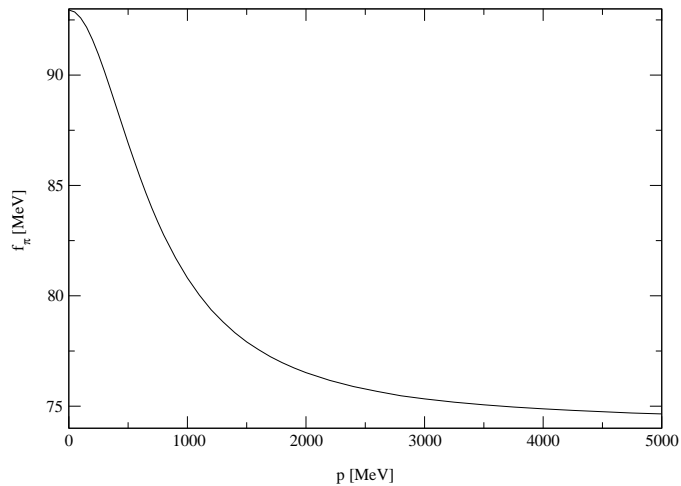}
\caption{\label{weak} The pion weak decay constant $f_{\pi}$ as a function of the pion momentum $p$. }
\end{figure}

The weak decay constants $(\ref{eq:formf2})$ and $(\ref{eq:formfK})$ can be used to fit the parameters $\mu_{\pi}$ and $\mu_{K}$. 
Because our model is not exactly Lorentz covariant, the weak decay constants become functions of the meson momentum and we choose them to be equal to their experimental values at rest. 
Thus, setting $f_{\pi}(0)=\,$93~MeV and $f_{K}(0)=\,$113~MeV leads to $\mu_{\pi}=\,$221~MeV and $\mu_{K}=\,$275~MeV. The momentum dependence of $f_{\pi}$ in our model for $m=306\mbox{ MeV}$ and $\mu_{\pi}=221\mbox{ MeV}$ is presented in Fig.~\ref{weak}, which shows the difference between the rest-value and the infinity-value at the level of $20\%$.

\begin{figure}[ht]
\centering
\vspace{0.2in}
\includegraphics[width=2.5in,height=3.5in,angle=270]{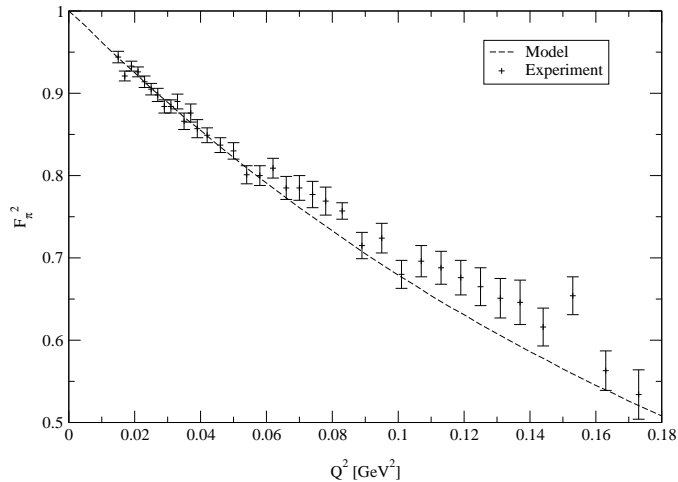}
\caption{\label{em} The pion electromagnetic form factor $F^{2}_{\pi}$ as a function of the momentum transfer square $Q^{2}$. }
\end{figure}

The strong coupling constant at this scale is approximately $g^{2}=10$, and for the effective mass of the gluon we will take $m_{g}=\,$500~MeV, following what we said in Chapter~3. 
In Fig.~\ref{em} we present $F^{2}(Q^{2})$ calculated with the same wave function parameters and compared with data~\cite{NA7}. The agreement is good for small momentum transfer, whereas the discrepancy for larger $Q^{2}$ indicates a missing, high momentum component of the wave function. Actually, we could constrain the quark mass $m$ from $f_{\pi}$ and $F_{\pi}$ but it appears that both quantities are not too sensitive to $m$. Thus its value taken from the averaged mass of the $\pi$ and $\rho$ mesons works pretty well.

For the $\rho$ and $b_{1}$ decays we will choose the values of all parameters $\mu$ to be equal to $\mu_{\pi}$.
The numerical predictions for the $\rho$ widths with $\Lambda=\mu_{\pi}$ are presented in Table~\ref{tab3}. As we discussed at the beginning of this chapter, these numbers correspond to each $q\bar{q}$ spin value in the hybrid component of a decaying meson. The experimental value of the width for $\rho\rightarrow2\pi$ is $149\mbox{ MeV}$ \cite{PDG}. This number can be used to fit the free parameter $\Lambda$, the coupling constant $g$, or the size parameter $\mu'_{ex}$ which need not to be of the same order as $\mu_{\pi}$. In order to do so, however, we need to know how each $J$ contributes to the total $\rho$ spin wave function.

\begin{table}[ht]
\centering
\vspace{0.2in}
\begin{tabular}{|c|c||r|r|r|r|}
\hline
\multicolumn{2}{|c||}{$\Gamma_{rel}(\Gamma_{nrel})$}&$^{3}P_{0}$&$a_{0}$&$a_{1}$&$a_{2}$\\
\hline\hline
$2\pi$&P& $59(195)$ & $7(15)$ & $12(45)$ & $58(75)$\\
\hline
\end{tabular}
\caption{\label{tab3} Relativistic (nonrelativistic) widths in MeV of the decay $\rho(770)\rightarrow2\pi$ for the $^{3}P_{0}$ model, and for the $^{3}S_{1}$ model with $\rho=a_{J}+g$ and $J=0,1,2$. }
\end{table}
 
The numerical predictions for the $b_{1}$ widths with $\Lambda=\mu_{\pi}$ are presented in Table~\ref{tab4}. 
The experimental value of the total width for the process $b_{1}\rightarrow\pi\omega$ is $142\mbox{ MeV}$, and for the ratio of the D-wave and S-wave width rates is $0.08$ \cite{PDG}.
Our predictions give a value less than $0.02$ for this ratio in the $^{3}S_{1}$ model, and close to $4$ for the $^{3}P_{0}$ decay. 
Therefore the real mechanism should lie somewhere in between, although the $^{3}S_{1}$ mechanism gives more accurate result. However, in the $^{3}P_{0}$ model the D/S ratio is very sensitive to the free parameters $\mu$ and for $\mu=$400 MeV one obtains this ratio on the order of the experimental value~\cite{bench1}. This value of $\mu$, however, is inconsistent with the weak decay constant and the elastic form factor for the pion.   
 
The $q{\bar q}$g wave function component of the $b_1$ wave functions used here is that of Eq.~(\ref{eq:exotb1}), corresponding to a $q{\bar q}$ pair with the $\pi$ quantum numbers. 
For a $q{\bar q}$ with the $\pi_2$ quantum numbers, the numeric value for the width is much smaller than $1\mbox{ MeV}$ for the S-wave and approximately $1\mbox{ MeV}$ for the D-wave. 
The ratio $D/S$ is respectively $230$. 
In the nonrelativistic limit we obtain similar results.

\begin{table}[ht]
\centering
\begin{tabular}{|c|c||r|r|}
\hline
\multicolumn{2}{|c||}{$\Gamma_{rel}(\Gamma_{nrel})$}&$^{3}P_{0}$&$\pi$\\
\hline\hline
$\pi\omega(782)$&S& $9(13)$ & $51(82)$\\
\cline{2-4}
 &D& $33(68)$ & $<1(<1)$\\
\hline
\end{tabular}
\caption{\label{tab4} Relativistic (nonrelativistic) widths in MeV of the decay $b_{1}(1235)\rightarrow\pi\omega(782)$ for the $^{3}P_{0}$ model, and for the $^{3}S_{1}$ model with $b_{1}=\pi+g$. }
\end{table}

We observe that treating the $b_{1}$ as the $\pi_{2}+g$ state increases dramatically the $D/S$ ratio. Therefore this may be an important component of the wave function. 
Relatively small values of the decay widths of $b_{1}$ with $\pi_{2}$ quantum numbers (L=2) compared to those of $b_{1}$ with pion quantum numbers resemble the situation for the process $\pi_{1}\rightarrow\pi b_{1}$, whose D-wave width was small compared to that in the S-wave.

\chapter{Decays of $\pi_{1}$}

\begin{figure}[h]
\centering
\includegraphics[width=4.0in]{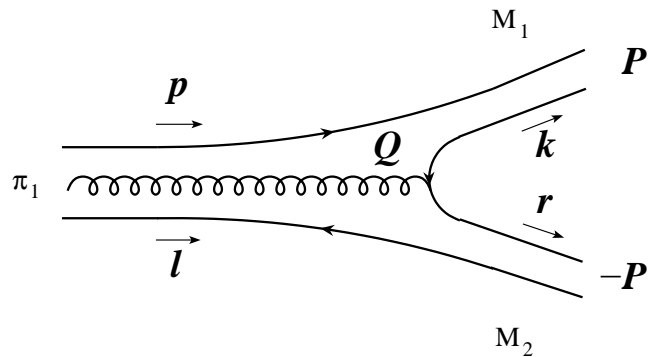}
\caption{\label{hybdec} The $^{3}S_{1}$ decay of a hybrid into two mesons. }
\end{figure}

In this chapter we will study the main subject of the presented work, i.e., a completely relativistic decay of the exotic meson $\pi_{1}$. 
In experiment we observe that most hadronic decays involve a minimal number of final state particles. This requires a small number of transitions at the quark level. 
Thus, an exotic meson is expected to decay into two normal mesons. In a constituent quark model the transverse gluon in the $\pi_{1}$ dissociates into a quark and an antiquark, and the two resulting quark-antiquark pairs rearrange themselves into two mesons, as shown in Fig.~\ref{hybdec}.
The possible decay modes of the $\pi_{1}$ for $m_{\pi_{1}}=$~1600 MeV are listed in Table~\ref{pi1}, where $L$ is the angular momentum between outgoing mesons. For brevity, we did not put there the decays into the antiparticles of the corresponding strange mesons. Hereinafter, such decays are understood to be included and have the same widths. In the following sections we will discuss the most important modes and calculate the corresponding widths.

\begin{table}
\centering
\begin{tabular}{|l|c|}
\hline
$\pi_{1}\rightarrow$ & $L$ \\
\hline\hline
$\pi\eta(547,1295,1440)$,\,$\pi\eta'(958)$,\,$\pi(1300)\eta$ & 1\\
\hline
$\pi b_{1}(1235)$ & 0,2\\
\hline
$\pi f_{1}(1285,1420)$ & 0,2\\
\hline
$\pi f_{2}(1270)$,\,$\pi f'_{2}(1525)$ & 2\\
\hline
$\pi\rho(770,1450,1700)$ & 1\\
\hline
$\pi\rho_{3}(1690)$ & 3\\
\hline
$\eta a_{1}(1260)$ & 0,2\\
\hline
$\eta a_{2}(1320)$ & 2\\
\hline
$\rho(770)\omega(782)$ & 1,3\\
\hline
$K{\bar{K}}_{1}(1400)$ & 0,2\\
\hline
$K{\bar{K}}_{1}(1270)$ & 0,2\\
\hline
$K{\bar{K}}_{2}^{\ast}(1430)$ & 2\\
\hline
$K{\bar{K}}^{\ast}(892,1410)$ & 1\\
\hline
\end{tabular}
\caption{\label{pi1} Possible decay modes and the angular momentum between the outgoing mesons $L$ for the $\pi_{1}$ with mass 2.0 GeV. }
\end{table}

\section{Decay of $\pi_{1}$ into $\pi\eta$ and $\pi b_{1}$}

The amplitudes for these modes are related to the matrix element 
\begin{equation}
\langle \pi({\bf P}),M(-{\bf P})|H|\pi_{1}\rangle,
\end{equation}
where $M$ is either $\eta$ or $b_{1}$. From rearranging the annihilation and creation operators one obtains two nonzero terms contributing to the total amplitude. We will show below that they are equal.
Summing over color gives a factor $4/3$, as for decays of the $q\bar{q}g$ component of a normal meson.
Summing over flavor leads to a trace of a product of Pauli matrices appearing in the isospin term (\ref{isospmatr}),
\begin{equation}
2^{-3/2}Tr(\sigma^{ex}_{i}\sigma^{\pi}_{j})\epsilon_{i}(I_{3}^{ex})\epsilon_{j}^{\ast}(I_{3}^{\pi})
\end{equation}
or
\begin{equation}
2^{-3/2}Tr(\sigma^{ex}_{i}\sigma^{\pi}_{j}\sigma^{b_{1}}_{k})\epsilon_{i}(I_{3}^{ex})\epsilon_{j}^{\ast}(I_{3}^{\pi})\epsilon_{k}^{\ast}(I_{3}^{b_{1}}),
\end{equation}
respectively (for all allowed isospin channels these factors are again equal to $\pm 1/\sqrt{2}$).
The spin factor is given by ($\ref{spinf1}$) with $S=J$ and $\lambda_{\rho}$ replaced by $\lambda_{ex}$, where the tensor $B^{\mu j}$ was introduced in ($\ref{eq:tr1}$) and $\psi_{\mu}^{j(S)}$ are given by ($\ref{eq:trs}$). The notation is shown in Fig.~\ref{hybdec}. Using ($\ref{eq:ss}$) we can write:
\begin{eqnarray} 
& & \psi_{\mu}^{\,\,j(S=0)}=-\sqrt{\frac{3}{8\pi}}g^{k}_{\mu}(\delta^{jk}-\bar{Q}^{j}\bar{Q}^{k})\bar{Q}^{l}\epsilon^{l}(\lambda_{ex}), \nonumber \\
& & \psi_{\mu}^{\,\,j(S=1)}=-\sqrt{\frac{3}{8\pi}}\Bigl[g^{k}_{\mu}-\frac{K_{\mu}K^{k}}{m_{q\bar{q}}(E_{q\bar{q}}+m_{q\bar{q}})}\Bigr](\delta^{jl}-\bar{Q}^{j}\bar{Q}^{l})\bar{Q}^{k}\epsilon^{l}(\lambda_{ex}), \nonumber \\
& & \psi_{\mu}^{\,\,j(S=2)}=\frac{3}{\sqrt{13}}(\psi_{\mu}^{\,\,j(S=1)}-\frac{2}{3}\psi_{\mu}^{\,\,j(S=0)}). 
\label{eq:25}
\end{eqnarray}
In the terms above we used the following notation:
\begin{equation}
m_{q\bar{q}}=m_{q\bar{q}}({\bf p},-{\bf P}-{\bf r}),\,\,\,\,{\bf Q}={\bf P}-{\bf p}+{\bf r},\,\,\,\,E_{q\bar{q}}=E(m_{q\bar{q}},{\bf Q}),\,\,\,\,{\bf K}=-{\bf Q},\,\,\,\,K^{0}=E_{q\bar{q}}+m_{q\bar{q}},
\label{eq:kl}
\end{equation}
and assumed all quarks being on-shell particles.

Integration over all momenta gives for the first term $(2\pi)^{3}\delta^{3}({\bf 0})A_{1}$, where the amplitude $A_{1}$ is given by
\begin{eqnarray}
& & A_{1(M)}^{(S)}({\bf P},\lambda_{M},\lambda_{ex})=\int\frac{d^{3}{\bf p}}{(2\pi)^{3}}\frac{d^{3}{\bf r}}{(2\pi)^{3}}\frac{(E(m,{\bf p})+E(m,{\bf k}))(E(m,{\bf r})+E(m,{\bf l}))}{4E(m,{\bf p})E(m,{\bf k})E(m,{\bf r})E(m,{\bf l})E(m_{g},{\bf Q})} \nonumber \\
& & \times\,(E(m,{\bf p})+E(m,{\bf l})+E(m_{g},{\bf Q}))\psi_{L}'(m_{q\bar{q}}({\bf p},{\bf l})/\mu_{ex},m_{q\bar{q}g}({\bf p},{\bf l},{\bf Q})/\mu'_{ex}) \nonumber \\
& & \times\,[N_{\pi}(P)N_{M}(P)N_{ex}]^{-1}\psi_{L}(m_{q\bar{q}}({\bf p},{\bf k})/\mu_{\pi})\psi_{L}(m_{q\bar{q}}({\bf r},{\bf l})/\mu_{M})Y^{\ast}_{J_{M}\lambda_{M}}({\bf q}/|{\bf q}|) \nonumber \\
& & \times\,g\frac{4}{3}\frac{1}{\sqrt{2}}B^{\mu}_{\,\,\,j}\psi_{\mu j}^{(S)}(\lambda_{ex}).
\label{eq:a1}
\end{eqnarray}
Here, $M$ denotes the second meson ($\eta$ or $b_{1}$), $J$ is its total spin, ${\bf q}={\bf q}({\bf r},-{\bf P})$ is given in ($\ref{eq:qq}$) and $P=|{\bf P}|$. If $M=\eta$, then the above expression must be proportional to $\epsilon^{i}(\lambda_{ex})P^{i}$ (or $Y_{1\lambda_{ex}}({\bf P}/P)$), being a vector function of the vector ${\bf P}$ (the outgoing mesons in the P-wave). However, if $M=b_{1}$ then the integral ($\ref{eq:a1}$) is proportional to
\begin{equation}
\epsilon^{i}(\lambda_{ex})\epsilon^{j\ast}(\lambda_{M})[P^{i}P^{j}+f(P)\delta^{ij}], 
\end{equation}
and can be represented as a superposition of spherical harmonics corresponding respectively to $l=0$ and $l=2$ (S-wave and D-wave).

The second term in the Hamiltonian matrix element $A_{2}^{(S)}$ is obtained from the first one by interchanging
\begin{equation}
{\bf p}\leftrightarrow{\bf r},\,\,{\bf P}\leftrightarrow-{\bf P},\,\,I_{3}^{(1)}\leftrightarrow I_{3}^{(2)} 
\end{equation}
everywhere in ($\ref{eq:a1}$), including ${\bf k},{\bf l},{\bf Q}$, except in the third line. For $\pi_{1}\rightarrow\pi\eta$ we can write schematically:
\begin{eqnarray}
& & {\bf A}_{1}({\bf P})=\int d^{3}{\bf p}\,d^{3}{\bf r}\,f_{\eta}(m_{q\bar{q}}({\bf p},{\bf P}-{\bf p}))f_{\pi}(m_{q\bar{q}}({\bf r},-{\bf P}-{\bf r}))f_{ex}(m_{q\bar{q}}({\bf p},-{\bf P}-{\bf r})) \nonumber \\
& & \times\,({\bf P}-{\bf p}+{\bf r}), \nonumber \\
& & {\bf A}_{2}({\bf P})=\int d^{3}{\bf p}\,d^{3}{\bf r}\,f_{\eta}(m_{q\bar{q}}({\bf p},{\bf P}-{\bf p}))f_{\pi}(m_{q\bar{q}}({\bf r},-{\bf P}-{\bf r}))f_{ex}(m_{q\bar{q}}({\bf r},{\bf P}-{\bf p})) \nonumber \\
& & \times\,(-{\bf P}-{\bf r}+{\bf p}).
\end{eqnarray}
We have not included here the energy and trace factors because they do not change under the above symmetry.
Using $m_{q\bar{q}}({\bf p},{\bf r})=m_{q\bar{q}}({\bf r},{\bf p})=m_{q\bar{q}}(-{\bf p},-{\bf r})$ one can show that both terms are equal (for the same $\lambda_{M}$ and $\lambda_{ex}$),
\begin{equation}
{\bf A}_{2}({\bf P})=-{\bf A}_{1}(-{\bf P})={\bf A}_{1}({\bf P}).
\end{equation}
Thus, we have $A=A_{1}+A_{2}=2A_{1}$. In similar fashion one can prove this equality (up to a sign) for all decays of any $q\bar{q}g$ particle into two mesons. 

If $\mu_{\eta}=\mu_{\pi}$, then each term ($A_{1}$ or $A_{2}$) is a product of a part that is symmetric under interchanging ${\bf p}\leftrightarrow{\bf r}+{\bf P}$ and a (vector) part that is antisymmetric. In this case the hybrid will not decay into $\pi$ and $\eta$. Neither can it decay into two pions because of a minus sign from interchanging the Pauli matrices in the isospin factor that makes both terms cancel. 
There is the same minus sign for the $\pi_{1}\rightarrow\pi b_{1}$, but now the amplitude is a scalar ($L=0$) or a tensor ($L=2$) function of ${\bf P}$ and again $A_{1}({\bf P})=A_{2}({\bf P})$. 
However, in this case $\mu_{b_{1}}=\mu_{\pi}$ does not imply $A=0$ because the orbital wave functions of these mesons are different. 
Thus we find that the $1^{-+}$ isovector does not decay into identical pseudoscalars. This is a relativistic generalization of a symmetry found in other nonrelativistic decay models~\cite{CP}.
One might think that a similar symmetry could occur for the decay $\rho\rightarrow2\pi$. In this case, however, the amplitude does not vanish because its vector part comes from a spherical harmonic associated not with the gluon momentum, but with the relative momentum in a $q\bar{q}$ pair. This harmonic is asymmetric under interchanging ${\bf p}\leftrightarrow{\bf r}+{\bf P}$ and the amplitude for the $\rho\rightarrow2\pi$ does not vanish. 

The width is given by
\begin{equation}
d\Gamma=\frac{1}{32\pi^{2}}\frac{P_{0}|A({\bf P}_{0})|^{2}}{m^{2}_{ex}}d\Omega,
\end{equation}
with ${\bf P}_{0}$ satisfying
\begin{equation}
E(m_{\pi},{\bf P}_{0})+E(m_{M},-{\bf P}_{0})=m_{ex}
\label{eq:p0}
\end{equation}
and $P_{0}=|{\bf P}_{0}|$. The amplitudes must be, according to the Wigner-Eckart theorem, of the form:
\begin{eqnarray}
& & A^{\pi\eta}({\bf P},\lambda_{ex})=a^{\pi\eta}_{1}(P)Y_{1\lambda_{ex}}({\bf P}/P), \nonumber \\
& & A^{\pi b_{1}}({\bf P},\lambda_{b_{1}},\lambda_{ex})=\sum_{L,l}a^{\pi b_{1}}_{L}(P)Y_{Ll}({\bf P}/P)\langle L,l;1,\lambda_{b_{1}}|1,\lambda_{ex}\rangle,
\label{eq:ameb}
\end{eqnarray}
which leads to Eq.~(\ref{eq:wid}) with $m_{b_{1}}$ replaced by $m_{ex}$.
The amplitudes $A({\bf P},\lambda_{M},\lambda_{ex})$ given in ($\ref{eq:a1}$) are multiplied by $2$, since there are two equal terms. The expression for the $\pi_{1}\rightarrow\pi\eta$ width is similar to (\ref{eq:widrho}), whereas for the $\pi_{1}\rightarrow\pi b_{1}$ we can use the ``parallel'' and ``perpendicular'' amplitudes introduced for the decay $b_{1}\rightarrow\pi\omega$. The final results are: 
\begin{eqnarray}
& & \Gamma_{\pi\eta}^{(p-wave)}=\frac{P_{0}}{24\pi m^{2}_{ex}}A^{2}_{\pi\eta}(P_{0}e_{z},0), \nonumber \\
& & \Gamma_{\pi b_{1}}^{(s-wave)}=\frac{P_{0}}{72\pi m^{2}_{ex}}[A_{||}^{\pi b_{1}}(P_{0})+2A_{\perp}^{\pi b_{1}}(P_{0})]^{2}, \nonumber \\
& & \Gamma_{\pi b_{1}}^{(d-wave)}=\frac{P_{0}}{36\pi m^{2}_{ex}}[A_{||}^{\pi b_{1}}(P_{0})-A_{\perp}^{\pi b_{1}}(P_{0})]^{2}.
\label{eq:ga1}
\end{eqnarray} 

The $\pi$ and $\eta$ mesons have the same quantum numbers (except isospin) so $\mu_{\pi}$ and $\mu_{\eta}$ should be almost equal. This equality is not exact because the $SU(3)_{f}$ is only an approximate symmetry and there is a contribution of the $s\bar{s}$ in $\eta$. Therefore the amplitude for the $\pi_{1}\rightarrow\pi\eta$ should be close to zero, and of two channels $\pi\eta$, $\pi b_{1}$ the latter will be favored. However, the free parameters $\mu$ need not to be close to each other for decays of the $\pi_{1}$ into two mesons with different radial quantum number, which would make such channels significant.

\section{Decay of $\pi_{1}$ into $\pi\rho$, $\pi f_{1}$, $\pi f_{2}$, $\eta a_{1}$ and $\eta a_{2}$}

For these channels the procedure is analogous to that in the preceding section.
The amplitude of this process is given by the matrix element $\langle M_{1}({\bf P}),M_{2}(-{\bf P})|H|\pi_{1}\rangle$ which is again a sum of two terms, and $M_{1}$ and $M_{2}$ denote the two outgoing mesons.
The color and flavor factors are again $4/3$ and $\pm 1/\sqrt{2}$, but the spin factor is given now by
\begin{equation}
C^{\mu\nu}_{\,\,\,\,\,\,j}\psi_{\mu\nu j}^{(S)},
\label{spinf2}
\end{equation}  
where (for the first term)
\begin{eqnarray}
& & C^{\mu\nu j}=-[2^{3/2}m_{q\bar{q}}({\bf p},{\bf k})m_{q\bar{q}}({\bf r},{\bf l})m_{q\bar{q}}({\bf p},{\bf l})]^{-1}Tr\Bigl[(\not{p}+m)\Bigl(\gamma^{\mu}-\frac{p^{\mu}-l^{\mu}}{m_{q\bar{q}}({\bf p},{\bf l})+2m}\Bigr) \nonumber \\
& & \times\,(\not{l}-m)\Bigl(\gamma^{\nu}-\frac{r^{\nu}-l^{\nu}}{m_{q\bar{q}}({\bf r},{\bf l})+2m}\Bigr)(\not{r}+m)\gamma^{j}(\not{k}-m)\gamma^{5}\Bigr],
\label{eq:tr2}
\end{eqnarray}
and
\begin{equation}
\psi_{\mu}^{\,\,\,\nu j(S)}(\lambda,\lambda_{ex})=\sum_{\lambda_{g}}\psi_{\mu(S)}({\bf Q},\lambda_{g},\lambda_{ex})\epsilon^{j}_{c}({\bf Q},\lambda_{g})\epsilon^{\nu\ast}(-{\bf P},\lambda)
\end{equation}
or equivalently:
\begin{eqnarray}
& & \psi_{\mu}^{\,\,\,\nu j(S=0)}=-\sqrt{\frac{3}{8\pi}}g^{k}_{\mu}(\delta^{jk}-\bar{Q}^{j}\bar{Q}^{k})\bar{Q}^{l}\epsilon^{l}(\lambda_{ex})\epsilon^{\nu\ast}(-{\bf P},\lambda), \nonumber \\
& & \psi_{\mu}^{\,\,\,\nu j(S=1)}=-\sqrt{\frac{3}{8\pi}}\Bigl[g^{k}_{\mu}-\frac{K_{\mu}K^{k}}{m_{q\bar{q}}(E_{q\bar{q}}+m_{q\bar{q}})}\Bigr](\delta^{jl}-\bar{Q}^{j}\bar{Q}^{l})\bar{Q}^{k}\epsilon^{l}(\lambda_{ex})\epsilon^{\nu\ast}(-{\bf P},\lambda), \nonumber \\
& & \psi_{\mu}^{\,\,\,\nu j(S=2)}=\frac{3}{\sqrt{13}}(\psi_{\mu}^{\,\,\,\nu j(S=1)}-\frac{2}{3}\psi_{\mu}^{\,\,\,\nu j(S=0)}).
\label{eq:simpl} 
\end{eqnarray}
The notation used above is the same as in ($\ref{eq:kl}$) and Fig.~\ref{hybdec}.

The amplitude for the first term is given by
\begin{eqnarray}
& & A_{1(M)}^{(S)}({\bf P},\lambda_{M},\lambda_{ex})=\sum_{\lambda,l}\int\frac{d^{3}{\bf p}}{(2\pi)^{3}}\frac{d^{3}{\bf r}}{(2\pi)^{3}}\frac{(E(m,{\bf p})+E(m,{\bf k}))(E(m,{\bf r})+E(m,{\bf l}))}{4E(m,{\bf p})E(m,{\bf k})E(m,{\bf r})E(m,{\bf l})E(m_{g},{\bf Q})} \nonumber \\
& & \times(E(m,{\bf p})+E(m,{\bf l})+E(m_{g},{\bf Q}))\psi_{L}'(m_{q\bar{q}}({\bf p},{\bf l})/\mu_{ex},m_{q\bar{q}g}({\bf p},{\bf l},{\bf Q})/\mu'_{ex}) \nonumber \\
& & \times[N_{\pi}N_{M}N_{ex}]^{-1}\psi_{L}(m_{q\bar{q}}({\bf p},{\bf k})/\mu_{\pi})\psi_{L}(m_{q\bar{q}}({\bf r},{\bf l})/\mu_{M})Y^{\ast}_{L_{q\bar{q}(M)}l}({\bf q}/|{\bf q}|) \nonumber \\ 
& & \times\,\langle 1,\lambda;L_{q\bar{q}(M)},l|J_{M},\lambda_{M}\rangle g\frac{4}{3}\frac{1}{\sqrt{2}}C^{\mu\nu}_{\,\,\,\,\,\,j}\psi_{\mu\nu j}^{(S)}(\lambda,\lambda_{ex}),
\label{eq:a2}
\end{eqnarray}
where $M$ denotes the second meson ($\rho$ with $L_{q\bar{q}}=0$ or $f_{J}$ with $L_{q\bar{q}}=1$), $J$ is its total spin, ${\bf q}={\bf q}({\bf r},-{\bf P})$ was defined in ($\ref{eq:qq}$) and $J=1,2$. Again one can show (for the same $\lambda$'s)
\begin{equation}
A_{2}^{(S)}({\bf P})=A_{1}^{(S)}({\bf P}).
\end{equation}
If $\mu_{\rho}=\mu_{\pi}$, then the amplitude of the decay into $\pi\rho$ is not symmetric under interchanging ${\bf p}\leftrightarrow{\bf r}$ and does not vanish, unlike for the $\pi\eta$ case. Therefore this mode may be significant. The same holds for the $\pi f_{J}$ channels. 

From the Wigner-Eckart theorem we have:
\begin{eqnarray} 
& & A^{\pi\rho}({\bf P},\lambda_{\rho},\lambda_{ex})=\sum_{L,l}a^{\pi\rho}_{L}(P)Y_{Ll}({\bf P}/P)\langle L,l;1,\lambda_{\rho}|1,\lambda_{ex}\rangle, \nonumber \\
& & A^{\pi f_{J}}({\bf P},\lambda_{f_{J}},\lambda_{ex})=\sum_{L,l}a^{\pi f_{J}}_{L}(P)Y_{Ll}({\bf P}/P)\langle L,l;J,\lambda_{f_{J}}|1,\lambda_{ex}\rangle.
\label{eq:amrf}
\end{eqnarray}
In formulae ($\ref{eq:amrf}$) and ($\ref{eq:wid}$) one must use the $A({\bf P},\lambda 's)$ given in ($\ref{eq:a2}$), multiplied by $2$. The mesons in the $\pi\rho$ channel go out in the P-wave, for the $\pi f_{1}$ it is either the S-wave or D-wave, whereas in the $\pi f_{2}$ they go out only in the D-wave. 
Taking ${\bf P}=Pe_{z}$ and $\lambda_{\rho}=\lambda_{ex}=+1$ leads to
\begin{equation}
a^{\pi\rho}_{1}(P)=\sqrt{\frac{8\pi}{3}}A^{\pi\rho}(Pe_{z},1,1),
\end{equation}
and therefore
\begin{equation} 
\Gamma_{\pi\rho}^{(p-wave)}=\frac{P_{0}}{12\pi m^{2}_{ex}}A^{2}_{\pi\rho}(P_{0}e_{z},1,1).
\label{eq:ga2}
\end{equation}
In the above ${\bf P}_{0}$ satisfies ($\ref{eq:p0}$) with $m_{M}=m_{\rho}$.
For decays into $\pi f_{J}$ one may follow the procedure with the parallel and perpendicular amplitudes described in the preceding section, or just use the general formula that is equivalent to (\ref{eq:amrf}),
\begin{equation}
a^{\pi f_{J}}_{L}(P)=\sum_{L,l}\int d\Omega_{{\bf P}}A^{\pi f_{J}}({\bf P},\lambda_{f_{J}},\lambda_{ex})Y^{\ast}_{Ll}({\bf P}/P)\langle L,l;J,\lambda_{f_{J}}|1,\lambda_{ex}\rangle,
\end{equation}
and substitute it into (\ref{eq:wid}) with $m_{b_{1}}$ replaced by $m_{ex}$. Here, ${\bf P}_{0}$ satisfies ($\ref{eq:p0}$) with $m_{M}=m_{f_{J}}$.
All results for the decays $\pi_{1}\rightarrow \pi f_{J}$ are valid also for the $\eta a_{J}$ modes (with different normalization constants, parameters $\mu$ and $P_{0}$). In this case, $\pi$ must be replaced with $\eta$ and $f$ with $a$.

\section{Decay $\pi_{1}\rightarrow\rho\omega$}

The amplitude of this process is given by the matrix element $\langle \rho({\bf P}),\omega(-{\bf P})|H|\pi_{1}\rangle$, being again a sum of two terms. The color and flavor factors are the same as before, and the spin factor is equal to
\begin{equation}
D^{\mu\nu\rho}_{\,\,\,\,\,\,\,\,\,\,j}\psi_{\mu\nu\rho j}^{(S)},
\label{spinf3}
\end{equation}
where (for the first term)
\begin{eqnarray}
& & D^{\mu\nu\rho j}=[2^{3/2}m_{q\bar{q}}({\bf p},{\bf k})m_{q\bar{q}}({\bf r}_{q_{2}},{\bf l})m_{q\bar{q}}({\bf p},{\bf l})]^{-1}\,Tr\Bigl[(\not{k}-m)\Bigl(\gamma^{\nu}-\frac{p^{\nu}-k^{\nu}}{m_{q\bar{q}}({\bf p},{\bf k})+2m}\Bigr) \nonumber \\
& & \times\,(\not{p}+m)\Bigl(\gamma^{\mu}-\frac{p^{\mu}-l^{\mu}}{m_{q\bar{q}}({\bf p},{\bf l})+2m}\Bigr)(\not{l}-m)\Bigl(\gamma^{\rho}-\frac{r^{\rho}-l^{\rho}}{m_{q\bar{q}}({\bf r},{\bf l})+2m}\Bigr)(\not{r}+m)\gamma^{j}\Bigr],
\label{eq:tr3}
\end{eqnarray}
and
\begin{equation}
\psi_{\mu}^{\,\,\,\nu\rho j(S)}=\sum_{\lambda_{g}}\psi_{\mu(S)}({\bf Q},\lambda_{g},\lambda_{ex})\epsilon^{j}_{c}({\bf Q},\lambda_{g})\epsilon^{\nu\ast}({\bf P},\lambda_{\rho})\epsilon^{\rho\ast}(-{\bf P},\lambda_{\omega})
\end{equation}
or equivalently:
\begin{eqnarray}
& & \psi_{\mu}^{\,\,\,\nu\rho j(S=0)}=-\sqrt{\frac{3}{8\pi}}g^{k}_{\mu}(\delta^{jk}-\bar{Q}^{j}\bar{Q}^{k})\bar{Q}^{l}\epsilon^{l}(\lambda_{ex})\epsilon^{\nu\ast}({\bf P},\lambda_{\rho})\epsilon^{\rho\ast}(-{\bf P},\lambda_{\omega}), \nonumber \\
& & \psi_{\mu}^{\,\,\,\nu\rho j(S=1)}=-\sqrt{\frac{3}{8\pi}}\Bigl[g^{k}_{\mu}-\frac{K_{\mu}K^{k}}{m_{q\bar{q}}(E_{q\bar{q}}+m_{q\bar{q}})}\Bigr](\delta^{jl}-\bar{Q}^{j}\bar{Q}^{l})\bar{Q}^{k}\epsilon^{l}(\lambda_{ex}) \nonumber \\
& & \times\,\epsilon^{\nu\ast}({\bf P},\lambda_{\rho})\epsilon^{\rho\ast}(-{\bf P},\lambda_{\omega}), \nonumber \\
& & \psi_{\mu}^{\,\,\,\nu\rho j(S=2)}=\frac{3}{\sqrt{13}}(\psi_{\mu}^{\,\,\,\nu\rho j(S=1)}-\frac{2}{3}\psi_{\mu}^{\,\,\,\nu\rho j(S=0)}). 
\end{eqnarray}
The notation used above is the same as in the preceding sections.

Again one can demonstrate that the second term in the amplitude is equal to the first one. Therefore the total amplitude is
\begin{eqnarray}
& & A^{(S)}({\bf P},\lambda_{\rho},\lambda_{\omega},\lambda_{ex})=2\int\frac{d^{3}{\bf p}}{(2\pi)^{3}}\frac{d^{3}{\bf r}}{(2\pi)^{3}}\frac{(E(m,{\bf p})+E(m,{\bf k}))(E(m,{\bf r})+E(m,{\bf l}))}{4E(m,{\bf p})E(m,{\bf k})E(m,{\bf r})E(m,{\bf l})E(m_{g},{\bf Q})} \nonumber \\
& & \times\,(E(m,{\bf p})+E(m,{\bf l})+E(m_{g},{\bf Q}))\cdot\psi_{L}'(m_{q\bar{q}}({\bf p},{\bf l})/\mu_{ex},m_{q\bar{q}g}({\bf p},{\bf l},{\bf Q})/\mu'_{ex}) \nonumber \\
& & \times\,[N_{\rho}N_{\omega}N_{ex}]^{-1}\psi_{L}(m_{q\bar{q}}({\bf p},{\bf k})/\mu_{\rho})\psi_{L}(m_{q\bar{q}}({\bf r},{\bf l})/\mu_{\omega})g\frac{4}{3}\frac{1}{\sqrt{2}}D^{\mu\nu\rho}_{\,\,\,\,\,\,\,\,\,\,j}\psi_{\mu\nu\rho j}^{(S)}.
\label{eq:a5}
\end{eqnarray}
If $\mu_{\rho}=\mu_{\omega}$, then the symmetry of the orbital wave functions causes $A=0$ and the hybrid will not decay into $\rho$ and $\omega$. Because both parameters $\mu$ are expected to be on the same order, the $\rho\omega$ mode will not be favored. The width is given by (\ref{eq:wid}) with $m_{b_{1}}$ replaced by $m_{ex}$,
where 
\begin{equation}
a^{\rho\omega}_{L}(P)=\sum_{L,l,J',\lambda'}\int d\Omega_{{\bf P}}A^{\rho\omega}({\bf P},\lambda_{\rho},\lambda_{\omega},\lambda_{ex})Y^{\ast}_{Ll}({\bf P}/P)\langle 1,\lambda_{\rho};1,\lambda_{\omega}|J',\lambda'\rangle\langle L,l;J',\lambda'|1,\lambda_{ex}\rangle, 
\end{equation}
and ${\bf P}_{0}$ is obtained from $E(m_{\rho},{\bf P}_{0})+E(m_{\omega},-{\bf P}_{0})=m_{ex}$. In this process we have either $L=1$ or $L=3$.

\section{Decay into strange mesons}

The above results can be straightforwardly generalized to the case where $m_{q}$ and $m_{\bar{q}}$ are different, for example to decays into mesons with one strange quark ($I=1/2$). The spin wave function for a quark-antiquark pair in a $J^{P}=0^{-}$ state is
\begin{equation}
\Psi_{q\bar{q}}({\bf l}_{q},{\bf l}_{\bar{q}},\lambda_{q},\lambda_{\bar{q}})=\frac{1}{\sqrt{2}{\tilde{m}}_{q\bar{q}}({\bf l}_{q},{\bf l}_{\bar{q}})}\bar{u}(m_{q},{\bf l}_{q},\lambda_{q})\gamma^{5}v(m_{\bar{q}},{\bf l}_{\bar{q}},\lambda_{\bar{q}}),
\label{eq:00s}
\end{equation}
and the $K$ states (understood to be both $K$ and $\bar{K}$) are given by
\begin{eqnarray}
& & |K({\bf P})\rangle =\sum_{all\,\,\lambda,c}\int\frac{d^{3}{\bf p}_{q}}{(2\pi)^{3}2E(m_{q},{\bf p}_{q})}\frac{d^{3}{\bf p}_{\bar{q}}}{(2\pi)^{3}2E(m_{\bar{q}},{\bf p}_{\bar{q}})}2(E(m_{q},{\bf p}_{q})+E(m_{\bar{q}},{\bf p}_{\bar{q}})) \nonumber \\
& & \times\,(2\pi)^{3}\delta^{3}({\bf p}_{q}+{\bf p}_{\bar{q}}-{\bf P})\frac{1}{\sqrt{3}}\delta_{c_{q}c_{\bar{q}}}\Psi_{q\bar{q}}({\bf p}_{q},{\bf p}_{\bar{q}},\lambda_{q},\lambda_{\bar{q}})\frac{1}{N(P)}\psi_{L}(m_{q\bar{q}}({\bf p}_{q},{\bf p}_{\bar{q}})/\mu) \nonumber \\
& & \times\,b^{\dag}_{{\bf p}_{q}\lambda_{q}f_{q}c_{q}}d^{\dag}_{{\bf p}_{\bar{q}}\lambda_{\bar{q}}f_{\bar{q}}c_{\bar{q}}}|0\rangle,
\label{eq:K}
\end{eqnarray}
where the operators satisfy the anticommutation relations:
\begin{eqnarray}
& & \{b_{{\bf p}\lambda fc},b^{\dag}_{{\bf p}'\lambda' f'c'}\}=(2\pi)^{3}2E(m_{q},{\bf p})\delta^{3}({\bf p}-{\bf p}')\delta_{\lambda\lambda'}\delta_{ff'}\delta_{cc'}, \nonumber \\
& & \{d_{{\bf p}\lambda fc},d^{\dag}_{{\bf p}'\lambda' f'c'}\}=(2\pi)^{3}2E(m_{\bar{q}},{\bf p})\delta^{3}({\bf p}-{\bf p}')\delta_{\lambda\lambda'}\delta_{ff'}\delta_{cc'}.
\end{eqnarray}
The invariant mass is defined as
\begin{equation}
m_{q\bar{q}}=m_{q\bar{q}}({\bf l}_{q},{\bf l}_{\bar{q}})=\sqrt{(E(m_{q},{\bf l}_{q})+E(m_{\bar{q}},{\bf l}_{\bar{q}}))^{2}-({\bf l}_{q}+{\bf l}_{\bar{q}})^{2}}
\label{eq:mis}
\end{equation}
and the ``modified'' invariant mass is
\begin{equation}
\tilde{m}_{q\bar{q}}=\sqrt{m_{q\bar{q}}^{2}-(m_{q}-m_{\bar{q}})^{2}}.
\end{equation}
The flavor indices $f_{q}$ and $f_{\bar{q}}$ give four combinations: $u\bar{s}$, $d\bar{s}$, $s\bar{u}$ and $s\bar{d}$, corresponding to appropriate mesons.

If in ($\ref{eq:K}$) $\Psi_{q\bar{q}}$ is given by
\begin{equation}
\Psi^{\lambda_{q\bar{q}}}_{q\bar{q}}({\bf l}_{q},{\bf l}_{\bar{q}},\lambda_{q},\lambda_{\bar{q}})=-\frac{1}{\sqrt{2}{\tilde{m}}_{q\bar{q}}}\bar{u}(m_{q},{\bf l}_{q},\lambda_{q})\Bigl[\gamma^{\mu}-\frac{l^{\mu}_{q}-l^{\mu}_{\bar{q}}}{m_{q\bar{q}}+m_{q}+m_{\bar{q}}}\Bigr]v(m_{\bar{q}},{\bf l}_{\bar{q}},\lambda_{\bar{q}})\epsilon_{\mu}({\bf l}_{q\bar{q}},\lambda_{q\bar{q}}),
\label{eq:01s}
\end{equation}
then one obtains the $K^{\ast}$ states ($J^{P}=1^{-}$). For a $q\bar{q}$ pair with spin $S_{q\bar{q}}=0$ and the orbital angular momentum $L=1$ we have
\begin{equation}
\Psi^{\lambda_{q\bar{q}}}_{q\bar{q}}({\bf l}_{q},{\bf l}_{\bar{q}},\lambda_{q},\lambda_{\bar{q}})=\frac{1}{\sqrt{2}{\tilde{m}}_{q\bar{q}}({\bf l}_{q},{\bf l}_{\bar{q}})}\bar{u}(m_{q},{\bf l}_{q},\lambda_{q})\gamma^{5}v(m_{\bar{q}},{\bf l}_{\bar{q}},\lambda_{\bar{q}})Y_{1\lambda_{q\bar{q}}}(\bar{{\bf q}}),
\label{eq:10s}
\end{equation}
which corresponds to a $K_{1}$ meson ($J^{P}=1^{+}$). For $S_{q\bar{q}}=1$ we get
\begin{eqnarray}
& & \Psi^{\lambda_{q\bar{q}}}_{q\bar{q}}({\bf l}_{q},{\bf l}_{\bar{q}},\lambda_{q},\lambda_{\bar{q}})=-\sum_{\lambda,l}\frac{1}{\sqrt{2}{\tilde{m}}_{q\bar{q}}}\bar{u}(m_{q},{\bf l}_{q},\lambda_{q})\Bigl[\gamma^{\mu}-\frac{l^{\mu}_{q}-l^{\mu}_{\bar{q}}}{m_{q\bar{q}}+m_{q}+m_{\bar{q}}}\Bigr]v(m_{\bar{q}},{\bf l}_{\bar{q}},\lambda_{\bar{q}}) \nonumber \\
& & \times\,\epsilon_{\mu}({\bf l}_{q\bar{q}},\lambda)Y_{1l}(\bar{{\bf q}})\langle 1,\lambda;1,l|J,\lambda_{q\bar{q}}\rangle,
\label{eq:11s}
\end{eqnarray}
where ${\bf q}$ was defined in ($\ref{eq:q}$). The last wave function characterizes mesons $K^{\ast}_{0}$, $K_{1}$ and $K^{\ast}_{2}$ ($J^{P}=0,1,2^{+}$). In a similar manner one can obtain states with higher orbital angular momenta $L$.

The decay of $\pi_{1}$ into two strange mesons proceeds through the dissociation of the gluon into a pair $s\bar{s}$.
For the processes $\pi_{1}\rightarrow K{\bar{K}}^{\ast}$ and $\pi_{1}\rightarrow K{\bar{K}}_{1}$ the Hamiltonian matrix element is $H=(2\pi)^{3}\delta^{3}({\bf 0})A$ (now there is only one term). The decay amplitude is given by
\begin{eqnarray}
& & A_{(M)}^{(S)}({\bf P},\lambda_{M},\lambda_{ex})=\sum_{\lambda,l}\int\frac{d^{3}{\bf p}}{(2\pi)^{3}}\frac{d^{3}{\bf r}}{(2\pi)^{3}}\frac{(E(m,{\bf p})+E(m',{\bf k}))(E(m',{\bf r})+E(m,{\bf l}))}{4E(m,{\bf p})E(m',{\bf k})E(m',{\bf r})E(m,{\bf l})E(m_{g},{\bf Q})} \nonumber \\
& & \times\,(E(m,{\bf p})+E(m,{\bf l})+E(m_{g},{\bf Q}))\psi_{L}'(m_{q\bar{q}}({\bf p},{\bf l})/\mu_{ex},m_{q\bar{q}g}({\bf p},{\bf l},{\bf Q})/\mu'_{ex}) \nonumber \\
& & \times\,[N_{K}N_{M}N_{ex}]^{-1}\psi_{L}(m_{q\bar{q}}({\bf p},{\bf k})/\mu_{\pi})\psi_{L}(m_{q\bar{q}}({\bf r},{\bf l})/\mu_{M})Y^{\ast}_{L_{M}l}({\bf q}/|{\bf q}|) \nonumber \\
& & \times\,\langle 1,\lambda;L_{M},l|J_{M},\lambda_{M}\rangle g\frac{4}{3}\frac{1}{\sqrt{2}}(spin)(\lambda,\lambda_{ex}).
\label{eq:as}
\end{eqnarray}
Masses of quarks $u$,$d$ (assumed to be equal) and $s$ are denoted respectively by $m$ and $m'$, whereas $M$ stands for either ${\bar{K}}^{\ast}$ or ${\bar{K}}_{1}$, and $S=S_{q\bar{q}g}$. The spin factor is given by
\begin{equation}
B^{\mu}_{\,\,\,j}\psi_{\mu j}^{(S)}(\lambda_{ex})\delta_{\lambda 0}
\end{equation}
for kaons with $S_{q\bar{q}}=1$, and
\begin{equation}
C^{\mu\nu}_{\,\,\,\,\,\,j}\psi_{\mu\nu j}^{(S)}(\lambda,\lambda_{ex})
\end{equation}
if $S_{q\bar{q}}=0$, where
\begin{equation}
B^{\mu j}=\frac{Tr\Bigl[(\not{k}-m')(\not{p}-m)\Bigl(\gamma^{\mu}+\frac{p^{\mu}-l^{\mu}}{m_{q\bar{q}}({\bf p},{\bf l})+2m}\Bigr)(\not{l}+m)(\not{r}+m')\gamma^{j}\Bigr]}{2^{3/2}m_{q\bar{q}}({\bf p},{\bf k})m_{q\bar{q}}({\bf r},{\bf l})m_{q\bar{q}}({\bf p},{\bf l})}
\label{eq:trs1}
\end{equation}
and
\begin{eqnarray}
& & C^{\mu\nu j}=-[2^{3/2}m_{q\bar{q}}({\bf p},{\bf k})m_{q\bar{q}}({\bf r},{\bf l})m_{q\bar{q}}({\bf p},{\bf l})]^{-1}\,Tr\Bigl[(\not{p}+m)\Bigl(\gamma^{\mu}-\frac{p^{\mu}-l^{\mu}}{m_{q\bar{q}}({\bf p},{\bf l})+2m}\Bigr) \nonumber \\
& & \times\,(\not{l}-m)\Bigl(\gamma^{\nu}-\frac{r^{\nu}-l^{\nu}}{m_{q\bar{q}}({\bf r},{\bf l})+m+m'}\Bigr)(\not{r}+m')\gamma^{j}(\not{k}-m')\gamma^{5}\Bigr].
\label{eq:trs2}
\end{eqnarray}
The notation used above is the same as that in ($\ref{eq:kl}$) and Fig.~\ref{hybdec}, with the invariant mass ($\ref{eq:mis}$) and some modifications due to difference between the masses of quarks:
\begin{equation}
p^{0}_{q_{1}}=E(m,{\bf p}),\,\,\,\,r^{0}=E(m',{\bf r}),\,\,\,\,k^{0}=E(m',{\bf k}),\,\,\,\,l^{0}=E(m,{\bf l}).
\end{equation}
The normalization constants for strange mesons are
\begin{eqnarray}
& & N^{2}_{M}(P)=(2E(m_{M},{\bf P}))^{-1}\int\frac{d^{3}{\bf k}}{(2\pi)^{3}}\frac{(E(m,{\bf k})+E(m',{\bf P}-{\bf k}))^{2}}{E(m,{\bf k})E(m',{\bf P}-{\bf k})}[Y_{L0}(\bar{{\bf q}}({\bf k},{\bf P}))]^{2} \nonumber \\
& & \times\,[\psi_{L}(m_{q\bar{q}}({\bf k},{\bf P}-{\bf k})/\mu_{M})]^{2},
\label{eq:normK}
\end{eqnarray}
where ${\bf q}={\bf q}({\bf k},{\bf P})$ is in ($\ref{eq:qq}$).
The kaon weak decay constant can be used to constrain the value of $\mu_{K}$:
\begin{equation}
f_{K}(P)=\frac{\sqrt{3}}{N_{K}(P)E(m_{K},{\bf P})}\int\frac{d^{3}{\bf p}}{(2\pi)^{3}}\frac{(p^{0}+q^{0})(m'p^{0}+mq^{0})}{p^{0}q^{0}m_{q\bar{q}}({\bf p},{\bf P}-{\bf p})}\,e^{-\frac{m_{q\bar{q}}^{2}}{8\mu_{K}^{2}}}.
\label{eq:formfK}
\end{equation}

\section{Nonrelativistic limit}

For the decays $\pi_{1}\rightarrow\pi\eta,\,\pi b_{1}$ the behaviour of the tensor (\ref{eq:tr1}) is given by Eq.~(\ref{tr1n}). Therefore the spin factor $B^{\mu}_{\,\,\,j}\psi^{(S)}_{\mu j}$ is:
\begin{eqnarray}
& & B^{\mu}_{\,\,\,j}\psi^{(S=0)}_{\mu j}\rightarrow\sqrt{\frac{3}{\pi}}m\bar{Q}^{l}\epsilon^{l}(\lambda_{ex}), \nonumber \\
& & B^{\mu}_{\,\,\,j}\psi^{(S=1)}_{\mu j}\rightarrow\sqrt{\frac{3}{4\pi}}m(\delta^{jl}-\bar{Q}^{j}\bar{Q}^{l})\bar{Q}^{j}\epsilon^{l}(\lambda_{ex})=0, \nonumber \\
& & B^{\mu}_{\,\,\,j}\psi^{(S=2)}_{\mu j}\rightarrow-\frac{2}{\sqrt{13}}B^{\mu}_{\,\,\,j}\psi^{(S=0)}_{\mu j}.
\label{eq:nr1}
\end{eqnarray}
From the above it follows
\begin{equation}
\Gamma_{(S=1)}\rightarrow0,\,\,\,\frac{\Gamma_{(S=2)}}{\Gamma_{(S=0)}}\rightarrow\frac{4}{13}. 
\label{eq:b1d}
\end{equation}
The amplitudes $A_{\pi\eta}(P)$ and $A_{\pi b_{1}}(P)$ given by ($\ref{eq:ameb}$) are in this limit (for $S=0$)
\begin{eqnarray}
& & A_{\pi M}(P)=-2\frac{4}{3}\frac{1}{\sqrt{2}}m\sqrt{\frac{3}{\pi}}\frac{2m+m_{g}}{m^{2}m_{g}}\frac{g}{N_{\pi}(P)N_{M}(P)N_{ex}}\int\frac{d^{3}{\bf p}}{(2\pi)^{3}}\frac{d^{3}{\bf r}}{(2\pi)^{3}}\bar{Q}_{z}Y^{\ast}_{J_{M}0}({\bf q}/|{\bf q}|) \nonumber \\
& & \times\,\psi_{L}(m_{q\bar{q}}({\bf p},{\bf k})/\mu_{\pi})\psi_{L}(m_{q\bar{q}}({\bf r},{\bf l})/\mu_{M})\psi_{L}'(m_{q\bar{q}}({\bf p},{\bf l})/\mu_{ex},m_{q\bar{q}g}({\bf p},{\bf l},{\bf Q})/\mu'_{ex}), 
\label{eq:n1}
\end{eqnarray}
where ${\bf q}={\bf p}_{q_{2}}+{\bf P}/2$, $\lambda_{ex}=0$ and ${\bf k},\,{\bf l},\,{\bf Q}$ were defined in ($\ref{eq:kl}$). As before, $M$ denotes $\eta$ or $b_{1}$, with $J=0$ or $1$. In the nonrelativistic limit only the $S=0$ and $S=2$ hybrid components of the $\pi_{1}$ state contribute to its decays into $\pi\eta$ and $\pi b_{1}$. Therefore one may expect the $S=1$ component not to be too large for a fully relativistic case. 

For $\pi_{1}\rightarrow\pi\rho,\,\pi f_{1,2}$ we have
\begin{equation}
C^{ikj}\rightarrow\sqrt{2}im\epsilon^{0ikj}
\end{equation}
and the other components are much smaller. This results in
\begin{eqnarray}
& & C^{\mu\nu}_{\,\,\,\,\,\,j}\psi_{\mu\nu j}^{(S=0)}\rightarrow0, \nonumber \\
& & C^{\mu\nu}_{\,\,\,\,\,\,j}\psi_{\mu\nu j}^{(S=1)}\rightarrow\sqrt{\frac{3}{4\pi}}im\bar{Q}^{i}\epsilon^{j}(\lambda_{ex})\epsilon^{k\ast}(\lambda)\epsilon^{ikj}, \nonumber \\
& & C^{\mu\nu}_{\,\,\,\,\,\,j}\psi_{\mu\nu j}^{(S=2)}\rightarrow\frac{3}{\sqrt{13}}C^{\mu\nu}_{\,\,\,\,\,\,j}\psi_{\mu\nu j}^{(S=1)}.
\end{eqnarray}
Therefore we have
\begin{equation}
\Gamma_{(S=0)}\rightarrow0,\,\,\,\frac{\Gamma_{(S=2)}}{\Gamma_{(S=1)}}\rightarrow\frac{9}{13}. 
\end{equation}
The amplitudes $A_{\pi\rho}(P)$ and $A_{\pi f_{1,2}}(P)$ given by ($\ref{eq:amrf}$) are in this limit (for $S=1$)
\begin{eqnarray}
& & A_{\pi M}(P)=2\frac{4}{3}\frac{1}{\sqrt{2}}m\sqrt{\frac{3}{4\pi}}\frac{2m+m_{g}}{m^{2}m_{g}}\frac{g}{N_{\pi}(P)N_{M}(P)N_{ex}}\int\frac{d^{3}{\bf p}}{(2\pi)^{3}}\frac{d^{3}{\bf r}}{(2\pi)^{3}}\bar{Q}_{z}Y^{\ast}_{L_{q\bar{q}(M)}0}({\bf q}/|{\bf q}|) \nonumber \\
& & \times\,\psi_{L}(m_{q\bar{q}}({\bf p},{\bf k})/\mu_{\pi})\psi_{L}(m_{q\bar{q}}({\bf r},{\bf l})/\mu_{M})\psi_{L}'(m_{q\bar{q}}({\bf p},{\bf l})/\mu_{ex},m_{q\bar{q}g}({\bf p},{\bf l},{\bf Q})/\mu'_{ex}) \nonumber \\
& & \times\,(CG)_{M},
\label{eq:n2}
\end{eqnarray}
with the notations from ($\ref{eq:kl}$) and $M$ standing for $\rho$ ($L_{q\bar{q}}=0$ and $(CG)=1$) or $f_{1,2}$ ($L_{q\bar{q}}=1$ and $(CG)=1/\sqrt{2}$). In the nonrelativistic limit only the $S=1$ and $S=2$ $q\bar{q}g$ component of the $\pi_{1}$ state contribute to its decays into $\pi\rho$ and $\pi f_{1,2}$. Therefore one may expect the $S=0$ component not to be too large for a relativistic case. Similar conlusions can be made for the $\eta a_{1,2}$ modes.

Finally, for $\pi_{1}\rightarrow\rho\omega$ we have
\begin{equation}
D^{ijkl}\rightarrow\sqrt{2}m(\delta^{ij}\delta^{kl}-\delta^{ik}\delta^{jl}+\delta^{il}\delta^{jk})
\end{equation}
(the other components are much smaller), which gives:
\begin{eqnarray}
& & D^{\mu\nu\rho}_{\,\,\,\,\,\,\,\,\,\,j}\psi_{\mu\nu\rho j}^{(S=0)}\rightarrow\sqrt{\frac{3}{\pi}}m\bar{Q}^{i}\epsilon^{i}(\lambda_{ex})\epsilon^{j\ast}(\lambda_{\rho})\epsilon^{j\ast}(\lambda_{\omega}), \nonumber \\
& & D^{\mu\nu\rho}_{\,\,\,\,\,\,\,\,\,\,j}\psi_{\mu\nu\rho j}^{(S=1)}\rightarrow\sqrt{\frac{3}{4\pi}}m(\bar{Q}^{i}\delta^{jk}-\bar{Q}^{j}\delta^{ik})\epsilon^{i\ast}(\lambda_{\rho})\epsilon^{j\ast}(\lambda_{\omega})\epsilon^{k}(\lambda_{ex}).
\end{eqnarray}
The amplitude $A_{\rho\omega}(P)$ in this limit is described by formula ($\ref{eq:n1}$) (for $S=0$) or ($\ref{eq:n2}$) (for $S=1$), where $\pi$ is replaced by $\rho$ and $M=\omega$.

We can straightforwardly understand the difference in amplitudes coming from the spin wave function. If we assume $m_{\eta}=m_{\rho},\,\mu_{\eta}=\mu_{\rho}$ and $\,m_{b_{1}}=m_{f_{1}}=m_{f_{2}},\,\mu_{b_{1}}=\mu_{f_{1}}=\mu_{f_{2}}$ (the second condition for masses is satisfied to a good approximation), then comparing ($\ref{eq:n1}$) with ($\ref{eq:n2}$) and ($\ref{eq:ga1}$) with ($\ref{eq:ga2}$) gives
\begin{eqnarray}
& & A_{\pi\rho}=\frac{1}{2}A_{\pi\eta}\,\rightarrow\,\Gamma_{\pi\rho}=\frac{1}{2}\Gamma_{\pi\eta} \nonumber \\
& & A_{\pi f_{1}}=\frac{1}{2\sqrt{2}}A_{\pi b_{1}}\,\rightarrow\,\Gamma_{\pi f_{1}}=\frac{1}{8}\Gamma_{\pi b_{1}}.
\end{eqnarray}
Here, $A_{\pi\eta}$, $A_{\pi b_{1}}$ were taken for $S=0$ and $A_{\pi\rho}$, $A_{\pi f_{1,2}}$ for $S=1$. The relation between $A_{\pi\eta}$ and $A_{\pi b_{1}}$ (or between $A_{\pi\rho}$ and $A_{\pi f_{1,2}}$) is more complicated and depends on the orbital angular momentum wave functions $\psi_{L}$ and $\psi'_{L}$. If $\mu_{\rho}=\mu_{\pi}$, then in the nonrelativistic limit $\pi_{1}$ would not decay into $\pi\rho$ since the corresponding amplitude becomes proportional to that of the $\pi\eta$ mode. Therefore the width of the $\pi\rho$ mode is expected to be much smaller than that of $\pi b_{1}$, assuming the parameters $\mu_{rho}$ and $\mu_{\pi}$ are very close to one another.

Analogous calculations can be made for decays of $\pi_{1}$ into strange mesons. The amplitudes $A_{K{\bar{K}}_{1}}$ ($S_{q\bar{q}}=0,1$) behave similarly to $A_{\pi b_{1}}$ and $A_{\pi f_{1}}$, whereas $A_{K{\bar{K}}^{\ast}}$ behaves like $A_{\pi\rho}$. Therefore the former will be dominant and the latter is expected to be much smaller. If $\mu_{K^{\ast}}=\mu_{K}$, then in the nonrelativistic limit $A_{K{\bar{K}}^{\ast}}=0$.

\section{Numerical results}

In Tables \ref{tab1} and \ref{tab2} we present the widths for various decay modes containing radially ground-state mesons. The numbers in parentheses correspond to calculations using the nonrelativistic formulae, and $S$ denotes the total spin of the $q\bar{q}g$ component in the $\pi_{1}$ wave function. As before, we are interested in relativistic effects in the $\pi_{1}$ decays. Thus it is sufficient to calculate the amplitudes for each value of $S$ separately. 
The procedure for obtaining the relative contributions for each $S$ was given at the beginning of Chapter~7. 
The values of parameters $\mu$ for all unflavored mesons and $\pi_{1}$ were taken equal to $\mu_{\pi}$, and for all strange mesons equal to $\mu_{K}$. This makes the width values for the modes $\pi\eta$, $\pi\eta'$ and $\rho\omega$ identically equal to zero.

\begin{table}
\centering
\begin{tabular}{|c|c||r|r|r|}
\hline
\multicolumn{2}{|c||}{$\Gamma_{rel}(\Gamma_{nrel})$}&$S=0$&$S=1$&$S=2$\\
\hline\hline
$\pi b_{1}(1235)$&S& $150(259)$ & $<1(0)$ & $44(80)$\\
\cline{2-5}
 &D& $<1(<1)$ & $<1(0)$ & $<1(<1)$\\
\hline
$\pi f_{1}(1285)$&S& $<1(0)$ & $20(33)$ & $14(23)$\\
\cline{2-5}
 &D& $<1(0)$ & $<1(<1)$ & $<1(<1)$\\
\hline
$\pi f_{2}(1270)$&D& $<1(0)$ & $<1(<1)$ & $<1(<1)$\\
\hline
$\pi\rho(770)$&P& $3(0)$ & $<1(0)$ & $1(0)$\\
\hline
$K{\bar{K}}^{\ast}(892)$&P& $1(0)$ & $<1(0)$ & $<1(0)$\\
\hline
\end{tabular}
\caption{\label{tab1} Relativistic (nonrelativistic) widths in MeV for the $\pi_{1}$ decay modes with mass 1.6 GeV, for all possible values of the total intrinsic spin $S$ of $\pi_{1}$. }
\vspace{0.3in}
\centering
\begin{tabular}{|c|c||r|r|r|}
\hline
\multicolumn{2}{|c||}{$\Gamma_{rel}(\Gamma_{nrel})$}&$S=0$&$S=1$&$S=2$\\
\hline\hline
$\pi b_{1}(1235)$&S& $48(70)$ & $<1(0)$ & $13(22)$\\
\cline{2-5}
 &D& $1(2)$ & $<1(0)$ & $<1(<1)$\\
\hline
$\pi f_{1}(1285)$&S& $<1(0)$ & $7(11)$ & $5(8)$\\
\cline{2-5}
 &D& $<1(0)$ & $2(<1)$ & $1(<1)$\\
\hline
$\pi f_{2}(1270)$&D& $<1(0)$ & $2(<1)$ & $1(<1)$\\
\hline
$\pi\rho(770)$&P& $2(0)$ & $<1(0)$ & $<1(0)$\\
\hline
$\eta a_{1}(1260)$&S& $<1(0)$ & $13(22)$ & $9(16)$\\
\cline{2-5}
 &D& $<1(0)$ & $1(<1)$ & $<1(<1)$\\
\hline
$\eta a_{2}(1320)$&D& $<1(0)$ & $1(<1)$ & $<1(<1)$\\
\hline
$K{\bar{K}}_{1}(1400)$&S& $127(45)$ & $<1(0)$ & $39(14)$\\
\cline{2-5}
 &D& $<1(<1)$ & $<1(0)$ & $<1(<1)$\\
\hline
$K{\bar{K}}_{1}(1270)$&S& $<1(0)$ & $11(4)$ & $7(3)$\\
\cline{2-5}
 &D& $<1(0)$ & $<1(<1)$ & $<1(<1)$\\
\hline
$K{\bar{K}}^{\ast}(892)$&P& $1(0)$ & $<1(0)$ & $<1(0)$\\
\hline
\end{tabular}
\caption{\label{tab2} Relativistic (nonrelativistic) widths in MeV for the $\pi_{1}$ decay modes with mass 2.0 GeV, for all possible values of the total intrinsic spin $S$ of $\pi_{1}$. }
\end{table}

The mode $\pi\rho$ was expected to be rather small because in the nonrelativistic limit the corresponding amplitude was proportional to that of the mode $\pi\eta$, assuming $\mu_{\rho}=\mu_{\eta}$. We also assumed $\mu_{\rho}=\mu_{\pi}$. Combining both equalities led to this conclusion. We also found that the $q\bar{q}g$ component of $\pi_{1}$ state with $S=0$ did not contribute to the hybrid decay in this limit identically. For the relativistic case, however, we observe that the largest contribution in the $\pi\rho$ channel comes from the $S=0$ component. Our other predictions are confirmed by numerical results, i.e., the $S=1$ component for the $\pi b_{1}$ mode and the $S=0$ component for the $\pi f_{1}$ mode are negligible in the relativistic model as well.   

The decay mode $\pi b_{1}$ is a dominant one. In the following, we will analyze how its widths depend on the free parameters, except the coupling constant $g$ which just multiplies the amplitude and $\mu_{ex}$ whose equality to $\mu_{\pi}$ seems to be justified. Because the D-wave value in this case is small as compared to the S-wave, we will only deal with the latter. Furthermore, in the following plots we will assume $S=0$ since $S=1$ gives negligible contribution (in the nonrelativistic limit it is zero) and thus the widths for $S=2$ are proportional to those of $S=0$ ($\ref{eq:b1d}$). The default values of the free parameters are those from Table~\ref{tab1}.

\begin{figure}[t]
\centering
\includegraphics[width=2.5in,height=3.5in,angle=270]{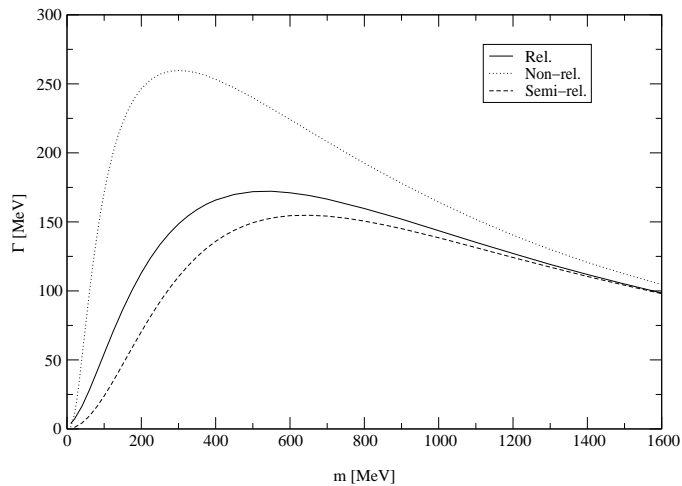}
\caption{\label{b1} Relativistic, nonrelativistic, and semirelativistic widths for the S-wave decay $\pi_{1}\rightarrow\pi b_{1}$ as a function of the quark mass $m$, for a $\pi_1$ with mass 1.6 GeV. }
\end{figure}

\begin{figure}[t]
\centering
\includegraphics[width=2.5in,height=3.5in,angle=270]{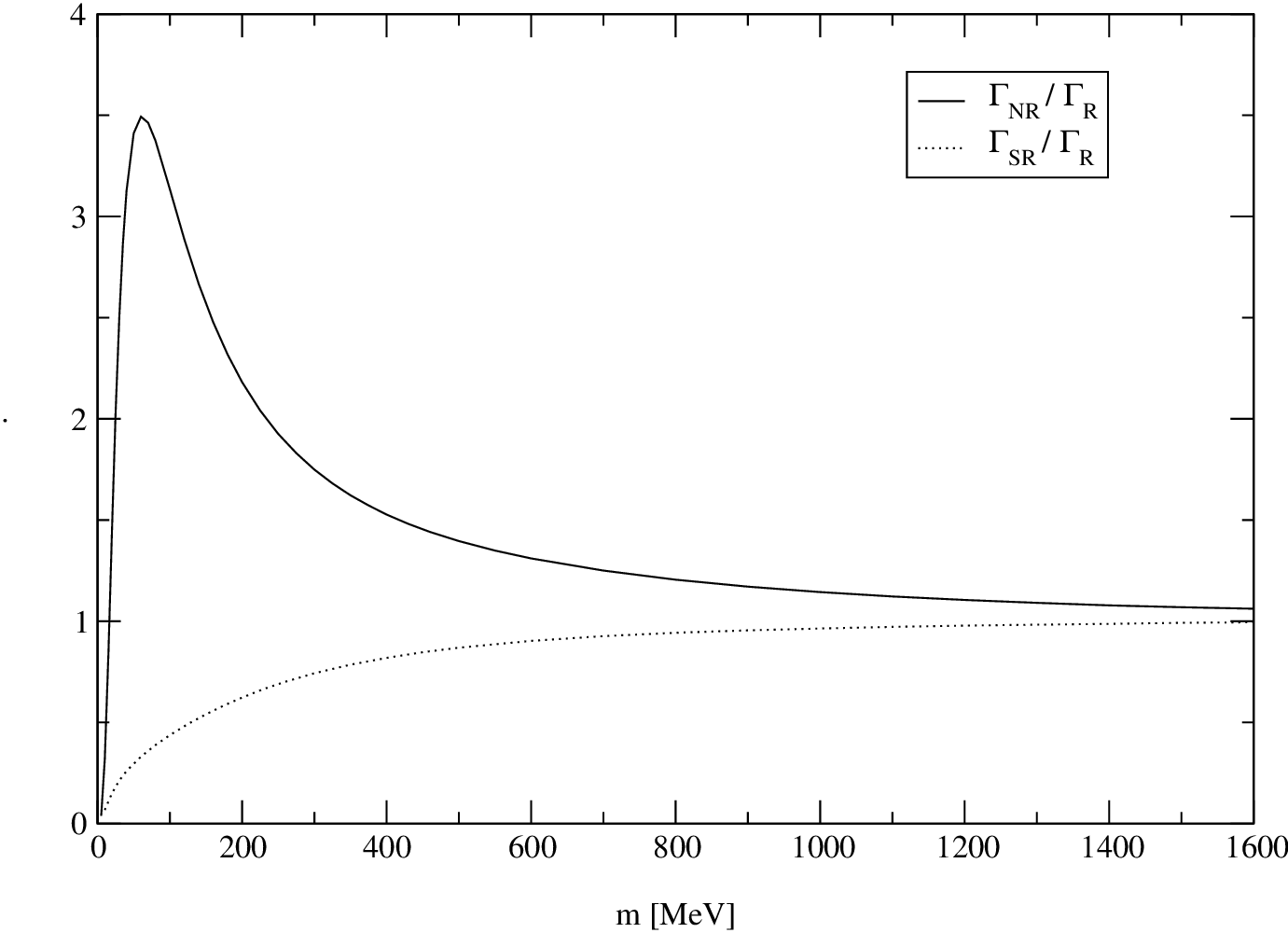}
\caption{\label{b1ratio} Width ratios NR/R and SR/R for $\pi_{1}\rightarrow\pi b_{1}$, as functions of the quark mass $m$, for a $\pi_1$ with mass 1.6 GeV. }
\vspace{0.27in}
\centering
\includegraphics[width=2.5in,height=3.5in,angle=270]{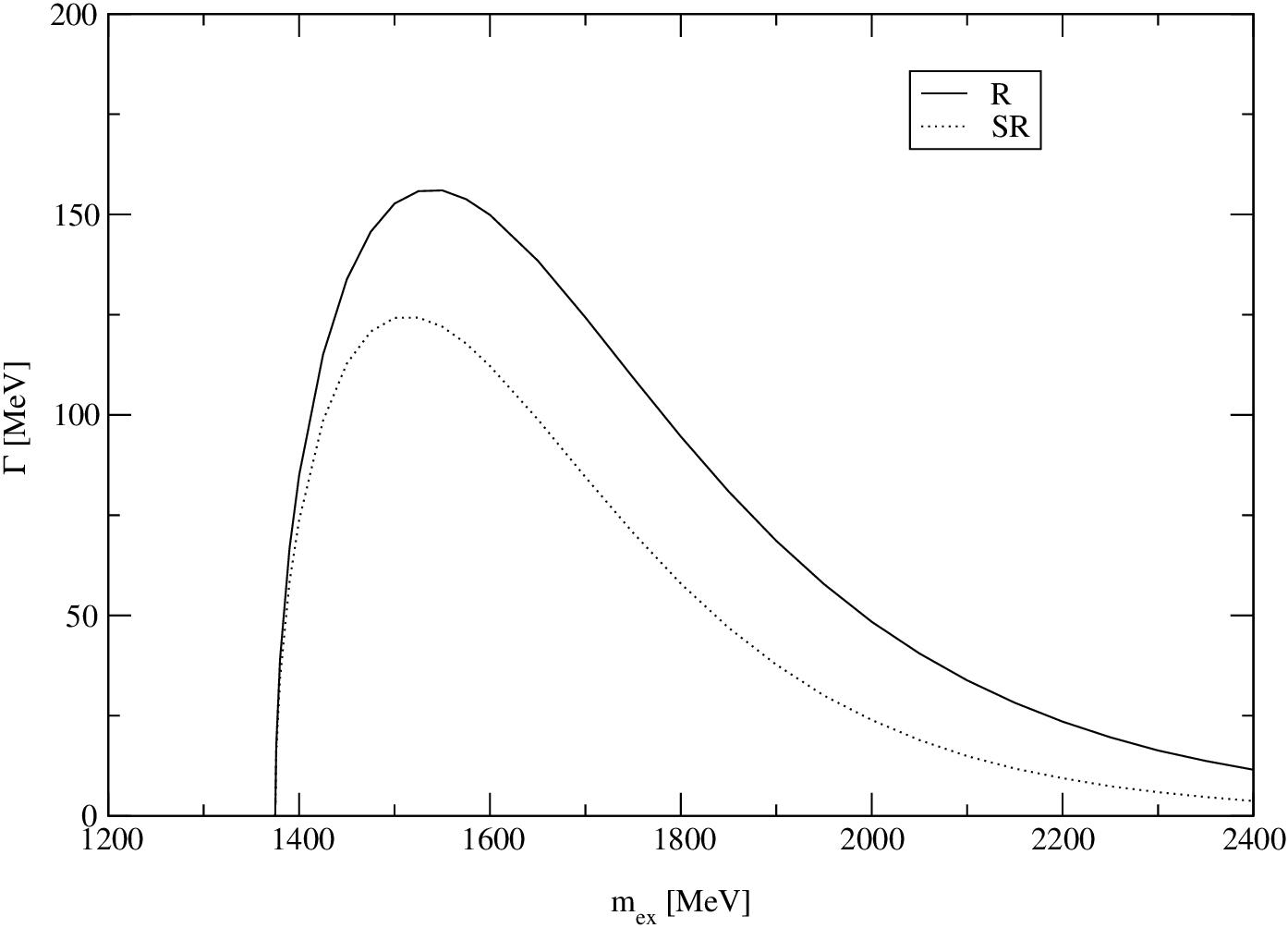}
\caption{\label{b1_mex} Decay widths for $\pi_{1}\rightarrow\pi b_{1}$ as functions of $\pi_{1}$. }
\end{figure}

In Fig.~\ref{b1} we compare relativistic and nonrelativistic predictions for the width of the decay $\pi_{1}\rightarrow\pi b_{1}$ as a function of the mass of the light quark $m$. It also shows the semirelativistic values which include relativistic phase space and orbital wave functions, but with no Wigner rotations. For larger values of $m$ all three curves R, NR and SR converge, as it should be. The corresponding ratios NR/R and SR/R are shown in Fig.~\ref{b1ratio}.

Let us proceed to analyze the contribution of the other free parameters. We show only relativistic and semirelativistic rates because the nonrelativistic ones correspond to different orbital wave functions and we are interested in corrections arising from Wigner rotations. The dependence of the width of $\pi_{1}\rightarrow\pi b_{1}$ on the mass of $\pi_{1}$ is presented in Fig.~\ref{b1_mex}. 
We observe that Wigner rotations contribute more for larger values of $m_{ex}$, and at 2000 MeV the ratio SR/R is already about 50\%. 
We also see that the maximal width occurs for $m_{ex}$ in the range $1500-1600$ MeV, i.e., for the values close to those experimentally reported. Thus, our predictions for the width are an upper limit with respect to $m_{ex}$.

\newpage 
\begin{figure}[h]
\centering
\includegraphics[width=2.5in,height=3.5in,angle=270]{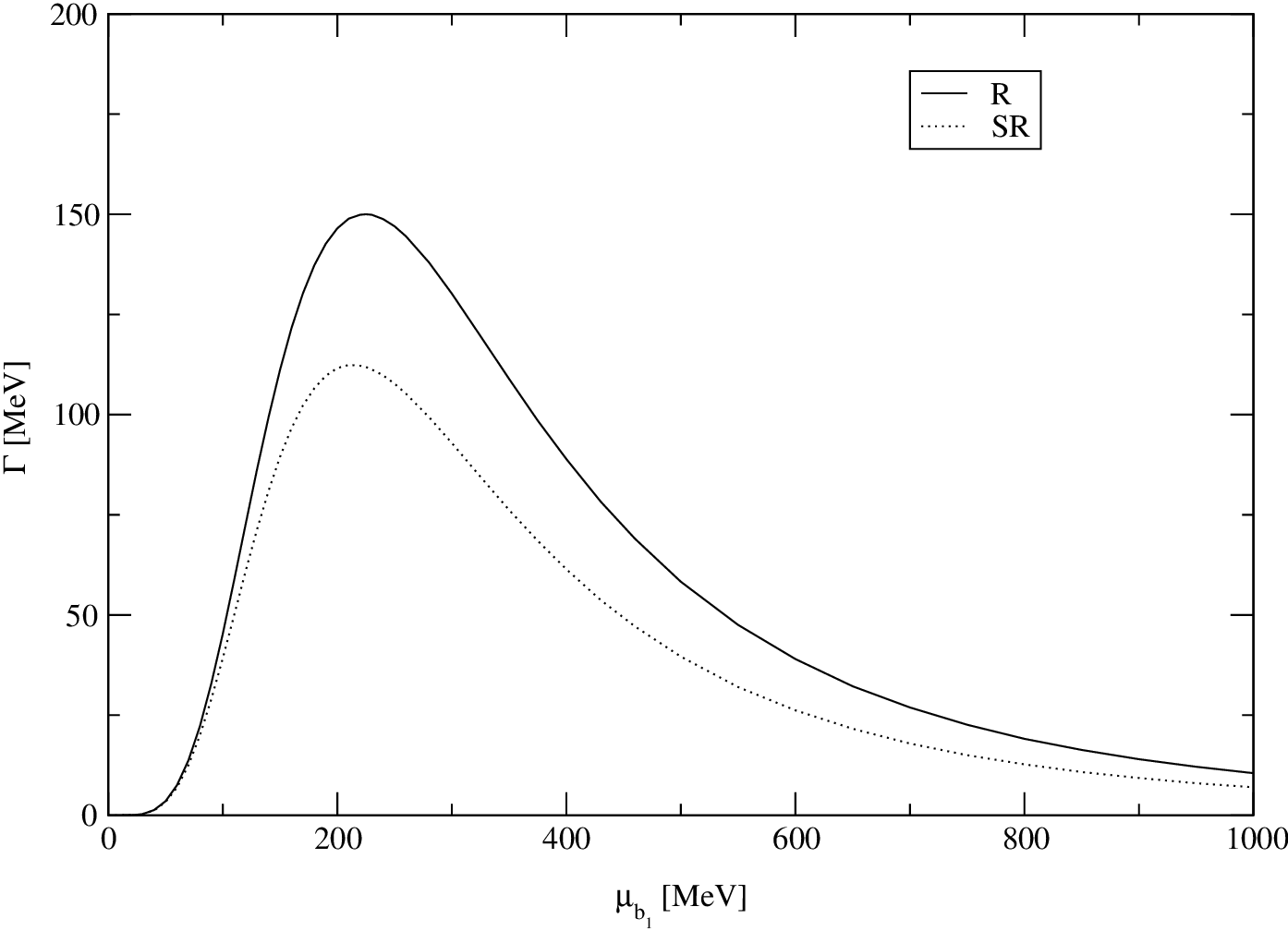}
\caption{\label{b1_mu} Decay widths for $\pi_{1}\rightarrow\pi b_{1}$ as functions of the parameter $\mu_{b_{1}}$, for a $\pi_1$ with mass 1.6 GeV. }
\vspace{0.27in}
\centering
\includegraphics[width=2.5in,height=3.5in,angle=270]{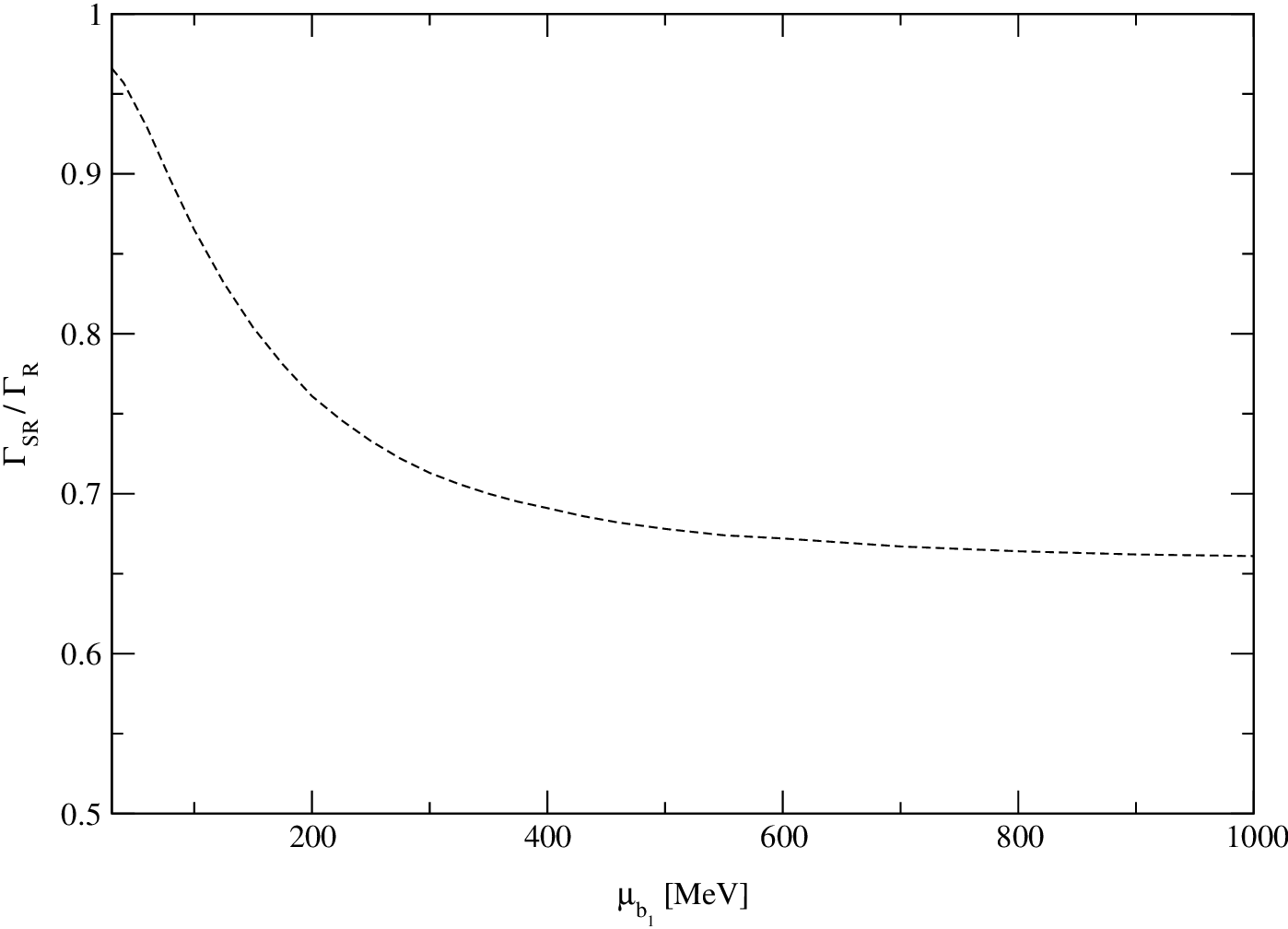}
\caption{\label{b1rat_mu} Ratio SR/R for $\pi_{1}\rightarrow\pi b_{1}$ as a function of the parameter $\mu_{b_{1}}$, for a $\pi_1$ with mass 1.6 GeV. }
\end{figure}

In Fig.~\ref{b1_mu} we show the width dependence on $\mu_{b_{1}}$. If the value of $\mu_{b_{1}}$ is equal to $\mu_{\pi}$, then the width is approximately at its maximum. Probably it is a coincidence, but it tells us that our prediction is an upper limit for this width with respect to $\mu_{b_{1}}$. The corresponding ratio SR/R is given in Fig.~\ref{b1rat_mu}. At masses around 1 GeV it already behaves like a constant. 

\begin{figure}[ht]
\vspace{0.27in}
\centering
\includegraphics[width=2.5in,height=3.5in,angle=270]{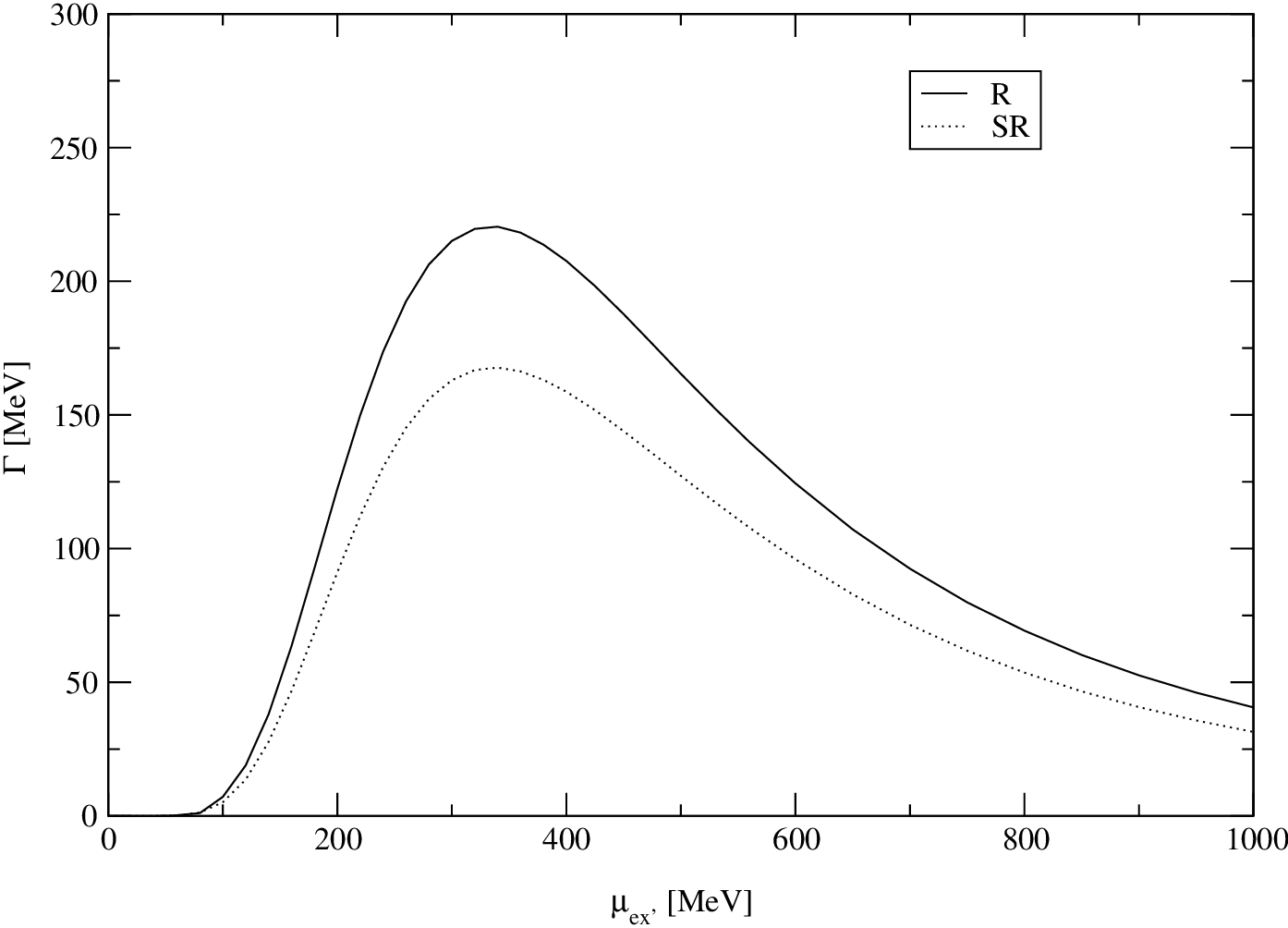}
\caption{\label{b1_exp} Decay widths for $\pi_{1}\rightarrow\pi b_{1}$ as functions of the parameter $\mu_{ex'}$, for a $\pi_1$ with mass 1.6 GeV. }
\vspace{0.27in}
\centering
\includegraphics[width=2.5in,height=3.5in,angle=270]{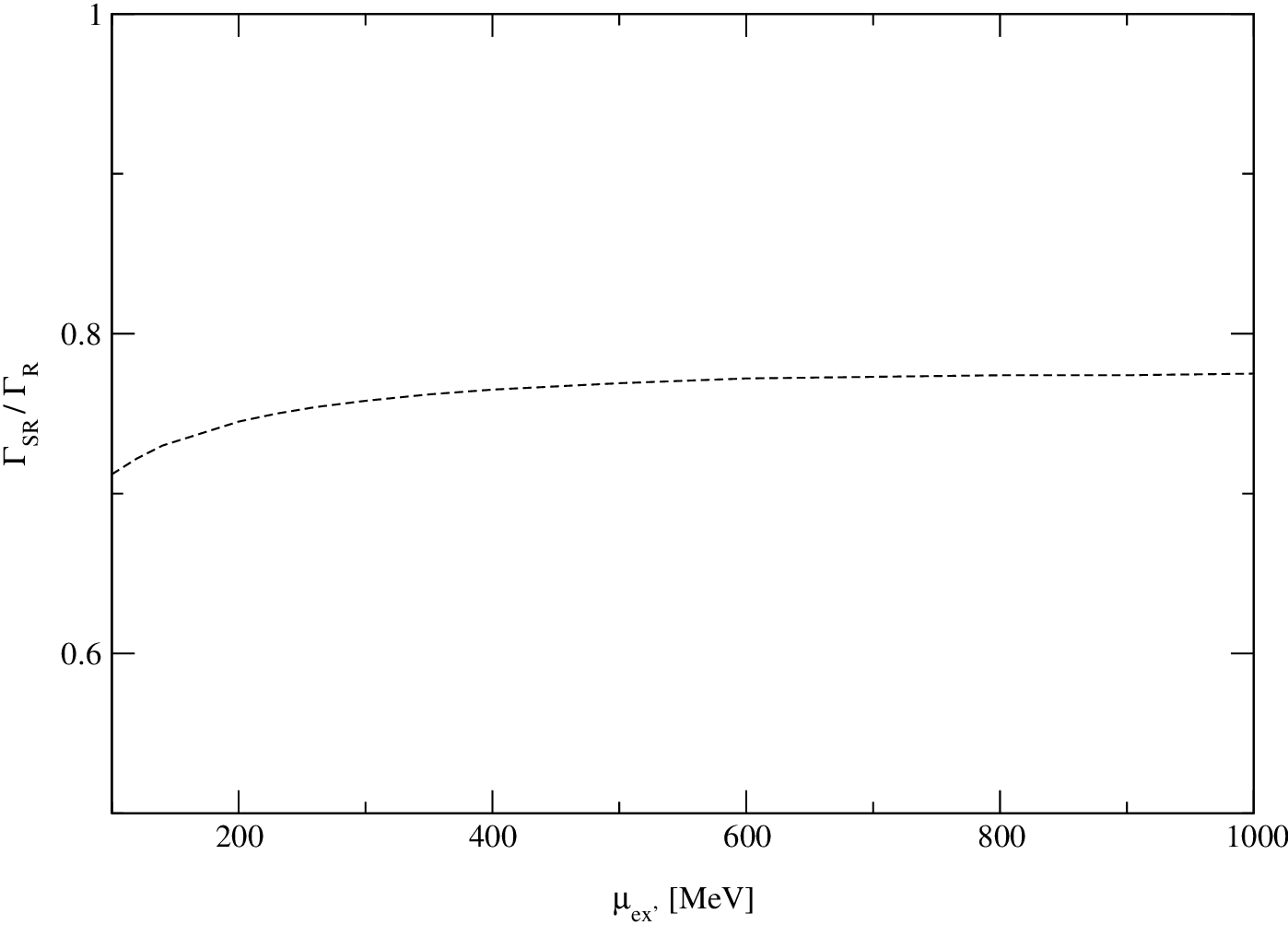}
\caption{\label{b1rat_exp} Ratio SR/R for $\pi_{1}\rightarrow\pi b_{1}$ as a function of the parameter $\mu_{ex'}$, for a $\pi_1$ with mass 1.6 GeV. }
\end{figure}

There is one parameter whose choice was not justified, $\mu_{ex'}$. This quantity divides the invariant mass of the $q\bar{q}g$ system, and there is no reason why it should be equal to $\mu_{\pi}$, as we assumed. Its value may be obtained from decays of normal mesons such as $\rho$ or $b_{1}$. In Fig.~\ref{b1_exp} we show how the width of $\pi_{1}\rightarrow\pi b_{1}$ as a function of $\mu_{ex'}$. The ratio SR/R is given in Fig.~\ref{b1rat_exp} and we see that it is close to a constant, i.e., the choice of $\mu_{ex'}$ is crucial for the widths but not for relative relativistic corrections. 

\begin{figure}[ht]
\centering
\includegraphics[width=2.5in,height=3.5in,angle=270]{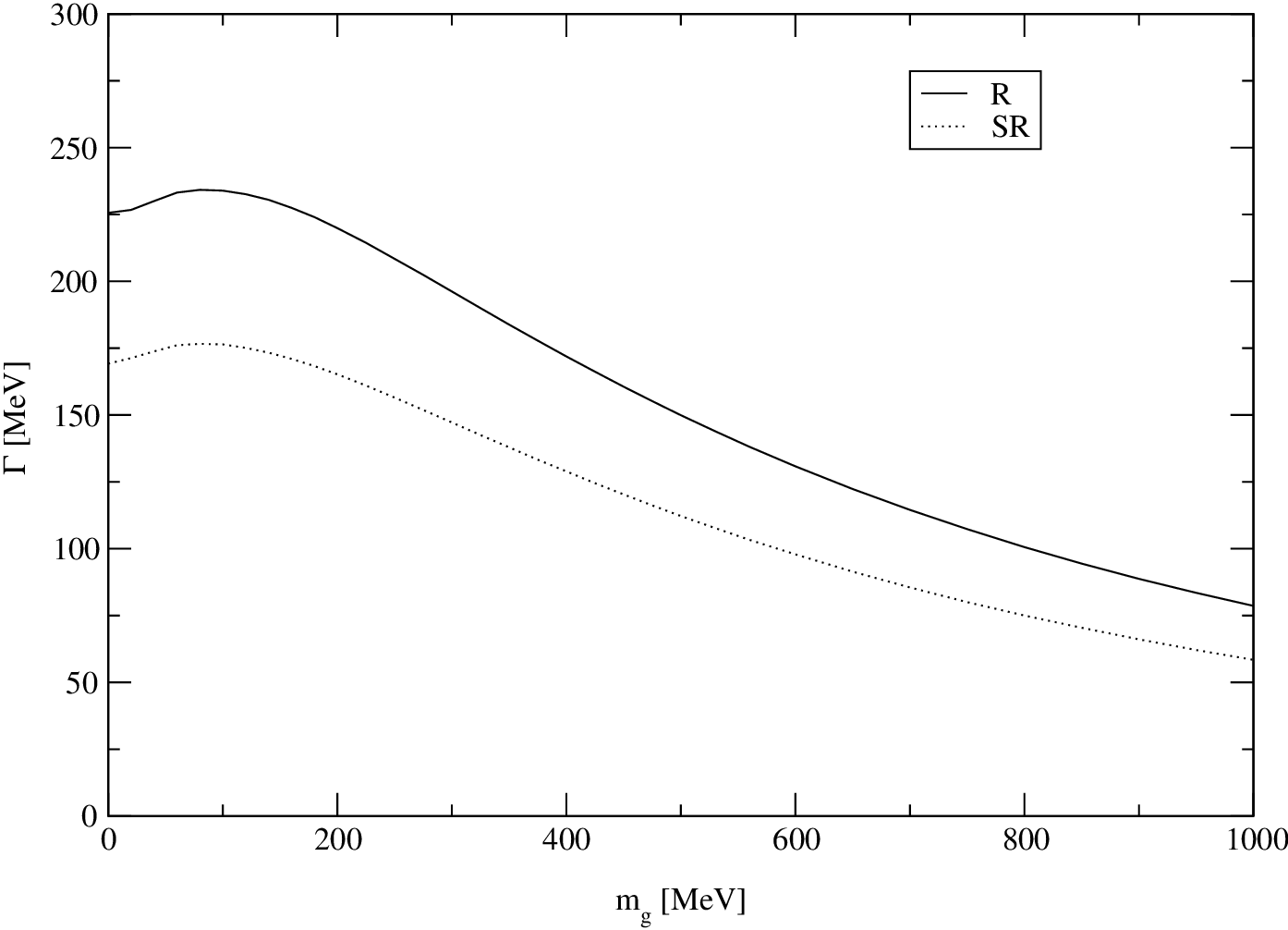}
\caption{\label{b1_mg} Decay widths for $\pi_{1}\rightarrow\pi b_{1}$ as functions of the effective mass of the gluon $m_{g}$, for a $\pi_1$ with mass 1.6 GeV. }
\centering
\vspace{0.35in}
\includegraphics[width=2.5in,height=3.5in,angle=270]{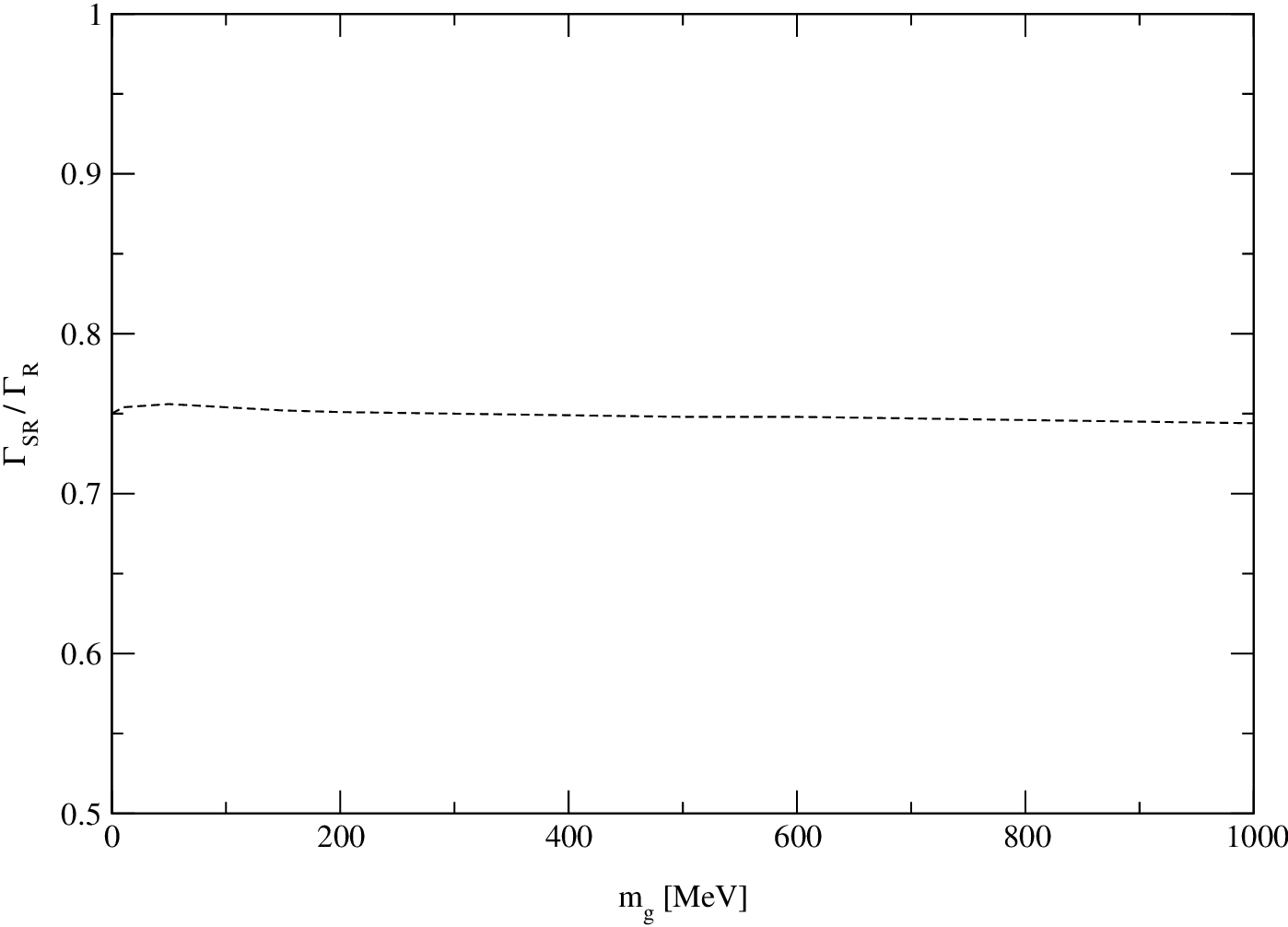}
\caption{\label{b1rat_mg} Ratio SR/R for $\pi_{1}\rightarrow\pi b_{1}$ as a function of the effective mass of the gluon $m_{g}$, for a $\pi_1$ with mass 1.6 GeV. }
\end{figure}

Finally, in Fig.~\ref{b1_mg} we show the width dependence on the effective mass of the gluon $m_{g}$, and in Fig.~\ref{b1rat_mg} the corresponding ratio SR/R. The latter displays clearly that the ratio is essentially independent of $m_{g}$. We observe that the relative relativistic corrections arising from Wigner rotations are insensitive to the free parameters referring to the $\pi_{1}$ at rest, i.e., $m_{ex'}$ and $m_{g}$. 

\newpage

\begin{figure}[ht]
\centering
\includegraphics[width=2.5in,height=3.5in,angle=270]{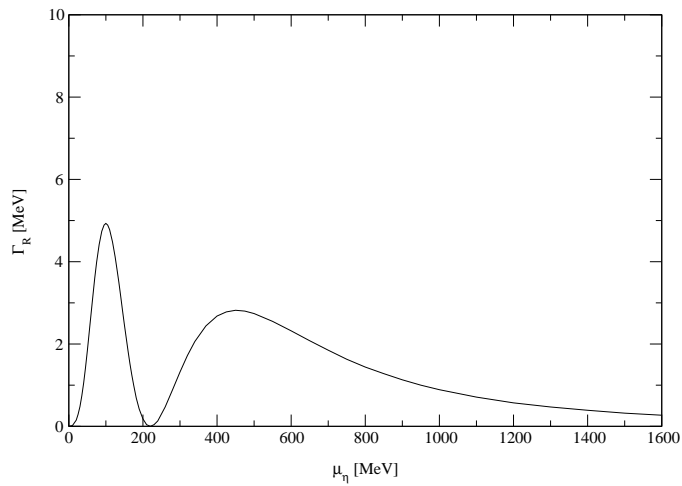}
\caption{\label{eta} Decay width for $\pi_{1}\rightarrow\pi\eta$ as a function of the parameter $\mu_{\eta}$, for a $\pi_1$ with mass 1.6 GeV. }
\end{figure}

\begin{figure}[ht]
\centering
\vspace{0.27in}
\includegraphics[width=2.5in,height=3.5in,angle=270]{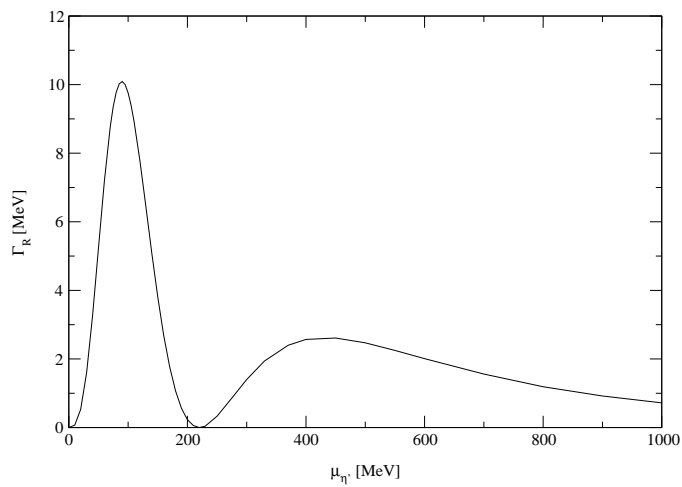}
\caption{\label{etap} Decay width for $\pi_{1}\rightarrow\pi\eta'$ as a function of the parameter $\mu_{\eta}$, for a $\pi_1$ with mass 1.6 GeV. }
\end{figure}

Now we move to the decays into $\pi$ and various $\eta$ mesons. Since the amplitude vanishes for $\mu_{\eta}=\mu_{\pi}$ we want to observe what happens if $\mu_{\eta}\neq\mu_{\pi}$. For brevity, we will consider only the relativistic case. 
In Fig.~\ref{eta} we show how the width of the decay $\pi_{1}\rightarrow\pi\eta$ depends on the value of $\mu_{\eta}$, while the other parameters are the defaults. The biggest value approached is only a few MeV. That confirms our previous predictions and is in a good agreement with other models \cite{CP}.
The same dependence for the decay $\pi_{1}\rightarrow\pi\eta'(958)$ is displayed in Fig.~\ref{etap}, and the maximum is around 10 MeV.

\newpage


\begin{figure}[ht]
\centering
\includegraphics[width=2.5in,height=3.5in,angle=270]{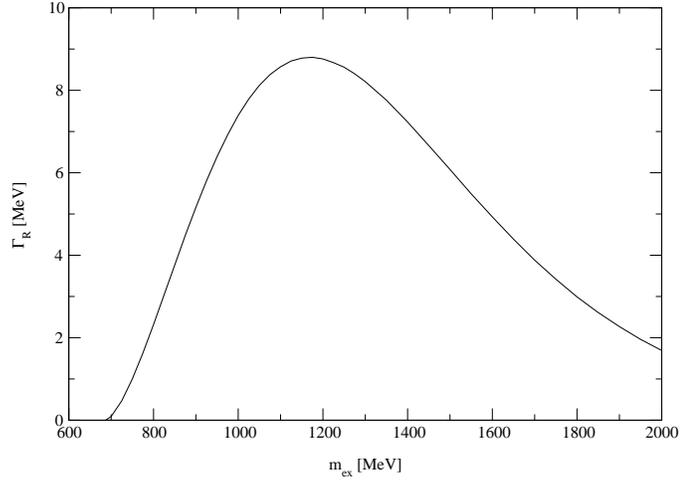}
\caption{\label{eta_ex} Decay width for $\pi_{1}\rightarrow\pi\eta$ at $\mu_{\eta}=\mu_{\eta}^{max}$ as a function of the mass of $\pi_{1}$. }
\end{figure}

\begin{figure}[ht]
\centering
\vspace{0.27in}
\includegraphics[width=2.5in,height=3.5in,angle=270]{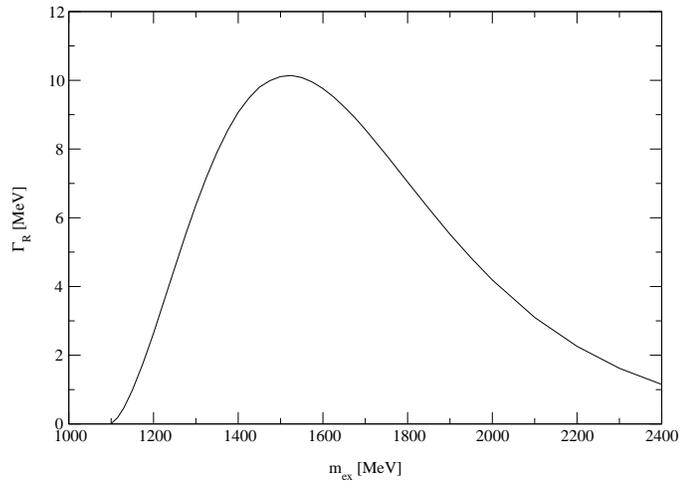}
\caption{\label{etap_ex} Decay width for $\pi_{1}\rightarrow\pi\eta'$ at $\mu_{\eta}=\mu_{\eta}^{max}$ as a function of the mass of $\pi_{1}$. }
\end{figure}

We see that the maximum of the width occurs when $\mu_{\eta}$ is approximately equal to 100 MeV. In Fig.~\ref{eta_ex} and Fig.~\ref{etap_ex} we show the width dependence on the mass of $\pi_{1}$ at this particular value $\mu_{\eta}^{max}$, for $\pi_{1}\rightarrow\pi\eta$ and $\pi_{1}\rightarrow\pi\eta'$ respectively. The reason for this is to see the largest values one can obtain for these decays. It turns out that these values have a maximum on the order of 10 MeV.
However, the value of $\mu_{\eta}$ should not differ much from $\mu_{\pi}$ and thus the above widths will be on the order of 1 MeV.


\begin{figure}
\centering
\includegraphics[width=2.5in,height=3.5in,angle=270]{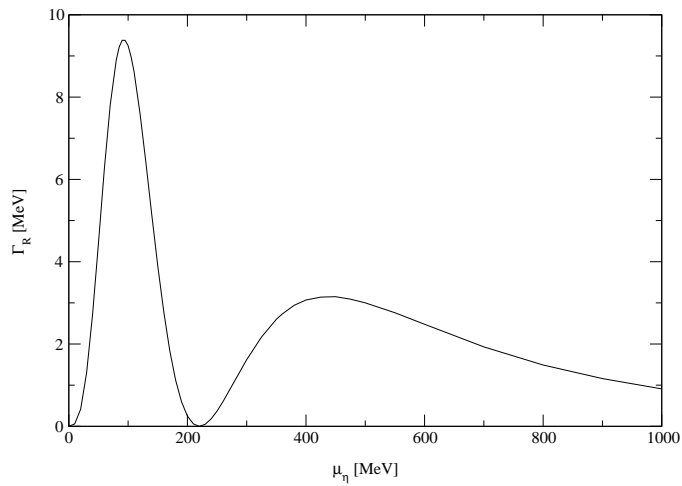}
\caption{\label{etah} Decay width for $\pi_{1}\rightarrow\pi\eta(1295)$ as a function of the parameter $\mu_{\eta}$, for a $\pi_1$ with mass 2.0 GeV. }
\end{figure}

In Tables \ref{tab1} and \ref{tab2} we presented the widths only for the modes including mesons with the lowest radial quantum number such as $\pi$ and $K$. It allowed us to choose the free parameters $\mu$ equal to either $\mu_{\pi}$ or $\mu_{K}$. This, however, cannot be justified for decays into mesons with higher values of the radial quantum number such as $\eta(1295)$ or $\rho(1450)$.
The dependence of the decay width on the value of $\mu_{\eta}$ for $\pi_{1}\rightarrow\pi\eta(1295)$ with $S=0$ and $m_{ex}=$ 2 GeV is shown in Fig.~\ref{etah}. This mode may be a significant one (the width is at most about 10 MeV), which agrees with the results obtained in \cite{PSS}. 
Unfortunately, there is not enough experimental data to constrain the free parameters $\mu$ for $\eta(1295)$ and other radially excited mesons.

From these results it is clear that fully relativistic values are significantly different from nonrelativistic ones. There are two sources of this: the Wigner rotation which introduces relativistic coupling between spin and spatial degrees of freedom in the wave function, and different relations between energy, momentum and the invariant masses (in the phase space and most of all in the orbital wave functions). Both corrections actually appear to introduce corrections as large as 10\% and thus should be included in phenomenological models.

We also observe that in our relativistic constituent model the $S+P$ selection rule, which was mentioned in Chapter~2, is obeyed. According to this rule the favored modes for the $\pi_{1}$ decay are $\pi b_{1}$, $\pi f_{1,2}$, $\eta a_{1,2}$, and both channels $K{\bar{K}}_{1}$. Our results support this prediction and agree with other nonrelativistic models \cite{IKP,CP}.

\chapter{Final state interactions}

In the previous chapters we constructed meson states and studied kinematical relativistic effects at the quark and gluon level. In this chapter we will estimate the size of corrections to the $\pi_{1}$ decays originating from the meson exchange forces between mesons. Since the particle number is not conserved and momenta of particles are on the same order as their masses, this problem should be treated in a relativistic formalism. We will begin with the Lippmann-Schwinger equation applied to the $\pi\rho$ and $\pi b_{1}$ states. Then we will describe the computational procedure of solving the resulting integral equations. At the end of this chapter the numerical results will be given.    

\section{The Lippmann-Schwinger equation}

\begin{figure}[h]
\centering
\includegraphics[width=3.0in]{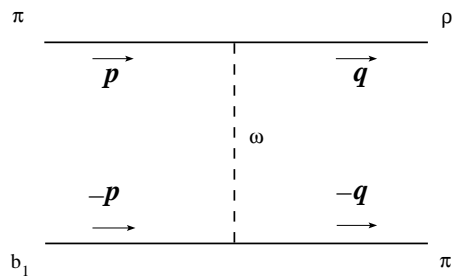}
\caption{\label{final} Final state interaction $\pi b_{1}\leftrightarrow\pi\rho$. }
\end{figure}

\begin{figure}
\centering
\includegraphics[width=1.5in,angle=270]{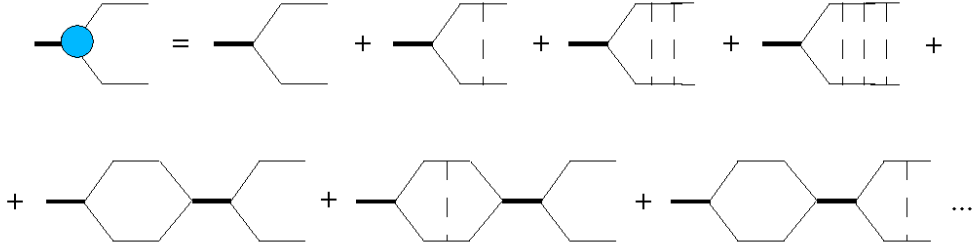}
\caption{\label{diag1} The lowest order FSI corrections to the $\pi_{1}$ decay from the interaction $\pi b_{1}\leftrightarrow\pi\rho$. }
\end{figure}

Other models \cite{IKP,CP,PSS} predict larger widths for the process $\pi_{1}\rightarrow\pi\rho$ than our model. The numerical values were given in Chapter~2. It is possible that this width is increased by final state interactions between the outgoing mesons.
The $b_{1}$ created in the process $\pi_{1}\rightarrow\pi b_{1}$ can subsequently decay into $\pi$ and $\omega$, and then the $\omega$ can absorb the other $\pi$ and create the $\rho$, as shown in Fig.~\ref{final}.
Of course the reverse $\pi\rho\rightarrow\pi b_{1}$ process can occur as well. An $\omega$ exchange may occur more than once and we must sum up all possible amplitudes; the lowest order diagrams are presented in Fig.~\ref{diag1}.
A bold horizontal solid line represents a hybrid and normal horizontal solid lines refer to mesons. A vertical dashed line corresponds to a single $\omega$ meson exchange. 
In order to describe a total contribution of the final state interactions to the decay widths of $\pi_{1}$, we need to solve the Lippmann-Schwinger equation 
\begin{equation}
T=V+VGT,
\end{equation}
which is equivalent to summing over all diagrams, including those with intermediate hybrid states.

\begin{figure}
\centering
\vspace{0.27in}
\includegraphics[width=1.8in,angle=270]{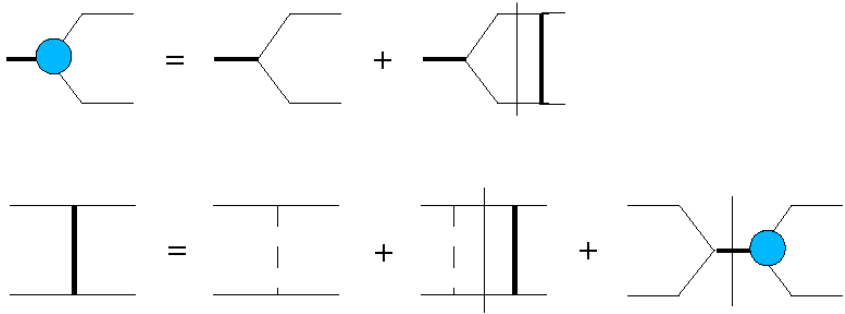}
\caption{\label{diag2} The Lippmann-Schwinger equation shown diagrammatically. } 
\end{figure}

Let the $\pi_{1}$ state be denoted by index $\alpha$, and the states $\pi\rho$ and $\pi b_{1}$ by Roman letters. For simplicity we will assume the $\pi_{1}$ spin wave function has only the $S=0$ $q\bar{q}g$ component; that gives the largest widths for both $\pi b_{1}$ and $\pi\rho$ modes. Thus our calculations should give the upper limit for FSI corrections. Let us introduce the matrix elements for the potential $V$:
\begin{equation}
V_{\alpha i}=\langle\alpha|V|j\rangle,\,\,V_{ij}=\langle i|V|j\rangle,
\end{equation}
and similarly for $T$. The elements $V_{\alpha i}$ are just the amplitudes of the corresponding decays of $\pi_{1}$, whereas $V_{ij}$ are related to the final state interaction potential.
In our state-space the Lippmann-Schwinger equation can be written as
\begin{eqnarray}
& & T_{\alpha i}=V_{\alpha i}+V_{\alpha j}G_{j}T_{ji}, \nonumber \\
& & T_{ji}=V_{ji}+V_{jk}G_{k}T_{ki}+V_{j\alpha}G_{\alpha}T_{\alpha i},
\label{LSE}
\end{eqnarray}
where 
\begin{equation}
G_{i}(E)=[E-H_{0}(i)+i\epsilon]^{-1}. 
\end{equation}
Both equations (\ref{LSE}) are represented diagrammatically in Fig.~\ref{diag2}.
A bold vertical solid line corresponds to the total amplitude of the interaction between two dimeson states, whereas a normal vertical solid line represents the sum over all intermediate states.  

\begin{figure}
\centering
\includegraphics[width=1.8in,angle=270]{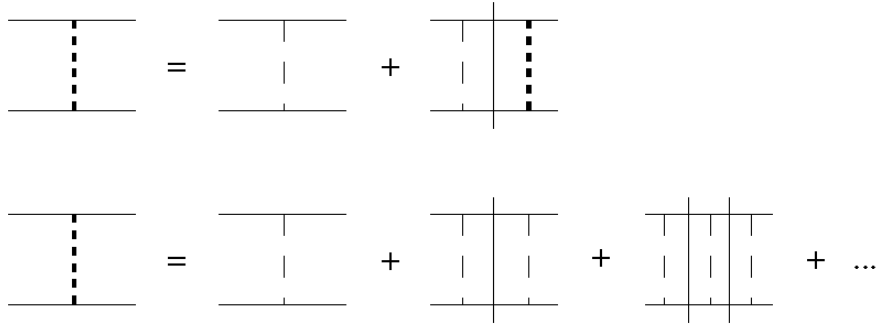}
\caption{\label{diag3} The Lippmann-Schwinger equation in a meson subspace. }
\end{figure}

Hereinafter, if an index appears twice or more, then summation over this index is implicitly assumed. The Hamiltonian of the state $|i\rangle$ is $H_{0}(i)$, $\epsilon\rightarrow 0^{+}$, and $E$ is the energy which will equal the mass of the $\pi_{1}$.
Calculating $T_{ij}$ from the second equation in ($\ref{LSE}$) and its substitution to the first one leads to
\begin{equation}
T_{\alpha i}=V_{\alpha i}+V_{\alpha j}G_{j}[1-VG]^{-1}_{jk}(V_{ki}+V_{k\alpha}G_{\alpha}T_{\alpha i}).
\end{equation}
Now we introduce the $T$ matrix acting only between states $|\pi\rho\rangle$ and $|\pi b_{1}\rangle$, denoted by $t$ and defined by
\begin{equation}
t=V+VGt,\,\,\,t_{ij}=[1-VG]^{-1}_{ik}V_{kj}.
\label{t}
\end{equation}
Diagrammatically this represents the sum of all diagrams without hybrid intermediate states, as shown in Fig.~\ref{diag3}.
Therefore
\begin{eqnarray}
& & T_{\alpha i}=V_{\alpha i}+V_{\alpha j}G_{j}t_{ji}+V_{\alpha j}G_{j}[1-VG]^{-1}_{jk}V_{k\alpha}G_{\alpha}T_{\alpha i}= \nonumber \\
& & =V_{\alpha i}+V_{\alpha j}G_{j}t_{ji}+V_{\alpha j}G_{j}[\delta_{jk}+(tG)_{jk}]V_{k\alpha}G_{\alpha}T_{\alpha i},
\end{eqnarray}
where we used $[1-VG]^{-1}=1+tG$ in the last term.

\begin{figure}
\centering
\includegraphics[width=1.0in,angle=270]{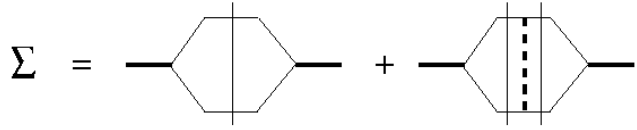}
\caption{\label{diag4} The self-energy of a hybrid. }
\end{figure}

\begin{figure}
\centering
\vspace{0.27in}
\includegraphics[width=1.0in,angle=270]{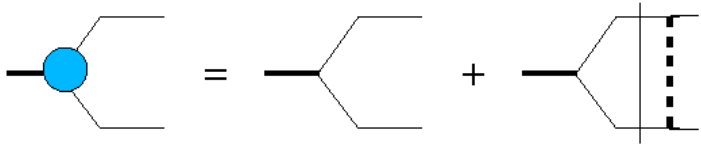}
\caption{\label{diag5} Renormalized FSI correction to the $\pi_{1}$ decay amplitude. }
\end{figure}

Using the identity $[1-GV]^{-1}=1+Gt$ on the first two terms on RHS leads to
\begin{equation}
T_{\alpha i}=(V[1-GV]^{-1})_{\alpha i}+[V(1+Gt)GV]_{\alpha\alpha}G_{\alpha}T_{\alpha i}.
\end{equation}
Solving for $T_{\alpha i}$ gives
\begin{equation}
T_{\alpha i}=[1-(V(1+Gt)GV)_{\alpha\alpha}G_{\alpha}]^{-1}(V[1-GV]^{-1})_{\alpha i}= G^{-1}_{\alpha}[G^{-1}_{\alpha}-\Sigma_{\alpha}]^{-1}(V[1-GV]^{-1})_{\alpha i},
\end{equation}
where $\Sigma_{\alpha}=(V(1+Gt)GV)_{\alpha\alpha}$ is the self-energy of the $\pi_{1}$, shown in Fig.~\ref{diag4}.
Both $G^{-1}_{\alpha}$ and $\Sigma_{\alpha}$ are numbers, and thus
\begin{equation}
T_{\alpha i}=\frac{E-m_{ex}}{E-m_{ex}-\Sigma_{\alpha}}(V+VGt)_{\alpha i},
\end{equation}
where we again used $[1-GV]^{-1}=1+Gt$.
Renormalization of the theory is related to cutting off the self-energy term. Therefore we obtain finally
\begin{equation}
T_{\alpha i}=V_{\alpha i}+V_{\alpha j}G_{j}t_{ji},
\label{T}
\end{equation}
with $t_{ji}$ defined by ($\ref{t}$).
This is equivalent to excluding an exotic intermediate state from the FSI corrections. Diagrammatic representation of Eq.~(\ref{T}) is given in Fig.~\ref{diag5}.
\vspace{0.3in}

Eq.~($\ref{t}$), written in a full notation in the rest frame of the $\pi_{1}$, is
\begin{equation}
t({\bf p},{\bf q},\lambda,\lambda')=V({\bf p},{\bf q},\lambda,\lambda')+\sum_{\lambda"}\int\frac{d^{3}{\bf k}}{(2\pi)^{3}4\omega_{1}(k)\omega_{2}(k)}V({\bf p},{\bf k},\lambda,\lambda")G(k)t({\bf k},{\bf q},\lambda",\lambda'),
\label{t2}
\end{equation}
where 
\begin{eqnarray}
& & \omega_{i}(k)=E(m_{i},{\bf k}),\,\,\,\,H_{0}(k)=\omega_{1}(k)+\omega_{2}(k), \nonumber \\
& & G(k)=[E-H_{0}(k)+i\epsilon]^{-1},
\end{eqnarray}
and
$m_{i}\,\,(i=1,2)$ are the meson masses in an intermediate two-meson state which is related to $G$.
In the above ${\bf p}$ and ${\bf q}$ are the relativistic relative momenta between two mesons, and the dependence on the center-of-mass momentum has been already separated.
Spins $\lambda$ refer to either $b_{1}$ or $\rho$.

We introduce the partial wave potentials
\begin{eqnarray}
& & V_{LL'}(p,q)=\sum_{M,M',\lambda,\lambda',j}\int d\Omega_{{\bf p}}d\Omega_{{\bf q}}\langle L,M;1,\lambda|J,j\rangle \langle L',M';1,\lambda'|J,j\rangle \nonumber \\
& & \times\,V({\bf p},{\bf q},\lambda,\lambda')Y_{LM}({\bf p})Y^{\ast}_{L'M'}({\bf q}),
\label{part}
\end{eqnarray}
where $d\Omega_{{\bf k}}$ is the element of the solid angle in the direction of the vector ${\bf k}$ and $k=|{\bf k}|$.
Similarly we define $t_{LL'}(p,q)$.
Substitution of $V_{LL'}(p,q)$ into ($\ref{t2}$) gives
\begin{equation}
t_{LL'}(p,q)=V_{LL'}(p,q)+\sum_{L"}\int\frac{k^{2}dk}{(2\pi)^{3}4\omega_{L",1}(k)\omega_{L",2}(k)}V_{LL"}(p,k)G_{L"}(k)t_{L"L'}(k,q),
\label{t3}
\end{equation}
where 
\begin{equation}
\omega_{L",i}(k)=E(m_{L",i},{\bf k}),\,\,\,\,G_{L"}(k)=[E-\omega_{L",1}(k)-\omega_{L",2}(k)+i\epsilon]^{-1}.
\end{equation}
In our state-space we can have $L=0,2$ (the relative angular momentum between $\pi$ and $b_{1}$) or $L=1$ (between $\pi$ and $\rho$). For a $\pi_{1}$ we also have $J=1$. Thus in Eq.~(\ref{t3}) we must substitute:
\begin{equation}
m_{L,1}=m_{\pi},\,\,\,\,m_{0,2}=m_{2,2}=m_{b_{1}},\,\,\,\,m_{1,2}=m_{\rho}.
\end{equation}
From the conservation of parity the only nonzero components of the final state interaction potential are $V_{01}$, $V_{10}$, $V_{12}$ and $V_{21}$. Moreover, from the CP conservation for strong interactions we have $V_{01}=V^{\ast}_{10}$ and $V_{12}=V^{\ast}_{21}$. The integral equation ($\ref{t3}$) cannot be solved in general analytically and one need to replace it with a set of matrix equations. The details will be given in the next section.
When this is done and $t_{LL'}$ are found, we may go back to Eq.~($\ref{T}$) which becomes
\begin{equation}
{\tilde{a}}_{L}(P)=a_{L}(P)+\sum_{L'}\int\frac{k^{2}dk}{(2\pi)^{3}(2J+1)4\omega_{1}(k)\omega_{2}(k)}a_{L'}(k)G(k)_{L'}t_{L'L}(k,P),
\label{T2}
\end{equation}
where $a_{L}(P)$ are the partial amplitudes defined in ($\ref{eq:ameb}$) and ($\ref{eq:amrf}$). In order to obtain the corrected widths, these amplitudes must be replaced by the corrected amplitudes ${\tilde{a}}_{L}(P)$ in Eq.~($\ref{eq:wid}$).

Finally, we proceed to the form of the final state interaction potential.
The Lagrangian of the $\rho\pi\omega$ vertex is
\begin{equation}
L_{\rho\pi\omega}=g_{\rho\pi\omega}\epsilon^{\mu\nu\lambda\sigma}\partial_{\mu}\omega_{\nu}\pi^{i}\partial_{\lambda}\rho^{i}_{\sigma},
\label{fsiL1}
\end{equation}
and for the $b_{1}\pi\omega$ vertex is
\begin{equation}
L_{b_{1}\pi\omega}=g_{b_{1}\pi\omega}\partial_{\mu}b_{1}^{\mu i}\pi^{i}\omega_{\mu},
\label{fsiL2}
\end{equation}
where $g_{\rho\pi\omega}$ and $g_{b_{1}\pi\omega}$ are the hadrodynamical coupling constants and the index $i$ corresponds to isospin. The first one lies between 0.01 MeV$^{-1}$ and 0.02 MeV$^{-1}$ and we will take a value 0.014 MeV$^{-1}$ \cite{KL}, whereas the second one can be obtained by calculating the width of the decay $b_{1}\rightarrow\pi\omega$ and comparing with its experimental value 142 MeV \cite{PDG}.
Consequently, the amplitude of $b_{1}\rightarrow\pi\omega$ is equal to
\begin{equation}
A({\bf P},\lambda_{b_{1}},\lambda_{\omega})=-g_{b_{1}\pi\omega}\epsilon^{i}(\lambda_{b_{1}})\epsilon^{i\ast}(\lambda_{\omega},{\bf P}),
\end{equation}
where ${\bf P}$ is given by
\begin{equation}
E(m_{\pi},{\bf P})+E(m_{\omega},{\bf P})=m_{b_{1}}.
\end{equation}
The width for the angular momentum $L$ between $\pi$ and $\omega$ is
\begin{equation}
\Gamma_{L}=\frac{P|A|^{2}}{8\pi(2L+1)m_{b_{1}}^{2}},
\end{equation}
where $|A|^{2}$ is summed over $\lambda_{\omega}$ and averaged with respect to $\lambda_{b_{1}}$. With the help of
\begin{equation}
\sum_{\lambda}\epsilon^{\mu}(\lambda,{\bf p})\epsilon^{\nu\ast}(\lambda,{\bf p})=-g^{\mu\nu}+\frac{p^{\mu}p^{\nu}}{p_{\rho}p^{\rho}},
\end{equation}
we have
\begin{equation}
|A|^{2}=g^{2}_{b_{1}\pi\omega}(1+\frac{P^{2}}{3m_{\omega}^{2}}),
\end{equation}
and from $\Gamma_{s-wave}=142$\,MeV$\cdot\,0.92=131$ MeV (the factor 0.92 comes from the partial wave D/S ratio) we get $g_{b_{1}\pi\omega}=3650$ MeV. Thus $g_{FSI}=g_{\rho\pi\omega}\cdot g_{b_{1}\pi\omega}$ is around 50.

The final state interaction potential can be obtained from Lagrangians ($\ref{fsiL1}$) and ($\ref{fsiL2}$), written in the momentum space and dressed with the instantaneous $\omega$ propagator,
\begin{equation}
V({\bf p},{\bf q},\lambda_{b_{1}},\lambda_{\rho})=g_{FSI}\epsilon_{\mu\nu\sigma\tau}p^{\mu}q^{\nu}\frac{1}{({\bf p}-{\bf q})^{2}+m^{2}_{\omega}}\epsilon^{\sigma}(\lambda_{b_{1}},-{\bf p})\epsilon^{\tau\ast}(\lambda_{\rho},{\bf q}),
\label{FSIpot}
\end{equation}
where ${\bf p}$ is the momentum of the $\pi$ in the $|\pi b_{1}\rangle$ state and ${\bf q}$ is the momentum of the $\rho$. For large values of $p$ and $q$ this potential grows with no limit which is a consequence of treating mesons as elementary particles. Therefore we must regulate this potential with an extra factor that tends to zero for large momenta. We choose an exponential function,
\begin{equation}
e^{-|{\bf p}_{\omega}|/\Lambda},
\end{equation}
of the quantity which is invariant under translations and rotations. Here $\Lambda$ is a scale parameter of order 1 GeV. This scale estimates a limit of treating mesons as elementary particles interacting via Lagrangians ($\ref{fsiL1}$) and ($\ref{fsiL2}$).

\section{Computational procedure}

The Lippmann-Schwinger integral equation (\ref{t3}) corresponds to outgoing wave boundary conditions. This means that the singularity of the term $G(k)$ is handled by giving the energy $E$ a small positive imaginary part $i\epsilon$. An integral of this form may be solved using the Cauchy principal-value prescription,
\begin{equation}
\int_{0}^{\infty}\frac{f(k)dk}{k-k_{0}+i\epsilon}=\wp\int_{0}^{\infty}\frac{f(k)dk}{k-k_{0}}-i\pi\,f(k_{0}), 
\label{Cau1}
\end{equation}  
where the principal value is defined by
\begin{equation}
\wp\int_{0}^{\infty}f(k)dk=\lim_{\epsilon\rightarrow0}\Bigl[\int_{0}^{k_{0}-\epsilon}f(k)dk+\int_{k_{0}+\epsilon}^{\infty}f(k)dk\Bigr],
\label{Cau2}
\end{equation}
with $k_{0}$ being the zero of the real function $f(k)$.
From the formula
\begin{equation}
\wp\int_{-\infty}^{+\infty}\frac{dk}{k-k_{0}}=0
\end{equation}
one obtains
\begin{equation}
\wp\int_{0}^{\infty}\frac{dk}{k^{2}-k^{2}_{0}}=0,
\end{equation}
that can be generalized to the so-called Hilbert transform of a function $f$:
\begin{equation}
\wp\int_{0}^{\infty}\frac{f(k)dk}{k^{2}-k^{2}_{0}}=\int_{0}^{\infty}\frac{[f(k)-f(k_{0})]dk}{k^{2}-k^{2}_{0}}. 
\label{hil}
\end{equation}
The integral in Eq.~(\ref{t3}) has a slightly different singular denominator term $[E-H_{0}(k)]^{-1}$, and in this case
\begin{eqnarray}
& & \int_{0}^{\infty}\frac{f(k)dk}{E-H_{0}(k)+i\epsilon}=\wp\int_{0}^{\infty}\frac{f(k)dk}{E-H_{0}(k)}\,-\,i\pi f(k_{0})\biggl(\frac{\partial H_{0}(k)}{\partial k}\biggr)^{-1}_{k=k_{0}}= \nonumber \\
& & =\wp\int_{0}^{\infty}\frac{f(k)dk}{k^{2}-k^{2}_{0}}\,\frac{k^{2}-k^{2}_{0}}{E-H_{0}(k)}\,-i\,\pi f(k_{0})\frac{\omega_{1}(k_{0})\omega_{2}(k_{0})}{k_{0}E}= \nonumber \\
& & =\int_{0}^{\infty}\frac{dk}{k^{2}-k^{2}_{0}}\biggl[\frac{f(k)(k^{2}-k^{2}_{0})}{E-H_{0}(k)}-f(k_{0})\lim_{k\rightarrow k_{0}}\frac{k^{2}-k^{2}_{0}}{E-H_{0}(k)}\biggr]\,-i\pi f(k_{0})\frac{\omega_{1}(k_{0})\omega_{2}(k_{0})}{k_{0}E}= \nonumber \\
& &  =\int_{0}^{\infty}\frac{dk}{k^{2}-k^{2}_{0}}\biggl[\frac{f(k)(k^{2}-k^{2}_{0})}{E-H_{0}(k)}-\frac{2f(k_{0})\omega_{1}(k_{0})\omega_{2}(k_{0})}{E}\biggr]-\,i\pi f(k_{0})\frac{\omega_{1}(k_{0})\omega_{2}(k_{0})}{k_{0}E}.
\end{eqnarray}
Making a transition
\begin{equation}
f(k)\rightarrow \frac{k^{2}V(p,k)t(k,q)}{4(2\pi)^{3}\omega_{1}(k)\omega_{2}(k)},
\end{equation}
and including the summation over partial waves, leads to a desired equation for $t_{LL'}(p,q)$ that no longer has a singularity:
\begin{eqnarray}
& & t_{LL'}(p,q)=V_{LL'}(p,q)+\sum_{L"}\int_{0}^{\infty}\frac{dk}{4(2\pi)^{3}(k^{2}-k^{2}_{0,L"})}\biggl[\frac{k^{2}(k^{2}-k^{2}_{0,L"})V_{LL"}(p,k)t_{L"L'}(k,q)}{[E-H_{0,L"}(k)]\omega_{L",1}(k)\omega_{L",2}(k)} \nonumber \\
& & +\frac{2k^{2}_{0,L"}}{E}V_{LL"}(p,k_{0,L"})t_{L"L'}(k_{0,L"},q)\biggr]\,-i\pi\sum_{L"}\frac{k_{0,L"}V_{LL"}(p,k_{0,L"})t_{L"L'}(k_{0,L"},q)}{4(2\pi)^{3}E},
\label{t4}
\end{eqnarray}
where 
\begin{equation}
H_{0,L}(k)=\omega_{L,1}(k)+\omega_{L,2}(k)
\end{equation}
and the quantities $k_{0,L}$ are defined by 
\begin{equation}
H_{0,L}(k_{0,L})=E.
\label{k0}
\end{equation}

Because in the decay $\pi_{1}\rightarrow\pi b_{1}$ the D-wave amplitude is negligible as compared to that of the S-wave, we may reduce our angular momentum space to $L=0,1$. Therefore we need only the expression for $V_{01}(p,q)$. From the definition (\ref{part}) and with the help of (\ref{app}) we obtain
\begin{equation}
V_{01}(p,q)=-\frac{i\sqrt{3}}{4\pi}\sum_{\lambda,\lambda'}\int d\Omega_{{\bf p}}d\Omega_{{\bf q}}V({\bf p},{\bf q},\lambda,\lambda')\epsilon^{ijk}\epsilon^{i\ast}(\lambda)\epsilon^{j}(\lambda')q^{k}/q,
\label{v01}
\end{equation}
where $V({\bf p},{\bf q},\lambda,\lambda')$ was given in (\ref{FSIpot}).
The S-matrix is obtained from the $t$ matrix via
\begin{equation}
S_{LL'}=\delta_{LL'}-i\frac{\sqrt{k_{0,L}k_{0,L'}}}{16\pi^{2}E}t_{LL'}(k_{0,L},k_{0,L'}),
\label{scat0}
\end{equation}
and for a $2\times2$ unitary matrix may be parametrized by scattering phase shifts $\delta_{0}$ and $\delta_{1}$,
\begin{equation}
S_{LL'}=\left( \begin{array}{cc}
\eta e^{2i\delta_{0}} & i\sqrt{1-\eta^{2}}e^{i(\delta_{0}+\delta_{1})} \\
i\sqrt{1-\eta^{2}}e^{i(\delta_{0}+\delta_{1})} & \eta e^{2i\delta_{1}} \end{array} \right),
\label{scat1}
\end{equation}
with $0\leq\eta\leq1$.
 
Eq.~(\ref{t4}) may be solved, as we already mentioned, by converting the integration over $k$ into the sum over $N$ integration points $k_{n},\,\,n=1,2,...\,N$ (determined by Gaussian quadratures) with weights $w_{n}$ \cite{num}: 
\begin{eqnarray}
& & t_{LL'}(p,q)=V_{LL'}(p,q)+\sum_{L",n}\frac{w_{n}}{4(2\pi)^{3}(k_{n}^{2}-k^{2}_{0,L"})}\biggl[\frac{k_{n}^{2}(k_{n}^{2}-k^{2}_{0,L"})V_{LL"}(p,k_{n})t_{L"L'}(k_{n},q)}{[E-H_{0,L"}(k_{n})]\omega_{L",1}(k_{n})\omega_{L",2}(k_{n})} \nonumber \\
& & +\frac{2k^{2}_{0,L"}}{E}V_{LL"}(p,k_{0,L"})t_{L"L'}(k_{0,L"},q)\biggr]\,-i\pi\sum_{L"}\frac{k_{0,L"}V_{LL"}(p,k_{0,L"})t_{L"L'}(k_{0,L"},q)}{4(2\pi)^{3}E}.
\label{nt1}
\end{eqnarray}
By taking $p,q$ equal to either $k_{n}$ or $k_{0,L}$ we can rewrite this equation as the $(2N+2)\times(2N+2)$ matrix equation
\begin{equation}
t_{mn}=V_{mn}+R_{mo}t_{on},\,\,\,\,m,n,o=1..2N+2,
\label{nt2}
\end{equation}
where 
\begin{eqnarray}
& V_{mn}=0 & \,\,m=1..N+1,\,\,n=1..N+1; \nonumber \\
& V_{mn}=V_{01}(l_{m},l_{n}) & \,\,m=1..N+1,\,\,n=N+2..2N+2; \nonumber \\
& V_{mn}=V_{01}(l_{m},l_{n}) & \,\,m=N+2..2N+2,\,\,n=1..N+1; \nonumber \\
& V_{mn}=0 & \,\,m=N+2..2N+2,\,\,n=N+2..2N+2 \end{eqnarray}
and
\begin{eqnarray}
& l_{n}=k_{n} & \,\,n=1..N; \nonumber \\
& l_{N+1}=k_{0,0}; \nonumber \\
& l_{n}=k_{n-N-1} & \,\,n=N+2..2N+1; \nonumber \\
& l_{2N+2}=k_{0,1}.
\end{eqnarray}
The matrix $R$ is defined by
\begin{equation}
R_{mn}=\delta_{mn}-V_{mn}D_{n}
\end{equation}
with no summation over $n$, where 
\begin{eqnarray}
& D_{n}=\frac{k^{2}_{n}w_{n}}{4(2\pi)^{3}\omega_{0,1}(l_{n})\omega_{0,2}(l_{n})[E-H_{0,0}(l_{n})]} & \,\,n=1..N; \nonumber \\
& D_{N+1}=\frac{k_{0,0}}{4(2\pi)^{3}E}\biggl[\sum_{j=n}^{N}\frac{2k_{0,0}w_{n}}{k^{2}_{n}-k^{2}_{0,0}}-i\pi\biggr]; \nonumber \\
& D_{n}=\frac{k^{2}_{n}w_{n}}{4(2\pi)^{3}\omega_{1,1}(l_{n})\omega_{1,2}(l_{n})[E-H_{0,1}(l_{n})]} & \,\,n=N+2..2N+1; \nonumber \\
& D_{2N+2}=\frac{k_{0,1}}{4(2\pi)^{3}E}\biggl[\sum_{n=1}^{N}\frac{2k_{0,1}w_{n}}{k^{2}_{n}-k^{2}_{0,1}}-i\pi\biggr].
\end{eqnarray}

Having inverted $R_{mn}$ and solved for $t_{mn}$, we subsitute it into (\ref{T2}) that in $2N+2$-dimensional space has the form
\begin{eqnarray}
& & {\tilde{a}}_{0}(k_{0,0})=a_{0}(k_{0,0})+\sum_{n=1}^{N+1}D_{n}a_{0}(l_{n})t_{n,N+1}+\sum_{n=N+2}^{2N+2}D_{n}a_{1}(l_{n})t_{n,N+1}, \nonumber \\
& & {\tilde{a}}_{1}(k_{0,1})=a_{1}(k_{0,1})+\sum_{n=1}^{N+1}D_{n}a_{0}(l_{n})t_{n,2N+2}+\sum_{n=N+2}^{2N+2}D_{n}a_{1}(l_{n})t_{n,2N+2}.
\label{nt3}
\end{eqnarray}
The quantities $k_{0,0}$ and $k_{0,1}$ are according to (\ref{k0}) the relative momenta of $\pi b_{1}$ and $\pi\rho$ systems, respectively. Thus Eq.~(\ref{nt3}) gives the FSI-corrected amplitudes of $\pi_{1}\rightarrow\pi b_{1}$ and $\pi_{1}\rightarrow\pi\rho$ as functions of the energy $E$ that is equal to $m_{ex}$.
The S-matrix in terms of $t_{mn}$ is
\begin{equation}
S_{ij}=\delta_{ij}-i\frac{\sqrt{l_{i(N+1)}l_{j(N+1)}}}{16\pi^{2}E}\,t_{i(N+1),j(N+1)},
\label{scat2}
\end{equation}
where $i,j=1,2$ refer to channels $\pi b_{1}$ and $\pi\rho$.

\section{Numerical results}

The partial wave potential (\ref{v01}) obtained from the $\pi b_{1}-\pi\rho$ final state interaction potential (\ref{FSIpot}) is not separable, i.e., cannot be represented as a product of functions of only one variable,
\begin{equation}
V(p,q)\neq f(p)g(q).
\end{equation}
This corresponds to non-locality of this potential in the position space and results from a finite structure of mesons. If they were elementary and therefore $V(p,q)$ was a separable potential, then Eq.~(\ref{t4}) could be solved analytically. In this section we will assume that this is the case and use it to test the accuracy of numeric results from the preceding section.

Consider the following symmetric and off-diagonal potential,
\begin{equation}
V_{LL'}(p,q)=\left( \begin{array}{cc}
0 & 1 \\
1 & 0 \end{array} \right)g(p)g(q).
\end{equation}
We seek a solution of (\ref{t4}) in the form
\begin{equation}
t_{LL'}(p,q)=\left( \begin{array}{cc}
\lambda_{00} & \lambda_{01} \\
\lambda_{01} & \lambda_{11} \end{array} \right)g(p)g(q).
\end{equation}
Substitution of the above matrices into (\ref{t4}) leads to
\begin{equation} 
\left( \begin{array}{cc}
\lambda_{00} & \lambda_{01} \\
\lambda_{01} & \lambda_{11} \end{array} \right)= \left( \begin{array}{cc}
0 & 1 \\
1 & 0 \end{array} \right)+\left( \begin{array}{cc}
I_{1}\lambda_{01} & I_{1}\lambda_{11} \\
I_{0}\lambda_{00} & I_{0}\lambda_{01} \end{array} \right),
\end{equation}
where the quantities $I_{0}$ and $I_{1}$ are given by the integrals:
\begin{eqnarray} 
& & I_{0}=\int_{0}^{\infty}\frac{k^{2}dk\,g^{2}(k)}{4(2\pi)^{3}\omega_{0,1}(k)\omega_{0,2}(k)[E-H_{0,0}(k)+i\epsilon]} \nonumber \\
& & I_{1}=\int_{0}^{\infty}\frac{k^{2}dk\,g^{2}(k)}{4(2\pi)^{3}\omega_{1,1}(k)\omega_{1,2}(k)[E-H_{0,1}(k)+i\epsilon]}.
\end{eqnarray}
For the $t$ matrix we obtain then
\begin{equation}
t_{LL'}(p,q)=\left( \begin{array}{cc}
I_{1} & 1 \\
1 & I_{0} \end{array} \right)\frac{g(p)g(q)}{1-I_{0}I_{1}},
\end{equation}
and the phase shifts can be computed from (\ref{scat0}) and (\ref{scat1}).
The integrals $I_{0}$ and $I_{1}$ may be easily calculated using the principal-value prescription described in the preceding section.

\begin{figure}
\centering
\includegraphics[width=2.5in,height=3.5in,angle=270]{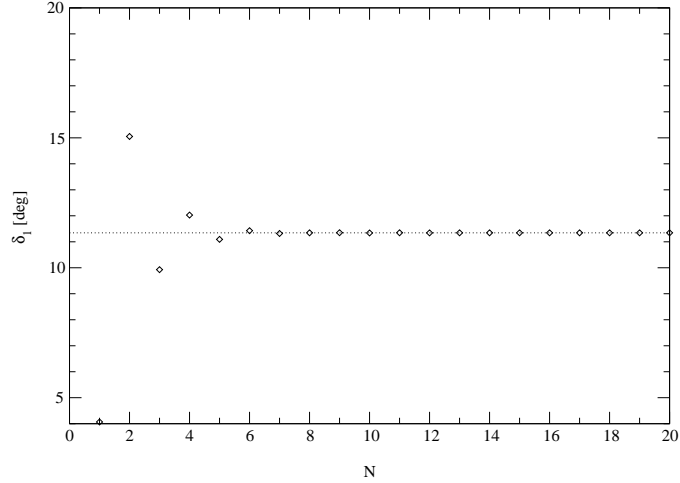}
\caption{\label{toy} Phase shift $\delta_{1}$ for a toy interaction $\pi b_{1}-\pi\rho$ as a function of the number of grid points $N$. }
\end{figure}

In Fig.~\ref{toy} we show the numerical values of the phase shift $\delta_{1}$ in degrees as a function of the number of grid points $N$, and compared to the corresponding analytical value. The toy potential was chosen such that
\begin{equation}
g(p)g(q)=-V_{0}e^{-(p^{2}+q^{2})/\mu^{2}},
\end{equation}
with $V_{0}=$1000 (the FSI potential is dimensionless) and $\mu=\,$1 GeV, whereas the energy $E=m_{ex}=\,$1600 MeV. We see that the numeric results for this phase shift converge pretty fast to a value that is in a good agreement with the analytical one. Accurate and stable numbers are obtained already for $N$ being approximately 10. 

\begin{figure}[ht]
\centering
\includegraphics[width=2.5in,height=3.5in,angle=270]{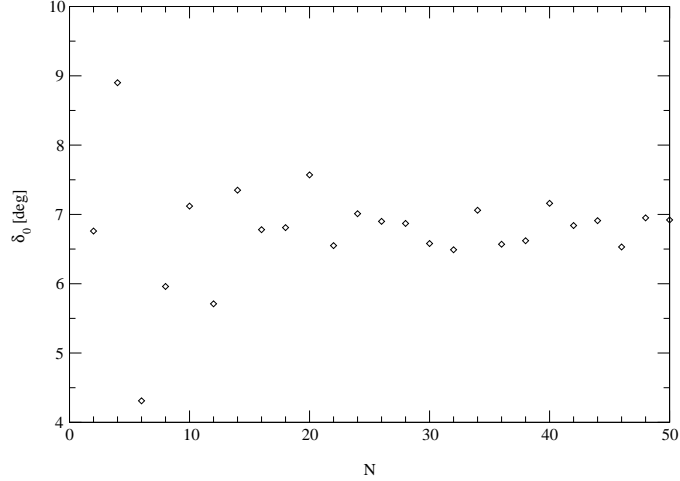}
\caption{\label{d0_N} Phase shift $\delta_{0}$ for the interaction $\pi b_{1}-\pi\rho$ as a function of the number of grid points $N$, for energy $E=1.6$ GeV and $g_{FSI}=200$. }
\end{figure}

\begin{figure}[ht]
\centering
\vspace{0.27in}
\includegraphics[width=2.5in,height=3.5in,angle=270]{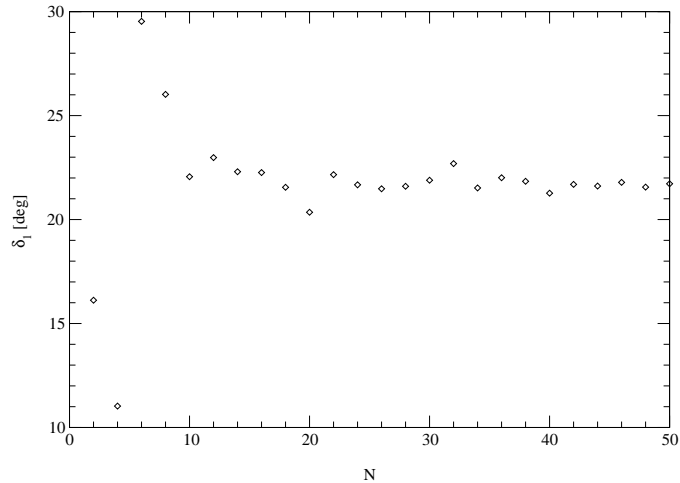}
\caption{\label{d1_N} Phase shift $\delta_{1}$ for the interaction $\pi b_{1}-\pi\rho$ as a function of the number of grid points $N$, for energy $E=1.6$ GeV and $g_{FSI}=200$. }
\end{figure}

Now we move to the original final state interaction and replace our toy potential with the real potential (\ref{v01}) obtained from Eq.~(\ref{FSIpot}). Fig.~\ref{d0_N} and Fig.~\ref{d1_N} show an increasing stability of phase shifts $\delta_{0}$ and $\delta_{1}$ (in degrees) for $g_{FSI}=\,$200 as the number of grid points $N$ grows. The default values in these plots are $E=\,$1600 MeV and $\Lambda=\,$2 GeV. We see that, for an unseparable potential, a larger $N$ is needed to obtain stable results. For $N=\,$20 the discrepancy is still about $1^{\circ}$. The unitarity condition is satisfied here with accuracy better than 0.1\%.

\begin{figure}[ht]
\centering
\includegraphics[width=2.5in,height=3.5in,angle=270]{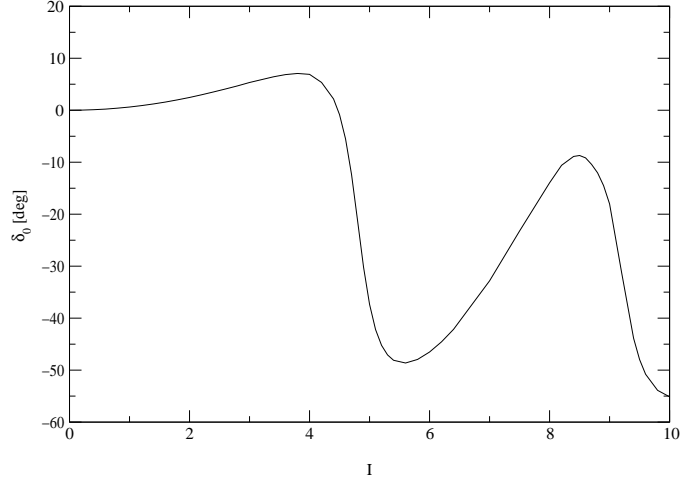}
\caption{\label{d0_v} Phase shift $\delta_{0}$ for the interaction $\pi b_{1}-\pi\rho$ as a function of the potential strength $I$, for energy $E=1.6$ GeV. }
\end{figure}

\begin{figure}[ht]
\centering
\vspace{0.27in}
\includegraphics[width=2.5in,height=3.5in,angle=270]{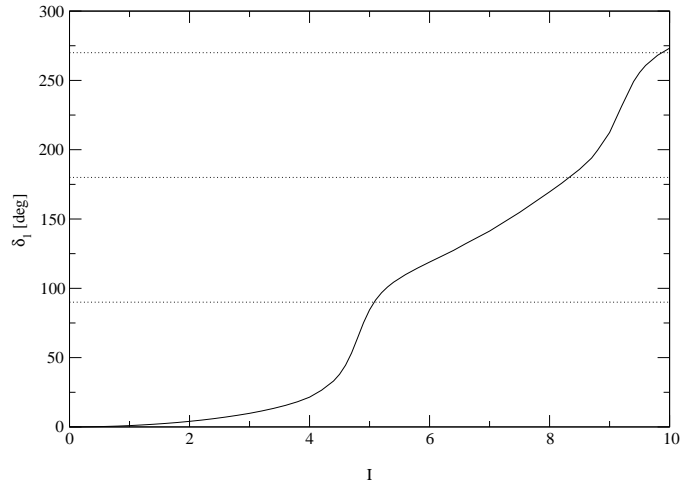}
\caption{\label{d1_v} Phase shift $\delta_{1}$ for the interaction $\pi b_{1}-\pi\rho$ as a function of the potential strength $I$, for energy $E=1.6$ GeV. }
\end{figure}

\begin{figure}[ht]
\centering
\includegraphics[width=2.5in,height=3.5in,angle=270]{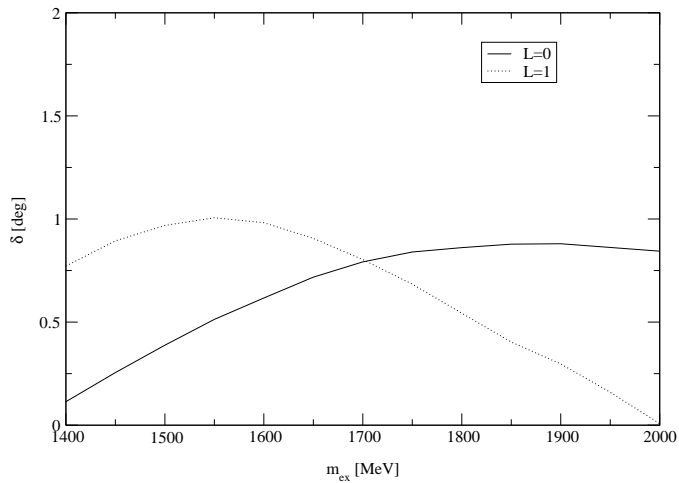}
\caption{\label{d01_mex} Phase shifts $\delta_{0}$ and $\delta_{1}$ for the interaction $\pi b_{1}-\pi\rho$ as functions of the energy $E$, for $I=1$. }
\end{figure}

Let us assume that the coupling constant $g_{FSI}$ is the variable and we introduce the potential strength $I=g_{FSI}/g^{(0)}_{FSI}$, where $g^{(0)}_{FSI}=50$. In Fig.~\ref{d0_v} and Fig.~\ref{d1_v} we present the phase shifts as functions of $I$. Their dependence on the energy $E$ for $I=1$ is shown in Fig.~\ref{d01_mex}.


Within an experimental range of the hybrid mass $m_{ex}=E$ and the hadrodynamical coupling constant $g$ the phase shifts for both channels $L=0$ ($\pi b_{1}$) and $L=1$ ($\pi\rho$) are rather small. Therefore one expects that final state interaction would not change much the widths of $\pi_{1}\rightarrow\pi b_{1}$ and $\pi_{1}\rightarrow\pi\rho$.    
 
\begin{figure}[ht]
\centering
\includegraphics[width=2.5in,height=3.5in,angle=270]{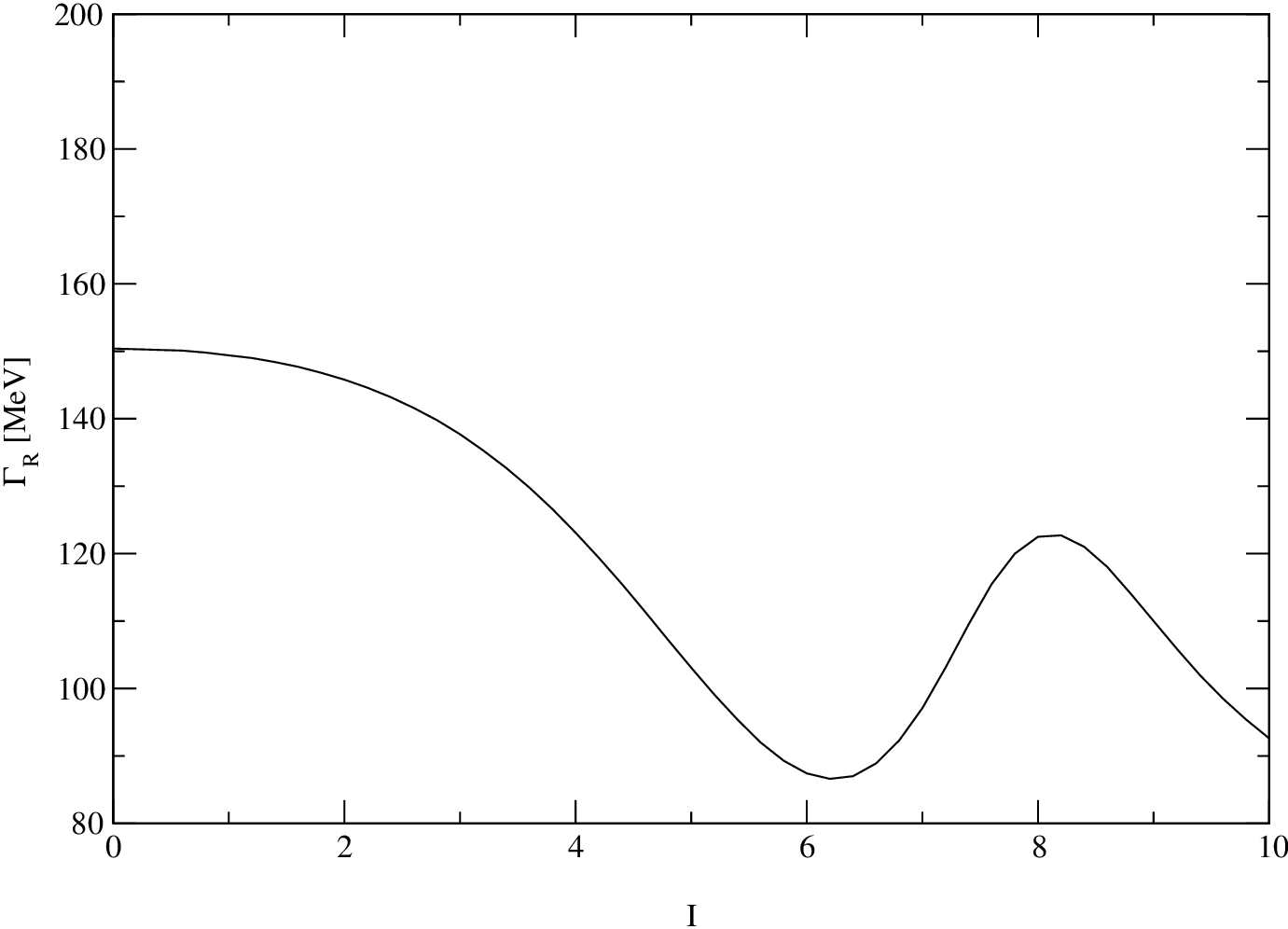}
\caption{\label{b1_fsi_v} FSI-corrected width of $\pi_{1}\rightarrow\pi b_{1}$ as a function of the potential strength $I$, for a $\pi_{1}$ with mass 1.6 GeV. }
\centering
\vspace{0.27in}
\includegraphics[width=2.5in,height=3.5in,angle=270]{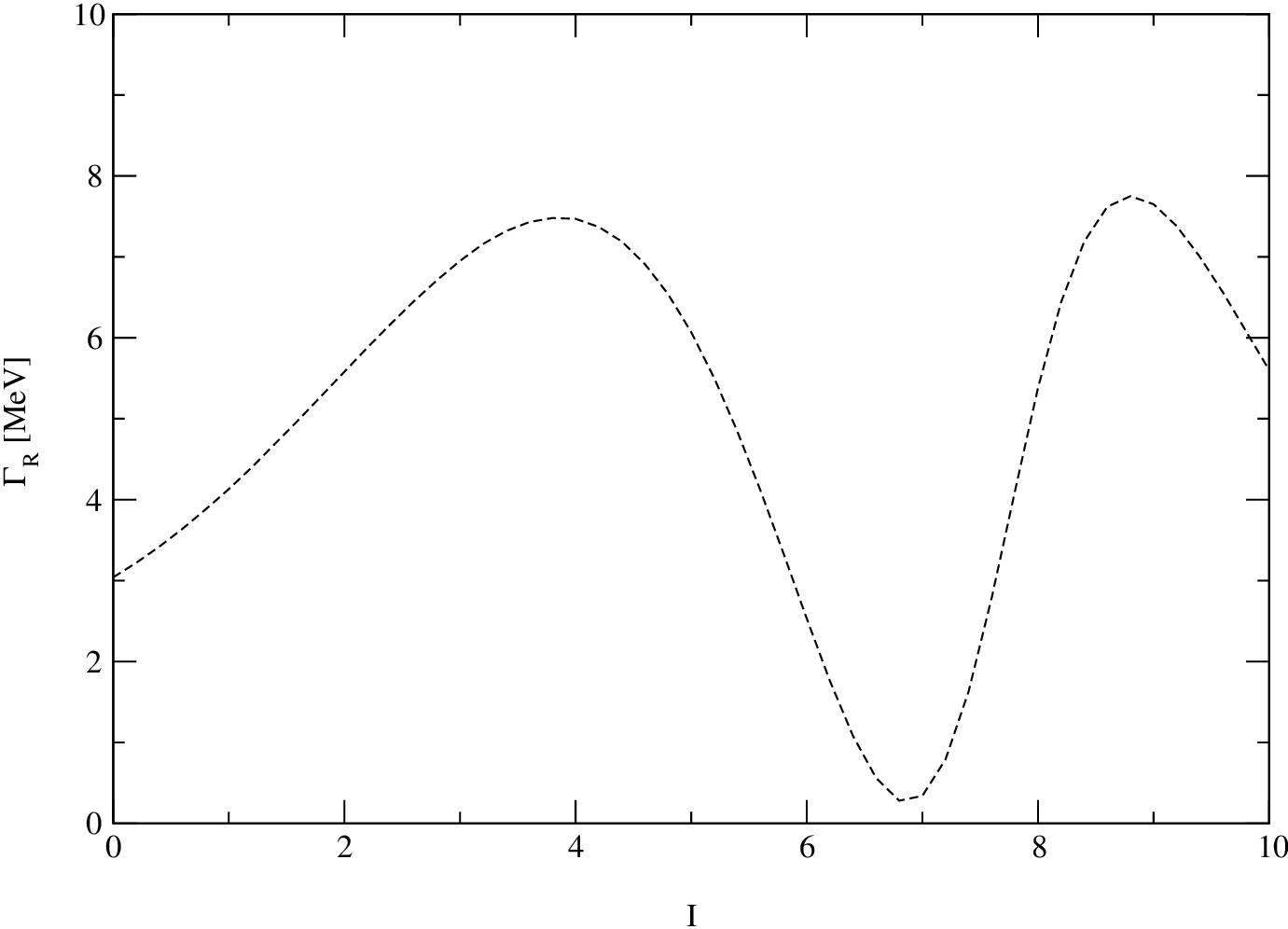}
\caption{\label{rho_fsi_v} FSI-corrected width of $\pi_{1}\rightarrow\pi\rho$ as a function of the potential strength $I$, for a $\pi_{1}$ with mass 1.6 GeV. }
\end{figure}

In Tables~\ref{fsi1} and~\ref{fsi2} we compare the original widths (in MeV) with the FSI-corrected ones for various values of $m_{ex}$ and $\Lambda$, for $I=2$. The strength was doubled in order to recompense damping caused by the regulating factor in the FSI potential. 
For $\Lambda$ other than 2 GeV, the coupling constant $g_{FSI}$ was renormalized such that the value of $V_{01}(p_{0},q_{0})$ remained equal to that for $\Lambda=$\,2 GeV. Here $p_{0}$ and $q_{0}$ are the physical values of the relative momenta between two mesons in the $\pi b_{1}$ and $\pi\rho$ channels, respectively. In Fig.~\ref{b1_fsi_v} and Fig.~\ref{rho_fsi_v} we show the width dependence on $I$ for $m_{ex}=1.6$\,GeV and $\Lambda=2$\,GeV. The behaviour of $\delta_{1}$ indicates the existence of the $\pi\rho$ resonances for the free parameters given (a resonance occurs when the phase increases over $n\cdot90^{\circ}$, where $n=1,2,3...$). 

The numerical results confirm our predictions. Only large FSI potentials may change the width values for both modes by more than few MeV. However, the $\pi\rho$ channel is subjected to a rather large relative correction. It may be caused by a big value of the original width for the $\pi b_{1}$ mode. In any event, the width for the $\pi_{1}\rightarrow\pi\rho$ is always on the order of a few MeV which agrees with \cite{CP} or \cite{PSS} and does not, for example, with \cite{CD}.

There is one constraint for the possible values of the cut-off parameter $\Lambda$ and the coupling constant $g_{FSI}$, given by the decay of the $\pi_{2}(1670)$. This meson almost does not decay into $\pi b_{1}$ but significantly does into $\pi\rho$. If the corrections from the final state interaction $\pi b_{1}\leftrightarrow\pi\rho$ were large, the $\pi b_{1}$ mode width could be increased by the $\pi\rho$ channel, resulting in a disagreement with experiment. Thus we arrive at the conclusion that FSI cannot introduce corrections that are too large. This agrees with our predictions for real $I$.

\begin{table}[ht]
\centering
\begin{tabular}{|c||r|r|r|r|}
\hline
$m_{ex}$\,[GeV] & no FSI & $\Lambda=0.5$\,GeV & $\Lambda=1$\,GeV & $\Lambda=2$\,GeV\\
\hline\hline
1.4 & 85 & 84 & 86 & 87\\
\hline
1.5 & 153 & 149 & 151 & 153\\
\hline
1.6 & 150 & 144 & 145 & 145\\
\hline
1.7 & 124 & 120 & 118 & 117\\
\hline
1.8 & 95 & 92 & 89 & 87\\
\hline
1.9 & 69 & 67 & 65 & 62\\
\hline
2.0 & 48 & 47 & 45 & 43\\
\hline
\end{tabular}
\caption{\label{fsi1} Original and FSI-corrected widths in MeV for the $\pi b_{1}$ mode of the $\pi_{1}$ decay, for various values of $m_{ex}$ and $\Lambda$. }
\end{table}

\begin{table}[ht]
\centering
\vspace{0.2in}
\begin{tabular}{|c||r|r|r|r|}
\hline
$m_{ex}$\,[GeV] & no FSI & $\Lambda=0.5$\,GeV & $\Lambda=1$\,GeV & $\Lambda=2$\,GeV\\
\hline\hline
1.4 & 3 & 8 & 8 & 8\\
\hline
1.5 & 3 & 6 & 7 & 7\\
\hline
1.6 & 3 & 4 & 5 & 6\\
\hline
1.7 & 2 & 2 & 3 & 4\\
\hline
1.8 & 2 & 1 & 2 & 3\\
\hline
1.9 & 2 & 1 & 1 & 2\\
\hline
2.0 & 1 & 1 & 1 & 1\\
\hline
\end{tabular}
\caption{\label{fsi2} Original and FSI-corrected widths in MeV for the $\pi\rho$ mode of the $\pi_{1}$ decay, for various values of $m_{ex}$ and $\Lambda$. }
\end{table}

\newpage

\chapter{Conclusions}

The quantitative description of the dynamics of gluons is complicated by the strong character of their interactions. Hadrons with excited gluonic degrees of freedom seem to be the only way to obtain information about the nature of low-energy gluons. Therefore they are of a great importance for our understanding of the quark-gluon interaction. In this work we focused on exotic mesons because they give insight into the dynamics of normal mesons.

The current candidates for the lightest exotic meson are $\pi_{1}(1400)$ and $\pi_{1}(1600)$. Most of the experimental resonances have large widths compared to normal mesons that decay strongly. Moreover, the observed decay channels should not be dominant according to the theory. Therefore, the reported signals do not completely agree with theoretical expectations and may originate from different phenomena such as rescattering. In order to resolve this problem it will be necessary to observe and measure exotic states with other quantum numbers. Searches for these states, as well as further exploration of the $\pi_{1}$ decays, are planned at Jefferson Lab (GlueX).
 
The dynamics of exotic mesons at the scale of $\sim1-2$ GeV should be described by a relativistic theory.
In this work we studied the size of relativistic effects in the decays of the $\pi_{1}$, and discussed a new picture of meson decays in the Coulomb gauge. Our considerations led to two important conclusions. 
First, numeric results showed significant relativistic corrections arising from the spin-orbit correlations introduced by Wigner rotation. 
The widths calculated using fully relativistic formulae are in general larger than the corresponding values calculated with no Wigner rotation (by a factor of order 10\%), but smaller than completely nonrelativistic values.
Some decays that are suppressed in the nonrelativistic limit (for example $\pi_{1}\rightarrow\pi\rho$ if the orbital wave functions of $\pi$ and $\rho$ are the same) acquire nonzero amplitudes in the relativistic case.

Second, the $\pi_{1}$ prefers to decay into two mesons, one of which has no orbital angular momentum and the other has $L=1$ (the $S+P$ selection rule).
Thus this selection rule, found in other models seems to be quite general.
Some decays ($\pi\eta$, $\rho\omega$, $K{\bar{K}}^{\ast}$) are suppressed by symmetries in the orbital wave functions, and the assumption that the parameters $\mu$ for mesons with the same radial quantum numbers should be almost equal.
We also noticed that the rates for the higher partial waves in decays of normal mesons where two waves are possible are larger in the $^3P_0$ then the $^3S_1$ model.

However, there is one problem which needs to be explored further. 
There are several components of the $q{\bar q}g$ normal and exotic meson wave functions, and in order to give accurate numerical predictions for the widths, one needs to know the relative amounts of these components. Since we were interested in relativistic effects and not in the absolute width values, we calculated the amplitudes for each component separately.  

We also studied the interaction between $\pi b_{1}$ and $\pi\rho$ final states. For the physical range of the parameters used in our analysis we observed rather small absolute corrections arising from FSI. The decay width for the process $\pi_{1}\rightarrow\pi b_{1}$ decreased only by a factor of order 1 MeV, whereas for the $\pi_{1}\rightarrow\pi\rho$ increased by a similar amount. Therefore, the latter process seems to be suppressed anyway.

The experiments have reported exotic resonances that are light and rather broad. This work showed that relativistic calculations give widths smaller than nonrelativistic ones, thus the discrepancy between theory and experiment becomes larger. For the dominant $\pi b_{1}$ mode of the $\pi_{1}(1600)$ decay the maximum width is $\sim200$ MeV as opposed to the experimental $\sim300$ MeV. Furthermore, the width in the $\pi\rho$ mode is on the order of a few MeV, and the modes $\pi\eta$ and $\pi\eta'$ are negligible. 

The experimental data for the $\pi_{1}$ exotic meson are not well understood yet. Further experiments are needed to clarify the nature of the reported signals and explain the dynamics of exotic mesons. This work showed that, in order to compare these data with the theory, one should use models which are relativistic.

\newpage
\thispagestyle{empty}
\singlespacing

\[ \]
\begin{center}
  \Large{CURRICULUM VITAE}\\
\end{center}

\begin{tabbing}
\\ \\
12\=123456789-12345678\= \kill

\>\bf{Personal} \\ \\
\>Name  :                \> Nikodem Janusz Poplawski \\ 
\>Date of Birth:         \> March 1, 1975 \\ 
\>Citizenship:           \> Polish \\  
\end{tabbing}
\begin{tabbing}
12\=123456789-12345678\= \kill

\>{\bf Education} \\ \\
\>Sep 29, 2004           \> PhD in Physics \\
\>                       \> Dissertation  -- \emph{A relativistic description of hadronic decays of the} \\
\>                       \> \emph{exotic meson $\pi_{1}$} \\
\>                       \> Supervisor -- Prof. Adam Szczepaniak
\end{tabbing}
\begin{tabbing}
12\=123456789-12345678\= \kill
\>Jan 31, 2004           \> MS in Physics 
\end{tabbing}
\begin{tabbing}
12\=123456789-12345678\= \kill
\>1999 -- 2004           \> Graduate studies -- Department of Physics, Indiana University, \\
\>                       \> Bloomington, IN
\end{tabbing}
\begin{tabbing}
12\=123456789-12345678\= \kill
\>Jul 14, 1999           \> MS in Astronomy \\
\>                       \> Thesis  -- \emph{A Michelson-Morley interferometer in the spacetime} \\
\>                       \> \emph{of a plane gravitational wave} \\
\>                       \> Supervisor -- Prof. Stanis\l aw Ba\.{z}a$\mathrm{\acute{n}}$ski
\end{tabbing}
\begin{tabbing}
12\=123456789-12345678\= \kill
\>1994 -- 1999           \> Undergraduate studies -- Center for Interfaculty Individual \\
\>                       \> Studies in Mathematical and Natural Sciences (MISMaP), \\
\>                       \> University of Warsaw, Poland 
\end{tabbing}
\begin{tabbing}
12\=123456789-12345678\= \kill
\>1993 -- 1994           \> Undergraduate studies -- Department of Physics, Jagiellonian \\
\>                       \> University, Cracow, Poland \\
\end{tabbing}
\begin{tabbing}
12\=123456789-12345678\= \kill

\>{\bf Work} \\ \\
\>2004 --                \> Postdoctoral Fellow -- Biocomplexity Institute, Indiana University \\
\>2001 -- 2004           \> Research Assistant -- Nuclear Theory Center, Indiana University \\
\>1999 -- 2001           \> Associate Instructor -- Department of Physics, Indiana University \\ 
\end{tabbing}

\end{document}